\documentclass[preprint,12pt,review]{elsarticle}

\usepackage{amssymb}
\usepackage{amsthm}
\usepackage{amsmath}
\usepackage{mathrsfs}
\usepackage{graphicx}
\usepackage{graphics}
\usepackage{epstopdf}
\usepackage{float}
\usepackage{caption}
\usepackage{subcaption}
\usepackage{bm}
\usepackage{bbm}
\usepackage{mathrsfs}
\usepackage{cleveref}
\usepackage{soul}
\usepackage{tikz} 
\usepackage{upgreek}
\usepackage{array}
\usepackage{nomencl}
\usepackage[margin=2cm]{geometry}
\usepackage{pdflscape}
\usepackage{adjustbox}
\usepackage{url}
\usepackage{makecell} 
\usetikzlibrary{shapes.geometric, arrows} 
\usepackage{adjustbox}
\biboptions{sort&compress}

\usepackage{nomencl}
\usepackage{framed} 

\tikzstyle{startstop} = [rectangle, rounded corners, minimum width=1cm, minimum height=1cm,text centered, draw=black, fill=red!5]
\tikzstyle{io} = [trapezium, trapezium left angle=70, trapezium right angle=110, minimum width=1cm, minimum height=1cm, text centered, draw=black, fill=blue!5]
\tikzstyle{process} = [rectangle, minimum width=1cm, minimum height=1cm, text centered, draw=black, fill=orange!5]
\tikzstyle{decision} = [diamond, minimum width=1cm, minimum height=0.5cm, text centered, draw=black, fill=green!5]
\tikzstyle{arrow} = [thick,->,>=stealth]

\journal{Engineering Fracture Mechanics}

\makeatletter
\def\@author#1{\g@addto@macro\elsauthors{\normalsize%
    \def\baselinestretch{1}%
    \upshape\authorsep#1\unskip\textsuperscript{%
      \ifx\@fnmark\@empty\else\unskip\sep\@fnmark\let\sep=,\fi
      \ifx\@corref\@empty\else\unskip\sep\@corref\let\sep=,\fi
      }%
    \def\authorsep{\unskip,\space}%
    \global\let\@fnmark\@empty
    \global\let\@corref\@empty  
    \global\let\sep\@empty}%
    \@eadauthor={#1}
}
\makeatother

\makeatletter
\def\thickhline{%
  \noalign{\ifnum0=`}\fi\hrule \@height \thickarrayrulewidth \futurelet
   \reserved@a\@xthickhline}
\def\@xthickhline{\ifx\reserved@a\thickhline
               \vskip\doublerulesep
               \vskip-\thickarrayrulewidth

             \fi
      \ifnum0=`{\fi}}
\makeatother

\newlength{\thickarrayrulewidth}
\begin{document}

\begin{frontmatter}



\title{A generalised framework for phase field-based modelling of coupled problems: application to thermo-mechanical fracture, hydraulic fracture, hydrogen embrittlement and corrosion}


\author{Yousef Navidtehrani\fnref{Uniovi}}
\author{Covadonga Beteg\'{o}n \fnref{Uniovi}}
\author{Emilio Mart\'{\i}nez-Pa\~neda\corref{cor1}\fnref{OXFORD}}
\ead{emilio.martinez-paneda@eng.ox.ac.uk}

\address[Uniovi]{Department of Construction and Manufacturing Engineering, University of Oviedo, Gij\'{o}n 33203, Spain}

\address[OXFORD]{Department of Engineering Science, University of Oxford, Oxford OX1 3PJ, UK}

\cortext[cor1]{Corresponding author.}

\begin{abstract}
We present a novel, generalised formulation to treat coupled structural integrity problems by combining phase field and multi-physics modelling. The approach exploits the versatility of the heat transfer equation and is therefore well suited to be adopted in commercial finite element packages, requiring only integration point-level implementation. This aspect is demonstrated here by implementing coupled, multi-variable phenomena through simple \texttt{UMAT} and \texttt{UMATHT} subroutines in the finite element package \texttt{Abaqus}. The generalised theoretical and computational framework presented is particularised to four problems of engineering and scientific relevance: thermo-mechanical fracture, hydraulic fracture, hydrogen-assisted cracking and metallic corrosion. 2D and 3D problems are considered. The results reveal a very good agreement with experimental data, and existing numerical and analytical solutions.The user subroutines developed are made freely available at \url{https://mechmat.web.ox.ac.uk/codes}.\\ 
\end{abstract}



\end{frontmatter}

\begin{framed}

\begin{thenomenclature} 
\nomgroup{A}
  \item [{$\alpha_b$}]\begingroup Biot's coefficient\nomeqref {0}\nompageref{1}
  \item [{$\alpha_{\text{r}}$}]\begingroup Biot's coefficient in the reservoir domain\nomeqref {0}\nompageref{1}
  \item [{$\alpha_{T}$}]\begingroup Thermal expansion coefficient\nomeqref {0}\nompageref{1}
  \item [{$\bm\epsilon_i^t$}]\begingroup Principal tensile strain tensor\nomeqref {0}\nompageref{1}
  \item [{$\bm{\epsilon}$}]\begingroup Principal strain tensor\nomeqref {0}\nompageref{1}
  \item [{$\bm{\sigma}$}]\begingroup Stress tensor\nomeqref {0}\nompageref{1}
  \item [{$\bm{\sigma}^{\text{eff}}$}]\begingroup Effective stress tensor\nomeqref {0}\nompageref{1}
  \item [{$\bm{\sigma}_0$}]\begingroup Stress tensor in the undamaged configuration\nomeqref {0}\nompageref{1}
  \item [{$\bm{\varepsilon}$}]\begingroup Strain tensor\nomeqref {0}\nompageref{1}
  \item [{$\bm{\varepsilon}_e$}]\begingroup Elastic strain tensor\nomeqref {0}\nompageref{1}
  \item [{$\bm{\varepsilon}_p$}]\begingroup Plastic strain tensor\nomeqref {0}\nompageref{1}
  \item [{$\bm{\varepsilon}_{T}$}]\begingroup Thermal strain tensor\nomeqref {0}\nompageref{1}
  \item [{$\bm{a}$}]\begingroup Transpose of the direction cosines matrix for the principal directions\nomeqref {0}\nompageref{1}
  \item [{$\bm{C}$}]\begingroup Elasticity tensor\nomeqref {0}\nompageref{1}
  \item [{$\bm{C}^{'}$}]\begingroup Tangential stiffness matrix in the principal direction\nomeqref {0}\nompageref{1}
  \item [{$\bm{C}_0$}]\begingroup Elastic stiffness tensor in the undamaged configuration\nomeqref {0}\nompageref{1}
  \item [{$\bm{C}_1^{\mathrm{M}}$}]\begingroup Elastic tensor of the first phase\nomeqref {0}\nompageref{1}
  \item [{$\bm{C}_2^{\mathrm{M}}$}]\begingroup Elastic tensor of the second phase\nomeqref {0}\nompageref{1}
  \item [{$(\bm{C}_2^{\mathrm{M}})'$}]\begingroup Material Jacobian in the principal direction of the second phase\nomeqref {0}\nompageref{1}  
  \item [{$\bm{I}$}]\begingroup Identity tensor\nomeqref {0}\nompageref{1}
  \item [{$\bm{K}$}]\begingroup Stiffness matrix\nomeqref {0}\nompageref{1}
  \item [{$\bm{K}_{\text{fl}}$}]\begingroup Permeability tensor\nomeqref {0}\nompageref{1}
  \item [{$\bm{K}_{\text{f}}$}]\begingroup Permeability tensor in the fracture domain\nomeqref {0}\nompageref{1}
  \item [{$\bm{K}_{\text{r}}$}]\begingroup Permeability tensor in the reservoir domain\nomeqref {0}\nompageref{1}
  \item [{$\boldsymbol{v}_i$}]\begingroup The $i$th of the principal vectors of the strain tensor\nomeqref {0}\nompageref{1}
  \item [{$\chi_{\mathrm{H}}$}]\begingroup Hydrogen damage coefficient\nomeqref {0}\nompageref{1}
  \item [{$\chi_{\text{f}}$}]\begingroup Fracture domain indicator field\nomeqref {0}\nompageref{1}
  \item [{$\chi_{\text{r}}$}]\begingroup Reservoir domain indicator field\nomeqref {0}\nompageref{1}
  \item [{$\Delta g_b^0$ }]\begingroup Gibbs free energy\nomeqref {0}\nompageref{1}
  \item [{$\delta_{ij}$}]\begingroup Kronecker delta\nomeqref {0}\nompageref{1}
  \item [{$\ell$}]\begingroup Phase field fracture length scale\nomeqref {0}\nompageref{1}
  \item [{$\ell_m$}]\begingroup Interface thickness\nomeqref {0}\nompageref{1}
  \item [{$\gamma$}]\begingroup Interface energy\nomeqref {0}\nompageref{1}
  \item [{$\mathbf{b}$}]\begingroup Body force vector\nomeqref {0}\nompageref{1}
  \item [{$\mathbf{f}$}]\begingroup Heat flux vector\nomeqref {0}\nompageref{1}
  \item [{$\mathbf{f}_{\xi}$}]\begingroup Flux vector of the diffusion field\nomeqref {0}\nompageref{1}
  \item [{$\mathbf{J}$}]\begingroup Flux of metal ions\nomeqref {0}\nompageref{1}
  \item [{$\mathbf{J}_{\mathrm{H}}$}]\begingroup Flux of hydrogen atoms\nomeqref {0}\nompageref{1}
  \item [{$\mathbf{R}$}]\begingroup Residual vector\nomeqref {0}\nompageref{1}
  \item [{$\mathbf{T}$}]\begingroup Surface traction\nomeqref {0}\nompageref{1}
  \item [{$\mathbf{u}$}]\begingroup Displacement vector\nomeqref {0}\nompageref{1}
  \item [{$\mathbf{v}_{\text{fl}}$}]\begingroup Fluid velocity vector\nomeqref {0}\nompageref{1}
  \item [{$\mathcal{F}_{\ell}$}]\begingroup Regularised energy functional\nomeqref {0}\nompageref{1}
  \item [{$\mathcal{H}$}]\begingroup History field in phase field fracture\nomeqref {0}\nompageref{1}
  \item [{$\mu$}]\begingroup Chemical potential of metal ions\nomeqref {0}\nompageref{1}
  \item [{$\mu_{\mathrm{H}}$}]\begingroup Chemical potential of hydrogen atoms\nomeqref {0}\nompageref{1}
  \item [{$\mu_{\text{fl}}$}]\begingroup Fluid viscosity\nomeqref {0}\nompageref{1}
  \item [{$\nabla^2$}]\begingroup Laplace operator\nomeqref {0}\nompageref{1}
  \item [{$\nu$}]\begingroup Poisson's ratio\nomeqref {0}\nompageref{1}
  \item [{$\Omega$}]\begingroup Domain of the system\nomeqref {0}\nompageref{1}
  \item [{$\omega$}]\begingroup Height of the double-well potential\nomeqref {0}\nompageref{1}
  \item [{$\Omega_{\text{f}}$}]\begingroup Fracture domain\nomeqref {0}\nompageref{1}
  \item [{$\Omega_{\text{r}}$}]\begingroup Reservoir domain\nomeqref {0}\nompageref{1}
  \item [{$\Omega_{\text{t}}$}]\begingroup Transition domain\nomeqref {0}\nompageref{1}
  \item [{$\partial \Omega$}]\begingroup Boundary of the domain\nomeqref {0}\nompageref{1}
  \item [{$\psi^+_0$}]\begingroup Positive part of the undamaged strain energy density\nomeqref {0}\nompageref{1}
  \item [{$\psi^-_0$}]\begingroup Negative part of the undamaged strain energy density\nomeqref {0}\nompageref{1}
  \item [{$\psi^{\mathrm{ch}}$}]\begingroup Chemical free energy density\nomeqref {0}\nompageref{1}
  \item [{$\psi^{\mathrm{ch}}_{\mathrm{L}}$}]\begingroup Chemical free energy density of the liquid phase\nomeqref {0}\nompageref{1}
  \item [{$\psi^{\mathrm{ch}}_{\mathrm{S}}$}]\begingroup Chemical free energy density of the solid phase\nomeqref {0}\nompageref{1}
  \item [{$\psi^{\mathrm{M}}$}]\begingroup Strain energy density\nomeqref {0}\nompageref{1}
  \item [{$\psi^{\mathrm{M}}_1$}]\begingroup Strain energy density of the first phase\nomeqref {0}\nompageref{1}
  \item [{$\psi^{\mathrm{M}}_2$}]\begingroup Strain energy density of the second phase\nomeqref {0}\nompageref{1}
  \item [{$\psi^{\mathrm{M}}_\mathrm{S}$}]\begingroup Undamaged strain energy of solid phase\nomeqref {0}\nompageref{1}
  \item [{$\psi_\text{S}^e$}]\begingroup Elastic energy density of the solid phase\nomeqref {0}\nompageref{1}
  \item [{$\psi_\text{S}^p$}]\begingroup Plastic energy density of the solid phase\nomeqref {0}\nompageref{1}
  \item [{$\rho$}]\begingroup Mass density\nomeqref {0}\nompageref{1}
  \item [{$\rho_{\text{fl}}$}]\begingroup Mass density of the fluid\nomeqref {0}\nompageref{1}
  \item [{$\sigma_f$}]\begingroup Flow stress\nomeqref {0}\nompageref{1}
  \item [{$\sigma_h$}]\begingroup Hydrostatic stress\nomeqref {0}\nompageref{1}
  \item [{$\sigma_y$}]\begingroup Yield stress\nomeqref {0}\nompageref{1}
  \item [{$\theta$}]\begingroup Hydrogen coverage\nomeqref {0}\nompageref{1}
  \item [{$\varepsilon^p$}]\begingroup Equivalent plastic strain\nomeqref {0}\nompageref{1}
  \item [{$\varepsilon_f$}]\begingroup Critical strain for film rupture\nomeqref {0}\nompageref{1}
  \item [{$\varepsilon_{\text{vol}}$}]\begingroup Volumetric strain\nomeqref {0}\nompageref{1}
  \item [{$\xi$}]\begingroup Diffusion field\nomeqref {0}\nompageref{1}
  \item [{$\zeta_{\text{fl}}$}]\begingroup Mass fluid content\nomeqref {0}\nompageref{1}
  \item [{$A$}]\begingroup Curvature of the free energy density\nomeqref {0}\nompageref{1}
  \item [{$a_0$}]\begingroup Length of the initial crack\nomeqref {0}\nompageref{1}
  \item [{$b$}]\begingroup Transient parameter\nomeqref {0}\nompageref{1}
  \item [{$c$}]\begingroup Normalised concentration of metal ions\nomeqref {0}\nompageref{1}
  \item [{$c_1$}]\begingroup First constant for domain indicator fields\nomeqref {0}\nompageref{1}
  \item [{$c_2$}]\begingroup Second constant for domain indicator fields\nomeqref {0}\nompageref{1}
  \item [{$c_m$}]\begingroup Concentration of dissolved ions\nomeqref {0}\nompageref{1}
  \item [{$c_{\mathrm{H}}$}]\begingroup Hydrogen concentration\nomeqref {0}\nompageref{1}
  \item [{$c_{\text{env}}$}]\begingroup Environmental hydrogen concentration\nomeqref {0}\nompageref{1}
  \item [{$C_{\text{fl}}$}]\begingroup Fluid compressibility\nomeqref {0}\nompageref{1}
  \item [{$c_{\text{Le}}$}]\begingroup Normalised equilibrium concentration for the liquid phase\nomeqref {0}\nompageref{1}
  \item [{$c_{\text{L}}$}]\begingroup Normalized concentration of the liquid phase\nomeqref {0}\nompageref{1}
  \item [{$c_{\text{sat}}$}]\begingroup Saturation concentration\nomeqref {0}\nompageref{1}
  \item [{$c_{\text{Se}}$}]\begingroup Normalised equilibrium concentration for the solid phase\nomeqref {0}\nompageref{1}
  \item [{$c_{\text{solid}}$}]\begingroup Concentration of atoms in the metal \nomeqref {0}\nompageref{1}
  \item [{$c_{\text{S}}$}]\begingroup Normalized concentration of the solid phase\nomeqref {0}\nompageref{1}
  \item [{$c_{T}$}]\begingroup Specific heat\nomeqref {0}\nompageref{1}
  \item [{$D_m$}]\begingroup Diffusion coefficient of metal ions\nomeqref {0}\nompageref{1}
  \item [{$D_{\mathrm{H}}$}]\begingroup Diffusion coefficient for hydrogen transport\nomeqref {0}\nompageref{1}
  \item [{$E$}]\begingroup Young's modulus\nomeqref {0}\nompageref{1}
  \item [{$E'$}]\begingroup Young's modulus for plane strain\nomeqref {0}\nompageref{1}
  \item [{$f_{\text{b1}}$}]\begingroup Bulk free energy density of the first phase\nomeqref {0}\nompageref{1}
  \item [{$f_{\text{b2}}$}]\begingroup Bulk free energy density of the second phase\nomeqref {0}\nompageref{1}
  \item [{$f_{\xi}$}]\begingroup Diffusion field flux\nomeqref {0}\nompageref{1}
  \item [{$G_c$}]\begingroup Critical energy release rate\nomeqref {0}\nompageref{1}
  \item [{$i$}]\begingroup Corrosion current density\nomeqref {0}\nompageref{1}
  \item [{$i_0$}]\begingroup Corrosion current density of the bare metal\nomeqref {0}\nompageref{1}
  \item [{$K$}]\begingroup Bulk modulus\nomeqref {0}\nompageref{1}
  \item [{$k$}]\begingroup Film stability coefficient\nomeqref {0}\nompageref{1}
  \item [{$k_0$}]\begingroup Thermal conductivity of the pristine material\nomeqref {0}\nompageref{1}
  \item [{$k_{\mathrm{m}}$}]\begingroup Mechanocorrosion coefficient\nomeqref {0}\nompageref{1}
  \item [{$k_{T}$}]\begingroup Thermal conductivity\nomeqref {0}\nompageref{1}
  \item [{$L$}]\begingroup Interface kinetics coefficient\nomeqref {0}\nompageref{1}
  \item [{$L_0$}]\begingroup Reference interface kinetics coefficient\nomeqref {0}\nompageref{1}
  \item [{$M$}]\begingroup Mobility coefficient\nomeqref {0}\nompageref{1}
  \item [{$N$}]\begingroup Strain hardening exponent\nomeqref {0}\nompageref{1}
  \item [{$N_i$}]\begingroup shape functions of node $i$\nomeqref {0}\nompageref{1}
  \item [{$n_{\text{pr}}$}]\begingroup Porosity in the reservoir domain\nomeqref {0}\nompageref{1}
  \item [{$n_{\text{p}}$}]\begingroup Porosity\nomeqref {0}\nompageref{1}
  \item [{$p$}]\begingroup Fluid pressure\nomeqref {0}\nompageref{1}
  \item [{$p_c$}]\begingroup Critical fluid pressure\nomeqref {0}\nompageref{1}
  \item [{$q_m$}]\begingroup External fluid source\nomeqref {0}\nompageref{1}
  \item [{$q_{\xi}$}]\begingroup Flux of the diffusion equation\nomeqref {0}\nompageref{1}
  \item [{$q_{\xi}$}]\begingroup Flux of the diffusion field\nomeqref {0}\nompageref{1}
  \item [{$q_{T}$}]\begingroup Heat source\nomeqref {0}\nompageref{1}
  \item [{$R$}]\begingroup Gas constant\nomeqref {0}\nompageref{1}
  \item [{$r$}]\begingroup Heat source\nomeqref {0}\nompageref{1}
  \item [{$S$}]\begingroup Storage coefficient\nomeqref {0}\nompageref{1}
  \item [{$T$}]\begingroup Temperature\nomeqref {0}\nompageref{1}
  \item [{$t$}]\begingroup Time\nomeqref {0}\nompageref{1}
  \item [{$T_0$}]\begingroup Initial temperature\nomeqref {0}\nompageref{1}
  \item [{$t_0$}]\begingroup Time interval before corrosion decay begins in a repassivated metal\nomeqref {0}\nompageref{1}
  \item [{$T_\mathrm{k}$}]\begingroup Absolute temperature\nomeqref {0}\nompageref{1}
  \item [{$T_a$}]\begingroup Ambient temperature\nomeqref {0}\nompageref{1}
  \item [{$t_f$}]\begingroup Drop time during a film rupture event\nomeqref {0}\nompageref{1}
  \item [{$U$}]\begingroup Internal heat energy\nomeqref {0}\nompageref{1}
  \item [{$U_{\xi}$}]\begingroup Internal energy of diffusion equation\nomeqref {0}\nompageref{1}
  \item [{$V_m$}]\begingroup Partial molar volume\nomeqref {0}\nompageref{1}
  \item [{$V_{\mathrm{H}}$}]\begingroup Partial molar volume of hydrogen\nomeqref {0}\nompageref{1}
  \item [{$V_{\text{b}}$}]\begingroup Bulk volume of porous medium\nomeqref {0}\nompageref{1}
  \item [{$V_{\text{p}}$}]\begingroup Volume of pores in porous medium\nomeqref {0}\nompageref{1}
  \item [{\(\beta_i\)}]\begingroup Representative field variable\nomeqref {0}\nompageref{1}
  \item [{\(\eta\)}]\begingroup Relaxation time constant\nomeqref {0}\nompageref{1}
  \item [{\(\kappa\)}]\begingroup Gradient energy coefficient\nomeqref {0}\nompageref{1}
  \item [{\(\mathbf{B}\)}]\begingroup Set of field variables\nomeqref {0}\nompageref{1}
  \item [{\(\mathbf{n}\)}]\begingroup Outward unit normal vector\nomeqref {0}\nompageref{1}
  \item [{\(\mathcal{F}\)}]\begingroup Free energy\nomeqref {0}\nompageref{1}
  \item [{\(\mathcal{F}_{\text{bulk}}\)}]\begingroup Bulk free energy\nomeqref {0}\nompageref{1}
  \item [{\(\mathcal{F}_{\text{int}}\)}]\begingroup Interface free energy\nomeqref {0}\nompageref{1}
  \item [{\(\nabla\)}]\begingroup Gradient operator in a spatial frame\nomeqref {0}\nompageref{1}
  \item [{\(\phi\)}]\begingroup Phase field\nomeqref {0}\nompageref{1}
  \item [{\(f_{\text{b}}\)}]\begingroup Bulk free energy density\nomeqref {0}\nompageref{1}
  \item [{\(f_{\text{int}}\)}]\begingroup Interface free energy density\nomeqref {0}\nompageref{1}
  \item [{\(g\)}]\begingroup Degradation function\nomeqref {0}\nompageref{1}
  \item [{\(V_1\)}]\begingroup Volume occupied by the first phase\nomeqref {0}\nompageref{1}
  \item [{\(V_{\text{RVE}}\)}]\begingroup Volume of a representative volume element\nomeqref {0}\nompageref{1}
  \item [{\(w\)}]\begingroup Double-well potential\nomeqref {0}\nompageref{1}

\end{thenomenclature}

\end{framed}

\section{Introduction}
\label{sec:Introduction}

Phase field models are enjoying remarkable popularity. Grounded on the foundational work by John W. Cahn and John E. Hilliard \cite{cahn1958free}, the phase field paradigm exploits a diffuse representation of otherwise sharp interfaces to capture complex morphologies and transitions, based on variational principles. The versatility of phase field models has led to their widespread adoption across science and engineering disciplines, including solidification and phase transformations \cite{wheeler1992phase}, general microstructural evolution problems \cite{Chen2002}, voiding in all-solid-state batteries \cite{zhao2022phase}, and fluid-structure interactions \cite{MOKBEL2018823}. This success has also reached the discipline of structural integrity, with phase field formulations opening new horizons in the modelling of fracture mechanics \cite{Bourdin2000,Bourdin2008,Karma2001} and corrosion \cite{Mai2016,JMPS2021}. In the case of phase field fracture models, the phase field order parameter $\phi$ regularises the crack - undamaged material interface, while for the corrosion ones, $\phi$ describes the evolution of the corrosion front (i.e., the corrosive electrolyte - metal interface). Corrosion and fracture phase field models have been developed independently although we will show here how both classes of phase field models can be encapsulated within a generalised framework. The success of phase field models in the area of structural integrity has been notable, spanning nearly all engineering and natural materials; these include fibre-reinforced composites \cite{Quintanas-Corominas2019,CST2021}, shape memory materials \cite{CMAME2021,lotfolahpour2023phase}, metals \cite{Wu2021,shishvan2021mechanism}, ice-sheets \cite{Sun2021,Clayton2022}, rocks \cite{Schuler2020,navidtehrani2024damage,AHMADIAN2024110417}, concrete \cite{narayan2019gradient,korec2024predicting}, and functionally graded materials \cite{CPB2019,Kumar2021}.\\

One of the key strengths of phase field models is their seamless integration into coupled multi-variable problems, as the phase field (interface) equation can be easily combined with differential equations describing various physical phenomena. Interfacial electro-thermo-chemo-mechanical phenomena can be captured within a single, thermodynamically-consistent framework. As such, phase field-based models have been developed to tackle a wide range of coupled fracture problems, including hydraulic fracture \cite{Bourdin2012,Heider2021}, Li-Ion battery degradation \cite{Boyce2022,Ai2022}, hydrogel fracture \cite{zhang2019phase,zheng2022phase}, concrete corrosion-induced cracking \cite{korec2023phase,fang2023multi}, thermo-mechanical fracture \cite{Bourdin2014,Miehe2015c}, hydrogen-assisted cracking \cite{MARTINEZPANEDA2018742,CUI2024315}, and electro-mechanical fracture \cite{abdollahi2015phase,quinteros2023electromechanical}. More recently - and independently - phase field models have been developed to tackle the long-standing challenge of predicting corrosion, both uniform and localised; see Ref. \cite{martinez2024phase} for a review. Different from their fracture counterparts, phase field corrosion models are built following the framework used for phase transformation problems (solidification, microstructural evolution). This hinders the coupling of phase field fracture and corrosion models, as needed to tackle important technological problems such as corrosion-fatigue and stress corrosion cracking \cite{cui2022generalised}. Hence, there is a need to establish a unified framework and move beyond \textit{ad hoc} numerical implementations; the aim of this work. Thus, the novelty of the work is two-fold. To start with, we present a theoretical and computational formulation that encapsulates a wide range of coupled phase field problems, for the first time providing a common root for phase field fracture and corrosion models. This generalised formulation enables a versatile and straightforward numerical implementation, allowing us to present the first integration point-level implementation of coupled problems such as thermo-mechanical fracture, corrosion, hydraulic fracture and hydrogen embrittlement. This is demonstrated in the commercial finite element package Abaqus through simple user material (\texttt{UMAT} and \texttt{UMATHT}) subroutines, which are freely provided. The remainder of this manuscript begins with the presentation of a generalised phase field formulation that encapsulates both fracture and corrosion models (Section \ref{sec:PhaseFieldModeling}). Then, this formulation is extended to general coupled (multi-physics) problems (Section \ref{sec:Multiphysics Phase-Field Modeling}). In Section \ref{sec:ThermalAnalogy}, details of the numerical implementation are provided, which is done at the integration point level by exploiting the thermal analogy, as demonstrated with user material subroutines in the commercial finite element package Abaqus. Representative results are presented in Section \ref{Sec:Examples} and concluding remarks are given in Section \ref{Sec:Conclusions}. The main body of text is complemented by Appendices aiming at providing additional theoretical and numerical details.

\section{A generalised phase field model}
\label{sec:PhaseFieldModeling}

We establish a generalised treatment of phase field fracture and phase field corrosion models from the Allen-Cahn equation, a fundamental mathematical framework in phase transition modelling \cite{cahn1958free,allen1979microscopic}. The Allen-Cahn equation typically models non-conserved order parameters, distinguishing it from the Cahn-Hilliard equation, which is used for conserved quantities.\\

The total free energy of a system involving two phases can be expressed as the sum of the bulk free energy $ \mathcal{F}_{\text{bulk}} $ and the interface free energy $ \mathcal{F}_{\text{int}} $. For a given body $ \Omega \subset \mathbb{R}^n $ (where $ n \in \{1, 2, 3\} $), the total free energy is given by:
\begin{equation}\label{Eq:FreeEnergy}
    \mathcal{F} = \mathcal{F}_{\text{bulk}} + \mathcal{F}_{\text{int}} = \int_{\Omega} \big( f_\text{b} (\{\beta_i\}, \phi) + f_{\text{int}} (\nabla \phi) \big) \, \mathrm{d}V,
\end{equation}

\noindent where $ f_\text{b}(\{\beta_i\}, \phi) $ is the bulk free energy density, depending on the phase field (order parameter) $ \phi $ and a set of field variables $ \mathbf{B} = \{ \beta_1, \beta_2, \dots, \beta_k \mid \beta_i \in \mathbb{B} \} $, related to mechanical or chemical processes. The interface free energy density $ f_{\text{int}}(\nabla \phi) $ depends on the gradient of the phase field $ \nabla \phi $. \\

The bulk free energy density $ f_\text{b}(\{\beta_i\}, \phi) $ can be defined as:
\begin{equation}\label{Eq:bulk_energy}
    f_\text{b}(\{\beta_i\}, \phi) = g(\phi) f_{\text{b1}}(\{\beta_i\}) + (1 - g(\phi)) f_{\text{b2}}(\{\beta_i\}) + w(\phi),
\end{equation}

\noindent where $ f_{\text{b1}} $ and $ f_{\text{b2}} $ represent the free energy densities for the first and second phases, respectively. By assuming that each material point is a mixture of the first and second phases, the function $ g(\phi) $, known as the degradation or interpolation function, defines the volume fraction of the first phase in a representative volume element  (RVE):
\begin{equation}\label{eq:g_phi}
    g(\phi) = \frac{V_1}{V_{\text{RVE}}},
\end{equation}

\noindent where $ V_1 $ is the volume occupied by the first phase and $ V_{\text{RVE}} $ is the volume of the RVE. The double-well potential $ w(\phi) $ ensures that the order parameter favors distinct phases by penalizing intermediate values.\\

The interface energy density is defined as:
\begin{equation}\label{Eq:gradient_energy}
    f_{\text{int}} (\nabla \phi) = \frac{\kappa}{2} \left| \nabla \phi \right|^2,
\end{equation}

\noindent where $ \kappa $ is the gradient energy coefficient, which controls the energetic cost associated with creating interfaces. Substituting Eqs.~\eqref{Eq:bulk_energy} and \eqref{Eq:gradient_energy} into the total free energy expression~\eqref{Eq:FreeEnergy}, we obtain:
\begin{equation}\label{Eq:FreeEnergy1}
    \mathcal{F} = \int_{\Omega} \Big( g(\phi) f_{\text{b1}}(\{\beta_i\}) + (1 - g(\phi)) f_{\text{b2}}(\{\beta_i\}) + w(\phi) + \frac{\kappa}{2} \left| \nabla \phi \right|^2 \Big) \, \mathrm{d}V.
\end{equation}

The last two terms in Eq. \eqref{Eq:FreeEnergy1} represent the Allen-Cahn energy, capturing the energetic cost of phase interfaces. This energy reflects the tendency of a system to minimize interface area during phase transitions~\cite{goldenfeld2018lectures}. The Allen-Cahn equation describes the evolution of the phase field by applying the following relaxation law, which drives the system towards equilibrium:
\begin{equation}\label{eq:PhaseFieldEvolution}
    \eta \frac{\partial \phi}{\partial t} = -\frac{\delta \mathcal{F}}{\delta \phi} = \kappa \nabla^2 \phi - w'(\phi) - g'(\phi) \big( f_{\text{b1}}(\{\beta_i\}) - f_{\text{b2}}(\{\beta_i\}) \big) \, .
\end{equation}

\noindent Here, $ \eta $ is the relaxation time constant, which characterizes the rate at which equilibrium is approached. The Neumann boundary condition is defined as:
\begin{equation}
    \nabla \phi \cdot \mathbf{n} = 0 \quad \text{on} \quad \partial \Omega,
\end{equation}

\noindent where $ \mathbf{n} $ is the outward unit normal vector on the surface of the domain $\Omega$.

\subsection{Phase field fracture method}
\label{sec:PhaseFieldFracture}

Let us now show how the Allen-Cahn-based framework introduced so far can encompass existing phase field fracture models. In phase field fracture modelling, the phase field variable $\phi$ is taken to be a representation of damage, in a continuum mechanics sense, and thus degrades the material stiffness. Consider a body $\Omega$ with two phases of materials undergoing mechanical deformation: the first phase represents the material with pristine stiffness ($\phi=0$), while the second phase corresponds to the fully damaged configuration ($\phi=1$). We shall then define a material toughness or critical energy release rate $G_c$, describing the material's resistance to fracture, and a characteristic length scale $\ell$ which governs the size of the interface thickness and the fracture process zone. Then, considering (for illustrative purposes) the original AT2 phase field fracture model \cite{Bourdin2008}, expressions for the double-well potential and the gradient energy coefficient in Eq. (\ref{Eq:FreeEnergy1}) can be found:
\begin{equation}\label{eq:wellPotetioal relation}
w(\phi)=\frac{G_c \, \phi^2}{2 \ell}, \quad \kappa=\frac{\ell \, G_c}{2}
\end{equation}

Let us denote the bulk free energy due to mechanical work as $\psi^{\mathrm{M}}$. Thus, for each phase we can write $f_{\text{b1}} (\mathbf{u})=\psi^{\mathrm{M}}_1 (\bm{\varepsilon}(\mathbf{u}))$ and $f_{\text{b2}}(\mathbf{u})=\psi^{\mathrm{M}}_2 (\bm{\varepsilon}(\mathbf{u}))$. Based on Eq. \eqref{Eq:FreeEnergy1}, the total potential energy of the deformation-fracture system can be written as:
\begin{equation}\label{Eq:FreEnerPhiFrac}
\mathcal{F}_{\ell} = \int_{\Omega} \left[G_c \left(\frac{\ell}{2}|\nabla \phi|^2 + \frac{\phi^2}{2 \ell}  \right) + g(\phi) \psi^{\mathrm{M}}_1(\bm{\varepsilon}(\mathbf{u})) + (1 - g(\phi)) \psi^{\mathrm{M}}_2 (\bm{\varepsilon}(\mathbf{u}))\right] \, \mathrm{d} V.
\end{equation}

In this expression, $\psi^{\mathrm{M}}_1$ and $\psi^{\mathrm{M}}_2$ denote the strain energy densities for the pristine (first phase) and fully degraded (second phase) materials, respectively. The displacement vector is denoted by $\mathbf{u}$, and the strain tensor is defined as $\bm{\varepsilon} = \left(\nabla \mathbf{u}^T + \nabla \mathbf{u}\right) / 2$. The volume fraction of the first phase, referred to as the degradation function, is given by the quadratic form $g(\phi) = (1 - \phi)^2$. Assuming that the second phase has no stiffness and consequently no strain energy ($\psi^{\mathrm{M}}_2 = 0$), Eq. \eqref{Eq:FreEnerPhiFrac} simplifies to:
\begin{equation}\label{Eq:FreEnerPhiFrac-1}
\mathcal{F}_{\ell} = \int_{\Omega} \left[G_c \left(\frac{\ell}{2}|\nabla \phi|^2 + \frac{\phi^2}{2 \ell} \right) + g(\phi) \psi^{\mathrm{M}}_1(\bm{\varepsilon}(\mathbf{u}))\right] \, \mathrm{d} V.
\end{equation}

The evolution equation for the phase field in the fracture model is derived from Eq. \eqref{eq:PhaseFieldEvolution} and the potential energy in Eq. \eqref{Eq:FreEnerPhiFrac-1}, assuming rate-independent damage evolution ($\eta = 0$):
\begin{equation}\label{Eq:phiEvolfrac-1}
G_c \left(-\ell \nabla^2 \phi + \frac{\phi}{\ell} \right) + g^{\prime}(\phi) \psi^{\mathrm{M}}_1(\bm{\varepsilon}(\mathbf{u})) = 0.
\end{equation}

Eq. (\ref{Eq:phiEvolfrac-1}) is arguably the most recognised form of the phase field evolution law, as it corresponds to the balance equation for the so-called AT2 phase field model. However, Eq. (\ref{Eq:phiEvolfrac-1}) does not distinguish between compressive and tensile stress states, and this has led to various extensions of the AT2 phase field model to ensure that damage only occurs under tension or to embed arbitrary failure surfaces \cite{Amor2009,Miehe2010a,Freddi2010,Lorenzis2021,Navidtehrani2022}. An asymmetric tension-compression behaviour can be captured by defining a non-zero stiffness for the second phase, such that
\begin{equation}\label{Eq:phiEvolfrac-2}
G_c \left(-\ell \nabla^2 \phi + \frac{\phi}{\ell} \right) + g^{\prime}(\phi) \left(\psi^{\mathrm{M}}_1(\bm{\varepsilon}(\mathbf{u})) - \psi^{\mathrm{M}}_2(\bm{\varepsilon}(\mathbf{u}))\right) = 0.
\end{equation}

\noindent whereby $\psi_1^M-\psi_2^M$ is the driving force for fracture. In the literature, the variable $\psi^+_0$ is often used to describe a fracture driving force based on the tensile part of a decomposed strain energy density (i.e., $\psi^+_0=\psi_1^M-\psi_2^M$). Adopting the notation most commonly found in the literature, the (undamaged) strain energy density can be decomposed into a tensile and a compressive part as $\psi_0=\psi_0^+ + \psi_0^-$ and, accordingly, the total strain energy density in the damaged configuration reads,
\begin{equation}\label{Eq:StDecomp}
\psi = g(\phi)\psi_0^+ + \psi_0^- = g(\phi)\psi_0 + (1 - g(\phi))\psi_0^- \, .
\end{equation}

Comparing Eqs. (\ref{Eq:StDecomp}) and (\ref{Eq:FreEnerPhiFrac}) one finds that the strain energy density of the undamaged configuration equals the strain energy of the first phase ($\psi_0 = \psi^{\mathrm{M}}_1$), while the strain energy of the second phase corresponds to the compressive part of the undamaged strain energy density ($\psi_0^- = \psi^{\mathrm{M}}_2$). For the formulation to be variationally consistent, this asymmetric degradation must also be considered in the deformation problem. Hence, the stress tensor $\bm{\sigma}$ reads,
\begin{equation}
\bm{\sigma} = g(\phi) \frac{\partial \psi^{\mathrm{M}}_1(\bm{\varepsilon}(\mathbf{u}))}{\partial \bm{\varepsilon}(\mathbf{u})} + (1 - g(\phi)) \frac{\partial \psi^{\mathrm{M}}_2(\bm{\varepsilon}(\mathbf{u}))}{\partial \bm{\varepsilon}(\mathbf{u})}.
\end{equation}

With the strong form of the coupled deformation-diffusion problem being given by,
\begin{align}
&\nabla \cdot \bm{\sigma} + \mathbf{b} = 0 \label{eq:LinearMomemtum} \\ 
&G_c \left(-\ell \nabla^2 \phi + \frac{\phi}{\ell} \right) + g^{\prime}(\phi) \mathcal{H} = 0. \label{eq:PhaseFieldFractFinal}
\end{align}

\noindent with $\mathbf{b}$ being a body force vector. Here, Eq. (\ref{eq:LinearMomemtum}) can be readily derived by taking the variation of Eq. \eqref{Eq:FreEnerPhiFrac} with respect to the displacement vector $\mathbf{u}$ and applying the divergence theorem. Also, a history field $\mathcal{H} = \max_{t \in [0, \tau]} (\psi^{\mathrm{M}}_1(t) - \psi^{\mathrm{M}}_2(t))$ has been defined to ensure damage irreversibility.

\subsection{Phase field corrosion}
\label{sec:CorrosionPhaseField}

The generalised framework presented before can also be particularised to the study of corrosion, the degradation of materials due to environmental chemical interactions. The dissolution of metals due to corrosion results in an evolving interface, separating the solid metal (electrode) from the liquid corrosive electrolyte. Defining the solid phase as the first phase and the liquid phase as the second one, the phase field variable is taken to be $\phi = 1$  in the metal and $\phi=0$ in the aqueous electrolyte. In its simplest form, a phase field model for corrosion needs to capture two phenomena: the dissolution of the metal (short-range interactions) and the subsequent transport of metal ions (long-range interactions). In terms of the generalised formulation presented above, this implies particularising the bulk free energy density to the chemical free energy density $\psi^{\mathrm{ch}}$ and considering an additional primary variable, the normalised concentration of metal ions, equal to $c=1$ in the metal phase and equal to $c=0$ in electrolyte regions very far from the corrosion interface. Accordingly, the energy functional in Eq. \eqref{Eq:FreeEnergy} is given by
\begin{equation}\label{Eq:FreeEnergy11}
    \mathcal{F} = \int_{\Omega} \big( \psi^{\mathrm{ch}} + \frac{\kappa}{2} \left| \nabla \phi \right|^2 \big) \, \mathrm{d}V.
\end{equation}

The chemical free energy density $\psi^{\mathrm{ch}}$ can be expressed as a weighted sum of the chemical free energy densities of the solid ($\psi^{\mathrm{ch}}_{\mathrm{S}}$) and liquid ($\psi^{\mathrm{ch}}_{\mathrm{L}}$) phases, using the interpolation function $g(\phi)$, as 
\begin{equation}\label{Eq:ChemicalDensity}
\psi^{\mathrm{ch}}(\phi ,c)=g(\phi)\psi^{\mathrm{ch}}_{\mathrm{S}} + (1-g(\phi)) \psi^{\mathrm{ch}}_{\mathrm{L}} +w(\phi).
\end{equation}

The interpolation function $g(\phi)$, characterising the volume fraction, as defined in Eq. \eqref{eq:g_phi}, is typically chosen to be $g(\phi) = -2 \phi^3 + 3 \phi^2$ in the phase field corrosion community, while the double-well potential $w(\phi)$ is typically defined as $w(\phi) = \omega \phi^2 (1 - \phi)^2$, with $\omega$ being the height of the double-well potential.\\

Following the literature, the chemical free energy densities of solid ($\psi^{\mathrm{ch}}_{\mathrm{S}}$) and liquid ($\psi^{\mathrm{ch}}_{\mathrm{L}}$) phases are defined as:
\begin{equation}\label{eq:chemical-energy-sq}
\psi^{\mathrm{ch}}_{\mathrm{S}}=A(c_\mathrm{S} - c_\mathrm{Se}), \quad \psi^{\mathrm{ch}}_{\mathrm{L}}= A(c_\mathrm{L} - c_\mathrm{Le}),
\end{equation}

\noindent where $A$ is the curvature of the free energy density, and $c_{\text{S}}$ and $c_{\text{L}}$ are the normalized concentrations of the solid and liquid phases. Also, $c_{\text{Se}} = c_{\text{solid}} / c_{\text{solid}} = 1$, and $c_{\text{Le}} = c_{\text{sat}} / c_{\text{solid}}$ are the normalised equilibrium concentrations for the solid and liquid phases, with $c_{\text{sat}}$ being the saturation concentration. Accordingly, Eq. (\ref{Eq:ChemicalDensity}) can be reformulated as, 
\begin{equation}
\psi^{\mathrm{ch}}(c, \phi)=A \left( c- g(\phi) (c_\mathrm{Se}-c_\mathrm{Le})-c_\mathrm{Le} \right)^2 + \omega \phi^2 (1 - \phi)^2.
\end{equation}

The interface energy density follows the same notation and definition as in Eq. \eqref{Eq:gradient_energy}. An interface energy $\gamma$ and interface thickness $\ell_m$ can be defined based $\omega$ and $\kappa$ as \cite{cui2023electro}:
\begin{equation}
\gamma=\sqrt{\frac{\kappa \omega}{18}}, \quad \ell_m= \sqrt{\frac{8 \kappa}{\omega}}.
\end{equation}

The corrosion phase field evolution equation can derived using the relaxation law, see Eq. \eqref{eq:PhaseFieldEvolution}, as:
\begin{equation}\label{Eq:phiEvol-Cor1}
\frac{1}{L} \dot{\phi} = \kappa \nabla^2 \phi - \frac{\partial \psi^{\mathrm{ch}}(c, \phi)}{\partial \phi}
\end{equation}

\noindent where $L$ is the interface kinetics coefficient. For more details, see Ref. \cite{JMPS2021}.\\

It remains to define the transport of the mass conserved quantity: the normalised concentration of metal ions $c=c_m (\textbf{x}, t)/c_{\text{solid}}$, with $c_{\text{solid}}$ being the concentration of atoms in the metal and $c_m (\textbf{x}, t)$ being the concentration of dissolved ions. The mass conservation law then reads
\begin{equation}\label{Eq:MassTraCor}
\dot{c} c_{\text{solid}} + \nabla \cdot \mathbf{J} = 0,
\end{equation}

As the medium is a mixture of solid and liquid phases, the normalized concentration $c$ can be expressed as a function of the normalised concentration in the solid ($c_S$) and liquid ($c_L$) phases,
\begin{equation}
c=g(\phi) c_{\text{S}}+ (1-g(\phi)) c_{\text{L}}.
\end{equation}

The mass transport is derived from the chemical potential using the relationships in Eq. \eqref{eq:chemical-energy-sq}, given as:
\begin{equation}
\mu = -\frac{1}{c_{\text{solid}}} \frac{\partial \psi^{\mathrm{ch}}}{\partial c} = -\frac{2 A}{c_{\text{solid}}} \left(c - g(\phi) (c_{\mathrm{Se}} - c_{\mathrm{Le}}) - c_{\mathrm{Le}}\right).
\end{equation}

Using a Fick’s law-type relation, the flux $\mathbf{J}$ can be expressed as:
\begin{equation}\label{Eq:Ficks}
\mathbf{J} = \frac{D_m}{2 A} \cdot c_{\text{solid}} \cdot \nabla \mu = -c_{\text{solid}} \cdot D_m \nabla\left(c - g(\phi) (c_{\mathrm{Se}} - c_{\mathrm{Le}}) - c_{\mathrm{Le}}\right),
\end{equation}

\noindent where $D_m$ is the diffusion coefficient of metal ions. Substituting Eq. \eqref{Eq:Ficks} into Eq. \eqref{Eq:MassTraCor}, the mass conservation law reads:
\begin{equation}\label{eq:corrosion-Con}
\dot{c} - \nabla \cdot \left[ D_m \nabla\left(c - g(\phi) (c_{\mathrm{Se}} - c_{\mathrm{Le}}) - c_{\mathrm{Le}}\right) \right] = 0.
\end{equation}

\section{Extension and particularisation to coupled problems}
\label{sec:Multiphysics Phase-Field Modeling}

Let us now extend this generalised formulation to the analysis of coupled problems, where solids undergo mechanical deformation, phase transitions, and a diffusion-type process. Considering as primary variables the displacement vector $\mathbf{u}$, the phase field $\phi$, and the diffusion field $\xi$, the coupled system of equations for a body $\Omega$ can be formulated as follows:
\begin{equation}\label{Eq:StrMulPhys11}
\nabla \cdot \bm{\sigma} (\mathbf{u},\phi,\xi ) + \mathbf{b} = 0, 
\end{equation}
\begin{equation}\label{Eq:StrMulPhys2}
\kappa \nabla^2 \phi - w'(\phi) - g'(\phi) \big( f_{\text{b1}}(\mathbf{u}, \xi) - f_{\text{b2}}(\mathbf{u},\xi) \big) - \eta \dot{ \phi}=0 ,
\end{equation}
\begin{equation}\label{Eq:StrMulPhys3}
\rho \dot{U}_{\xi}(\xi, \nabla \xi, \mathbf{u}, \phi) + \nabla \cdot \mathbf{f}_{\xi} (\xi, \nabla \xi, \mathbf{u}, \phi) - q_{\xi} = 0,
\end{equation}

\noindent Here, Eq. (\ref{Eq:StrMulPhys11}) represents the linear momentum equation, Eq. (\ref{Eq:StrMulPhys2}) is the phase field evolution equation, and Eq. (\ref{Eq:StrMulPhys3})
corresponds to a diffusion-type field equation. In the following subsections, we explore four specific cases for multiphysics phase field modelling, each one addressing different coupling mechanisms and physical phenomena. More specifically, this general framework will be particularised to the analysis of thermal fracture ($\xi = T$, Section. \ref{sec:Thermal Fracture}), hydraulic fracture ($\xi = p$, Section. \ref{sec:Hydraulic-Fracture}), hydrogen embrittlement ($\xi = c_{\mathrm{H}}$, Section. \ref{sec: Hydrogen-Embrittlement}), and stress-assisted corrosion ($\xi = c$, Section. \ref{sec:Corrosion-Stress}).

\subsection{Thermal fracture}
\label{sec:Thermal Fracture}

Thermal fractures are commonplace in a wide range of engineering applications and sectors, from aerospace \cite{PETROVA2020102605} to nuclear energy \cite{LI2021793}. Changes in temperature lead to thermal strains, which can result in fractures \cite{RUAN2024105756}.\\ 

In thermoelasticity, the strain tensor is decomposed into an elastic part, $\bm{\varepsilon}_e$, and a thermal part, $\bm{\varepsilon}_{T}$, as follows:
\begin{equation}
\bm{\varepsilon} = \bm{\varepsilon}_e + \bm{\varepsilon}_{T} \, ,
\end{equation}

\noindent where the thermal strain, $\bm{\varepsilon}_{T}$, is defined in terms of the thermal expansion coefficient $\alpha_{T}$:
\begin{equation}
\bm{\varepsilon}_{T} = \alpha_{T} (T - T_0) \bm{I} \, ,
\end{equation}

\noindent with $T_0$ representing the initial temperature and $\bm{I}$ being the identity tensor. Since only the elastic (stored) strain contributes to stress, Eqs. (\ref{Eq:StrMulPhys11}) and (\ref{Eq:StrMulPhys2}) can be respectively rewritten as:
\begin{align}
\label{Eq:Momentom}
\nabla \cdot \bm{\sigma} (\bm{\varepsilon}_e, \phi) + \mathbf{b} = \mathbf{0} \quad \text{in} \quad \Omega, \\
\label{Eq:PhaseField}
G_c \left(\frac{\phi}{\ell} - \ell \nabla^2 \phi\right) - 2(1 - \phi) \mathcal{H}(\bm{\varepsilon}_e) = 0 \quad \text{in} \quad \Omega.
\end{align}

While Eq. (\ref{Eq:StrMulPhys3}) is particularised to the heat transfer equation, given by:
\begin{equation}\label{Eq:SHeat}
\rho c_{T} \dot{T} -\color{black} k_{T} \nabla^2 T = q_{T} \, ,
\end{equation}

\noindent where $\rho$ is the material density, $c_{T}$ is the specific heat, $k_{T}$ is the thermal conductivity, and $q_{T}$ is the heat source. Often, heat transfer and phase field fracture are assumed to interact through a damage-dependent thermal conductivity, such that
\begin{equation}\label{Eq:SHeatPhase}
\rho c_{T} \dot{T} \color{blue}{-}\color{black} k_{T}(\phi) \nabla^2 T = q_{T} \, ,
\end{equation}

\noindent where the thermal conductivity $k_{T}(\phi)$ is defined as $k_{T}(\phi) = g(\phi) k_0$, with $k_0$ representing the thermal conductivity of the pristine material.

\subsection{Hydraulic fracture}
\label{sec:Hydraulic-Fracture}

Hydraulic fracture is a process of crack nucleation and propagation caused by changes in pore pressure within a solid body due to fluid injection or natural forces. This phenomenon occurs in both natural settings and engineering applications, with hydraulic fracture modelling being widely used in geotechnical engineering, oil and gas extraction, environmental management, and other fields to predict and control fracture behaviour for resource extraction and infrastructure stability \cite{10.1115/1.3231067,chen2021review,Sun2021,Clayton2022}.\\

Particularising the above-presented generalised formulation to the study of hydraulic fracture requires defining a balance equation for the evolution of fluid pressure $p$. This is achieved by particularising Eq. (\ref{Eq:StrMulPhys3}) to the mass balance equation governing fluid flow within a porous medium. When considering an external fluid source $q_m$, the fluid mass balance equation is defined as:
\begin{equation}\label{Eq:MassConsFL}
\dot{\zeta}_{\text{fl}} + \nabla \cdot (\rho_{\text{fl}} \mathbf{v}_{\text{fl}}) = q_m,
\end{equation}

\noindent where $\zeta_{\text{fl}}$ is the mass fluid content, corresponding to the mass of fluid per unit bulk volume and $\mathbf{v}_{\text{fl}}$ is the fluid velocity vector. This can be defined using porosity ($n_{\text{p}}$) and the density of the fluid ($\rho_{\text{fl}}$) as:
\begin{equation}\label{Eq:MassContent0}
\zeta_{\text{fl}} = \rho_{\text{fl}} n_{\text{p}},
\end{equation}

\noindent where porosity $n_{\text{p}} = V_{\text{p}} / V_{\text{b}}$ is the ratio of the volume of pores ($V_{\text{p}}$) to the bulk volume ($V_{\text{b}}$) of the medium. Using Biot's theory of poroelasticity and Darcy's law, the mass balance equation can be rewritten as:
\begin{equation}\label{Eq:dMassContent3}
\rho_{\text{fl}} \left(S \dot{p} + \alpha_b \dot{\varepsilon}_{\text{vol}} \right) + \nabla \cdot \left(-\rho_{\text{fl}} \frac{\bm{K}_{\text{fl}}}{\mu_{\text{fl}}} \nabla p \right) = q_m,
\end{equation}

\noindent where $\alpha_b$ is Biot's coefficient, $\dot{\varepsilon}_{\text{vol}}$ is the rate of volumetric strain, $\bm{K}_{\text{fl}}$ is the permeability tensor, $\mu_{\text{fl}}$ is the fluid viscosity, and $S$ is the storage coefficient, defined as:
\begin{equation}\label{Eq:dMassContent2}
S = \frac{(1 - \alpha_b)(\alpha_b - n_{\text{p}})}{K} + n_{\text{p}} C_{\text{fl}},
\end{equation}

\noindent with $K$ being the bulk modulus and $C_{\text{fl}}$ the fluid compressibility.\\

Next, Biot's theory of poroelasticity is utilised to establish a constitutive relationship between stress and strain. In a saturated porous medium, the total strain arises from the stress acting on the solid skeleton (effective stress $\bm{\sigma}^{\text{eff}}$), and the pore pressure of fluid. Under static conditions or slow fluid flow, the pore pressure contributes only to changes in volumetric strain. Consequently, the total stress is expressed as the sum of the effective stress $\bm{\sigma}^{\text{eff}} = \bm{C} : \bm{\varepsilon}$, and the pore fluid pressure scaled by Biot's coefficient $\alpha_b$, resulting in:
\begin{equation}\label{Eq:TotalStress}
\bm{\sigma} = \bm{C} : \bm{\varepsilon} - \alpha_b p \bm{I} = \bm{\sigma}^{\text{eff}} - \alpha_b p \bm{I},
\end{equation}

\noindent where $\bm{I}$ is the identity tensor. In the context of phase field hydraulic fracture, and considering a decomposition of the strain energy density, the effective stress is defined as:
\begin{equation}\label{eq:EffStress}
\bm{\sigma}^{\text{eff}} = g(\phi) \frac{\partial \psi^+_0(\bm{\varepsilon})}{\partial \bm{\varepsilon}} + \frac{\partial \psi^-_0(\bm{\varepsilon})}{\partial \bm{\varepsilon}}.
\end{equation}

Thus, Eq. \eqref{Eq:StrMulPhys11} can be reformulated as:
\begin{equation}
\nabla \cdot \left[ \bm{\sigma}^{\text{eff}} - \alpha_b p \bm{I} \right] = \mathbf{0} \quad \text{in} \quad \Omega.
\end{equation}

One must also consider the interplay between cracking phenomena and fluid flow. Building on the work of Lee et al. \cite{Lee2016}, we couple the fluid and phase field equations by dividing the domain into three distinct regions: the reservoir ($\Omega_{\text{r}}$), fracture ($\Omega_{\text{f}}$), and transition ($\Omega_{\text{t}}$) domains. These regions are distinguished using domain indicator fields as functions of the phase field variable $\phi$ and material constants $c_1$ and $c_2$. The domain indicator fields (${\chi}_{\text{r}},{\chi}_{\text{f}}$) can be defined as:
\begin{equation}\label{Eq:Axu}
{\chi}_{\text{r}}\left({\phi}\right)=
\begin{cases}
    1 & {\phi} \le {c}_1 \\
    \frac{{c}_2 - {\phi}}{{c}_2 - {c}_1} & {c}_1 < {\phi} < {c}_2 \\
    0 & {c}_2 \le {\phi} 
\end{cases}, \quad
{\chi}_{\text{f}}\left({\phi}\right)=
\begin{cases}
    0 & {\phi} \le {c}_1 \\
    \frac{{{\phi} - {c}_1}}{{c}_2 - {c}_1} & {c}_1 < {\phi} < {c}_2 \\
    1 & {c}_2 \le {\phi} 
\end{cases},
\end{equation}

Thus, the fluid and solid parameters between the reservoir and fracture domains are defined as:
\begin{equation}
\alpha_b = \chi_{\text{r}} \alpha_{\text{r}} + \chi_{\text{f}},
\end{equation}
\begin{equation}
n_{\text{p}} = \chi_{\text{r}} n_{\text{pr}} + \chi_{\text{f}},
\end{equation}
\begin{equation}\label{Eq:Kzhou}
\bm{K}_{\text{fl}} = \chi_{\text{r}} \bm{K}_{\text{r}} + \phi^{b} \chi_{\text{f}} \bm{K}_{\text{f}},
\end{equation}

\noindent where $\alpha_{\text{r}}$, $n_{\text{pr}}$, and $\bm{K}_{\text{r}}$ denote Biot's coefficient, porosity, and the permeability tensor of the reservoir domain, respectively. $\bm{K}_{\text{f}}$ is the permeability tensor of the fracture domain and $b$ is a transient parameter. Finally, we can write the fluid flow equation in a form valid across all domains using the domain indicator fields $\chi_{\text{r}}$ and $\chi_{\text{f}}$:
\begin{equation}\label{Eq:dMassContent5}
\rho_{\text{fl}} \left(S(\alpha_b(\phi), n_{\text{p}}(\phi)) \dot{p} + \alpha_b(\phi) \chi_{\text{r}}(\phi) \dot{\varepsilon}_{\text{vol}} \right) + \nabla \cdot \left(-\rho_{\text{fl}} \frac{\bm{K}_{\text{fl}}(\phi)}{\mu} \nabla p \right) = q_m.
\end{equation}

\subsection{Hydrogen embrittlement}
\label{sec: Hydrogen-Embrittlement}

When metallic materials are exposed to hydrogen-containing environments, such as seawater or hydrogen gas, they experience a phenomenon known as \emph{hydrogen embrittlement}, whereby the absorption of hydrogen atoms into the metal results in a dramatic degradation of their ductility, fracture toughness and fatigue crack growth resistance \cite{Gangloff2012,CHEN2024}. Hydrogen embrittlement phenomena can be captured in a phase field setting, as first achieved by Mart\'{\i}nez-Pa\~neda and co-workers \cite{CMAME2018,JMPS2020,CupertinoMalheiros2024}.\\ 

Predicting hydrogen-assisted failures requires solving a three-field system, modelling deformation, hydrogen diffusion and fracture. The transport of hydrogen can be simulated considering an equation of the type  (\ref{Eq:StrMulPhys3}), with the primary variable being the diffusible hydrogen concentration $c_{\mathrm{H}}$. Considering the balance of mass, hydrogen transport can be described as,
\begin{equation}\label{eq:Htransport}
\dot{c}_{\mathrm{H}} + \nabla \cdot \mathbf{J}_{\mathrm{H}} = 0,
\end{equation}

\noindent where the hydrogen flux, $\mathbf{J}_{\mathrm{H}}$, is defined based on the gradient of the chemical potential $\mu_{\mathrm{H}}$:
\begin{equation}\label{eq:HtransportFlux}
\mathbf{J}_{\mathrm{H}} = -\frac{D_{\mathrm{H}} c_{\mathrm{H}}}{R T_\mathrm{k}} \nabla \mu_{\mathrm{H}} = -D_{\mathrm{H}} \nabla c_{\mathrm{H}} + \frac{D_{\mathrm{H}}}{R \, T_\mathrm{k}} c_{\mathrm{H}} V_{\mathrm{H}} \nabla \sigma_h,
\end{equation}

\noindent with $D_{\mathrm{H}}$ being the diffusion coefficient, $R$ the gas constant, $T_\mathrm{k}$ the absolute temperature, $V_{\mathrm{H}}$ the partial molar volume of hydrogen in solid solution, and $\sigma_h$ the hydrostatic stress. Hydrogen accumulates in regions of high hydrostatic stress, providing a source of coupling between the solid mechanics and diffusion problems.\\

An important source of coupling is how hydrogen degrades the fracture resistance of metals. This is naturally captured in phase field by defining the material toughness as a function of the hydrogen content: $G_c (c_{\mathrm{H}})$. Among the multiple ways available for defining this relationship, we follow Ref. \cite{CMAME2018}, and use an atomistically-informed linear degradation law:
\begin{equation}
G_c(c_{\mathrm{H}}) = G_c(\theta) = (1 - \chi_{\mathrm{H}} \theta) G_c(0),
\end{equation}

\noindent where $\chi_{\mathrm{H}}$ is an atomistically-estimated damage coefficient that quantifies the reduction in fracture energy due to the presence of hydrogen and $\theta$ is the hydrogen coverage. The latter can be related to the hydrogen concentration through Oriani's equilibrium or the Langmuir–McLean's isotherm,
\begin{equation}
\theta=\frac{c_{\mathrm{H}}}{c_{\mathrm{H}}+\exp \left(\frac{-\Delta g_b^0}{R T_\mathrm{k}}\right)},
\end{equation}

\noindent where the hydrogen content is here expressed in units of impurity mole fraction, and $\Delta g_b^0$ is the Gibbs free energy difference between the decohering interface and the surrounding material. 

\subsection{Stress-assisted corrosion}
\label{sec:Corrosion-Stress}

The interplay between mechanical deformation and corrosion is what underpins localised corrosion failures. Mechanical stresses can rupture the protective passive film and accelerate corrosion kinetics, while metal dissolution alters stress distributions - a two-way coupling problem. These can be coupled by extending the formulation in Section \ref{sec:CorrosionPhaseField} to define a bulk free energy density that encompasses both mechanical and chemical contributions:
\begin{equation}
f_\text{b}(\mathbf{u},\phi,c)=\psi^{\mathrm{M}}(\mathbf{u}, \phi)+\psi^{\mathrm{ch}} (c, \phi),
\end{equation}

\noindent where $\psi^{\mathrm{M}}(\mathbf{u})$ is the mechanical free energy (i.e., the strain energy density), and $\psi^{\mathrm{ch}} (c)$ is the chemical free energy density, as defined in Eq. \eqref{Eq:ChemicalDensity}. Since the liquid phase is assumed not to carry stress, the mechanical free energy can be written as:
\begin{equation}\label{eq:psiMep}
\psi^{\mathrm{M}} =g(\phi)\psi^{\mathrm{M}}_\mathrm{S}(\mathbf{u})= g(\phi)(\psi_\mathrm{S}^e + \psi_\text{S}^p)
\end{equation}

\noindent where $\psi^{\mathrm{M}}_\mathrm{S}$ is the undamaged strain energy of the solid phase and $g(\phi)$ is the degradation function, as defined in Section \ref{sec:CorrosionPhaseField}, which satisfies $g(0) = 0$ for the electrolyte phase $(\phi = 0)$ and $g(1) = 1$ for the undissolved solid $(\phi = 1)$. In Eq. (\ref{eq:psiMep}), both elastic and plastic contributions to the strain energy density are considered, as respectively denoted by the $e$ and $p$ superscripts. Assuming J2 plasticity, the elastic and plastic strain energy densities are given by:
\begin{align}
& \psi_\text{S}^e\left(\bm{\varepsilon}_e\right) = \frac{1}{2}\left(\bm{\varepsilon}_e\right)^T : \bm{C}_0 : \bm{\varepsilon}_e \\
& \psi_\text{S}^p = \int_0^t \bm{\sigma_0} : \dot{\bm{\varepsilon}}_p \, \mathrm{d} t,
\end{align}

\noindent where $\bm{\varepsilon}_e$ and $\bm{\varepsilon}_p$ are the elastic and plastic parts of the strain tensor, and $\bm{C}_0$ is the elastic stiffness matrix. Accordingly, the undamaged stress $\bm{\sigma}_0$ and total stress $\bm{\sigma}$ are defined as:
\begin{equation}
\bm{\sigma} = g(\phi) \bm{\sigma}_0=g(\phi) \, \boldsymbol{C}_0 : (\boldsymbol{\varepsilon}_e-\boldsymbol{\varepsilon}_p).
\end{equation}

Work hardening is considered by means of an isotropic power law relationship:
\begin{equation}
\sigma_f = \sigma_y \left(1 + \frac{E \varepsilon^p}{\sigma_y}\right)^N ,
\end{equation}

\noindent where $E$ is Young's modulus, $\sigma_f$ is the flow stress, $\sigma_y$ is the initial yield stress and $N$ is the strain hardening exponent ($0\leq N \leq 1)$. The equivalent plastic strain is defined as $\varepsilon^p=\sqrt{ (2/3) \, \boldsymbol{\varepsilon}_p : \boldsymbol{\varepsilon}_p}$.\\

Using the aforementioned definitions of the mechanical and chemical free energy densities, the total free energy can then be written as:
\begin{equation}\label{Eq:FreeEnergy-corr}
    \mathcal{F} = \int_{\Omega} \left( g(\phi) (\psi_\mathrm{S}^e + \psi_\text{S}^p) + \psi^{\mathrm{ch}}(c, \phi) + \frac{\kappa}{2} \left| \nabla \phi \right|^2 \right)   \, \mathrm{d}V.
\end{equation}

From Eq. \eqref{Eq:FreeEnergy-corr}, the linear momemtum balance equation can be readily obtained by taking the stationary of the functional with respect to the displacement field:
\begin{equation}
\nabla \cdot \bm{\sigma} (\mathbf{u},\phi) = \nabla \cdot \left[g(\phi) \, \boldsymbol{C}_0 : (\boldsymbol{\varepsilon}_e-\boldsymbol{\varepsilon}_p) \right]=0
\end{equation}

However, as extensively discussed in Ref. \cite{martinez2024phase}, defining the phase field evolution equation in a variationally consistent way results in non-physical behaviour, with mechanical strains impacting corrosion not only during activation-controlled corrosion, as observed experimentally, but also during diffusion-controlled corrosion. To work around this, Cui and co-workers \cite{JMPS2021} suggested instead to enrich the description of the phase field mobility coefficient $L$ to make it a function of mechanical fields, such that the phase field corrosion evolution equation reads,
\begin{equation}\label{Eq:phiEvol-Cor11}
\frac{1}{L \left( \sigma_h, \varepsilon^p \right)} \dot{\phi} = \kappa \nabla^2 \phi - \frac{\partial \psi^{\mathrm{ch}}(c, \phi)}{\partial \phi}
\end{equation}

\noindent where $\sigma_h$ is the hydrostatic stress. The precise definition of $L \left( \sigma_h, \varepsilon^p \right)$ is aimed at incorporating the two main mechanisms by which mechanics interplays with corrosion: (i) enhancement of corrosion kinetics, and (ii) film rupture.\\

To incorporate the role of mechanical fields in enhancing corrosion kinetics, we define the mobility coefficient as,
\begin{equation}\label{eq:Lgutman}
L=k_{\mathrm{m}}\left(\varepsilon^p, \sigma_h\right) L_0=\left(\frac{\varepsilon^p}{\varepsilon_y}+1\right) \exp \left(\frac{\sigma_h V_m}{R T_{\text{k}}}\right) L_0,
\end{equation}

\noindent where $L_0$ is the reference mobility coefficient, which can be quantitatively related to the corrosion current density \cite{JMPS2021}, $V_m$ is the molar volume, and $k_m$ is the so-called mechanochemical coefficient. Eq. (\ref{eq:Lgutman}) incorporates the impact on corrosion kinetics of both lattice expansion and dislocation phenomena through the hydrostatic stress and the effective plastic strain rate, respectively.\\

The interplay between mechanics and film rupture and re-passivation is an important one. Mechanical strains lead to localised rupture of the protective film in corrosion-resistant materials, leading to localised corrosion phenomena (pitting, stress corrosion cracking), which are very detrimental and difficult to predict. The process is a cyclic one, with film rupture being followed by material dissolution and subsequent repassivation (film formation). This phenomenon is known as the film-rupture-dissolution-repassivation (FRDR) mechanism \cite{Parkins1987}, and for a corrosion current density $i$, can be expressed as,
\begin{equation}
i\left(t_i\right)=\begin{cases}
i_0, & \text { if } 0<t_i \leqslant t_0 \\
i_0 \exp \left(-k\left(t_i-t_0\right)\right), & \text { if } t_0<t_i \leqslant t_0+t_f
\end{cases} , 
\end{equation}

\noindent where $i_0$ is the corrosion current density of the bare metal, $t_0$ is the time interval before decay begins, $t_f$ is the drop time during a film rupture event, $t_i$ is the current time (within a specific cycle) and $k$ is a parameter that characterises the sensitivity of the corrosion rates to the stability of the passive film, as dictated by the material and the environment. Following experimental observations, the decay in corrosion current density with the improvement of the film stability in time is characterised by an exponential function. The time $t_f$ at which a rupture event will happen is dictated by straining kinetics. This is often described by considering the accumulated plastic strain - over a FRDR cycle $\varepsilon_i^p$ (i.e., undergone by the newly developed oxide layer) - with failure occurring when a critical value is reached ($\varepsilon_i^p=\varepsilon_f \approx 0.001$). As elaborated in Ref. \cite{cui2022generalised}, this is effectively captured by exploiting the proportionality relationship between the phase field mobility coefficient and the corrosion current density. Hence, considering as well Eq. (\ref{eq:Lgutman}), the mobility coefficient can be defined as,
\begin{equation}\label{eq:Lmech}
L = \begin{cases}
k_{\mathrm{m}}\left(\varepsilon^p, \sigma_h\right) L_0, & \text{if } 0 < t_i \leqslant t_0 \\
k_{\mathrm{m}}\left(\varepsilon^p, \sigma_h\right) L_0 \exp\left(-k(t_i - t_0)\right), & \text{if } t_0 < t_i \leqslant t_0 + t_f
\end{cases}.
\end{equation}

\section{Numerical implementation exploiting the thermal analogy}
\label{sec:ThermalAnalogy}

In Sections \ref{sec:Thermal Fracture}-\ref{sec:Corrosion-Stress}, we have illustrated that a wide range of physical phenomena can be modelled using a diffusion type equation, Eq. (\ref{Eq:StrMulPhys3}). One such phenomenon is heat transfer, which has been widely implemented across in-house and commercial finite element packages. Therefore, leveraging the thermal analogy of diffusion-type equations, these physical processes can be incorporated into these codes with minimal effort. We demonstrate this here, exploiting this heat transfer analogy to straightforwardly implement the coupled problems discussed before in the commercial finite element package Abaqus.\\ 

The heat transfer equation can be expressed in a general form as:
\begin{equation}\label{Eq:GHEAT1}
\rho \dot{U} + \nabla \cdot \mathbf{f} - r = 0 \, , 
\end{equation}

\noindent where $U$ denotes internal heat energy, $\mathbf{f}$ is the heat flux vector, and $r$ is the heat source. Using Eq. \eqref{Eq:GHEAT1} as a foundation, any diffusion-type equation can be reformulated analogously, as shown below.\\

\noindent \textbf{General phase field}. Consider the general phase field evolution equation given in Eq. \eqref{eq:PhaseFieldEvolution}. Taking the phase field variable as the temperature ($\phi \equiv T$), Eq. \eqref{eq:PhaseFieldEvolution} can be reformulated to resemble Eq. (\ref{Eq:GHEAT1}) as, 
\begin{equation}\label{eq;thermal-general-corrosion}
\underbrace{1}_{\rho} \underbrace{\left(- \eta \frac{\partial \phi}{\partial t}- w'(\phi) - g'(\phi) \big( f_{b1}(\beta) - f_{b2}(\beta) \big)\right)}_{\dot{U}} + \nabla \cdot \underbrace{(\kappa \nabla \phi)}_{\mathbf{f}}  = 0.
\end{equation}

\noindent with $r=0$. \\

\noindent \textbf{Phase field fracture}. The specific case of phase field fracture, given by Eq. (\ref{eq:PhaseFieldFractFinal}), can also readily be expressed in a way that resembles the heat transfer PDE:
\begin{equation}\label{Eq:phasefieldfracture}
\underbrace{1}_{\rho} \underbrace{\left(\frac{\phi}{\ell^2} + g^{\prime}(\phi) \frac{\mathcal{H}}{G_c \ell}\right)}_{\dot{U}} + \nabla \cdot \underbrace{(-\nabla \phi)}_{\mathbf{f}} = 0
\end{equation}

\noindent with the heat source term being also $r=0$ and $\phi \equiv T$.\\

\noindent \textbf{Phase field corrosion}. The modelling of stress-assisted corrosion requires defining a phase field equation to describe material dissolution and an equation to describe the long-range transport of metal ions, in addition to considering mechanical equilibrium. The heat transfer analogy can be exploited to model both short-range interactions (corrosion) and long-range interactions (metal ion transport). The latter is given by Eq. (\ref{eq:corrosion-Con}), which can be reformulated as,
\begin{equation}\label{Eq:Coros}
\underbrace{1}_{\rho} \underbrace{\dot{c}}_{\dot{U}} + \nabla \cdot \underbrace{\left[-D_m \nabla \left(c - h(\phi)(c_{\mathrm{Se}} - c_{\mathrm{Le}}) - c_{\mathrm{Le}}\right)\right]}_{\mathbf{f}} = 0.
\end{equation}

\noindent with $r=0$ and the primary variable, the normalised concentration of metal ions, being analogous to the temperature ($c \equiv T$). On the other side, the phase field evolution equation for the corrosion problem, Eq. (\ref{Eq:phiEvol-Cor11}), can be expressed as,
\begin{equation}\label{Eq:phasefieldCoro}
\underbrace{1}_{\rho} \underbrace{\left(-\frac{1}{ L} \dot{\phi} - \frac{\partial \psi^{\mathrm{ch}}(c, \phi)}{\partial \phi}\right)}_{\dot{U}} + \nabla \cdot \underbrace{(\kappa \nabla \phi)}_{\mathbf{f}}  = 0.
\end{equation}

\noindent where $r=0$ and $\phi \equiv T$.\\

\noindent \textbf{Hydraulic fracture}. In addition to using the thermal analogy to implement the phase field fracture equation, tackling hydraulic fracture problems requires determining the evolution of the fluid pressure $p$. This can readily be achieved by establishing $p \equiv T$, and reformulating the mass conservation equation (\ref{Eq:dMassContent5}) as:
\begin{equation}\label{Eq:dMassContent6}
\underbrace{\rho_{\text{fl}}}_{\rho} \underbrace{\left(S(\alpha(\phi), n_{\text{p}}(\phi)) \dot{p} + \alpha(\phi) \chi_{\text{r}}(\phi) \dot{\varepsilon}_{\text{vol}} \right)}_{\dot{U}} + \nabla \cdot \underbrace{\left(-\rho_{\text{fl}} \frac{\bm{K}_{\text{fl}}(\phi)}{\mu_{\text{fl}}} \nabla p \right)}_{\mathbf{f}} - \underbrace{q_m}_{r} = 0.
\end{equation}

\noindent \textbf{Hydrogen embrittlement}. The prediction of hydrogen-assisted fractures using phase field requires solving the phase field evolution equation, as discussed above, but also considering the role of hydrogen in degrading $G_c$, and the hydrogen transport equation. The latter can also be made analogous to the heat transfer problem by taking $c_{\mathrm{H}} \equiv T$ and reformulating the hydrogen transport problem, Eqs. (\ref{eq:Htransport})-(\ref{eq:HtransportFlux}), as
\begin{equation}\label{Eq:HydrogenAna}
\underbrace{1}_{\rho} \underbrace{\dot{c}_{\mathrm{H}}}_{\dot{U}} + \nabla \cdot \underbrace{\left(-D_{\mathrm{H}} \nabla c_{\mathrm{H}} + \frac{D_{\mathrm{H}}}{R T_{\mathrm{k}}} c_{\mathrm{H}} V_{\mathrm{H}} \nabla \sigma_h\right)}_{\mathbf{f}} = 0.
\end{equation}

\noindent with $r=0$.\\

Finally, we also provide for generality how the balance equation of the heat transfer problem, Eq. \eqref{Eq:SHeatPhase}, can be given in the form of Eq. (\ref{Eq:GHEAT1}); i.e.,
\begin{equation}\label{Eq:SHeatEQ}
\rho \underbrace{c_{T} \dot{T}}_{\dot{U}} + \nabla \cdot \underbrace{(-\color{black}k_{T} \nabla T)}_{\mathbf{f}} - \underbrace{q_{T}}_{r} = 0 \, ,
\end{equation}

For numerical implementation purposes, one also needs to build the appropriate terms of the stiffness matrix, which requires defining the variation of internal thermal energy per unit mass with respect to temperature $\partial U / \partial T$, the variation of internal thermal energy per unit mass with respect to the spatial gradients of temperature $\partial U / \partial (\nabla T)$, the variation of the heat flux vector with respect to temperature $\partial \mathbf{f} / \partial T$, and the variation of the heat flux vector with respect to the spatial gradients of temperature $\partial \mathbf{f} / \partial (\nabla T)$. These are provided in \ref{Sec:AppDerivativesUMATHT} for each of the physical phenomena considered here. For the sake of generality, complement implementation details (including definition of discretised residuals and stiffness matrix components) are given in \ref{App:FEdetails}, although these details are not used here, where the implementation is carried out entirely at the integration point level.

\subsection{Abaqus implementation}

Exploiting the thermal analogy can significantly simplify the numerical implementation. This is here demonstrated in the context of the commercial software Abaqus, showing how complex multi-field coupled problems can be implemented at the integration point level. In the past, we showed how the coupled deformation-phase field fracture problem could be implemented into Abaqus using only a user material (\texttt{UMAT}) subroutine \cite{Navidtehrani2021a} or, for versions older than 2020,  a user material (\texttt{UMAT}) subroutine and an internal heat generation (\texttt{HETVAL}) user subroutine \cite{Navidtehrani2021}. A notable advantage of these methods is their ability to operate at the integration point level, thereby eliminating the need for element-level implementation. This approach also enables the use of Abaqus's built-in features, such as various element types and contact interactions. However, a new paradigm is needed here, as the heat analogy is exploited to treat simultaneously two or more equations and only one temperature degree of freedom can be defined in Abaqus.\\ 

First, the use of a user material heat transfer (\texttt{UMATHT}) subroutine is suggested, in combination with a \texttt{UMAT}. This provides greater flexibility as it enables to conduct transient analyses with history-dependent variables. This is, for example, required to implement the phase field corrosion equation, Eq. (\ref{Eq:phiEvol-Cor11}), as exploiting the heat transfer analogy using a \texttt{UMAT} (or \texttt{HETVAL}) only allows to define the quantity $\rho c_p$ as the coefficient multiplying $\partial T/\partial t$; $\rho c_p$ is a constant quantity, with $c_p$ being the specific heat,
but the equivalent term $1/L(\sigma_h, \varepsilon^p)$ is not constant in time. In terms of computational efficiency, both approaches (UMAT/HETVAL vs UMAT/UMATHT) are equivalent.\\

\begin{figure}[H]
    \centering
    \includegraphics[width=1\linewidth]{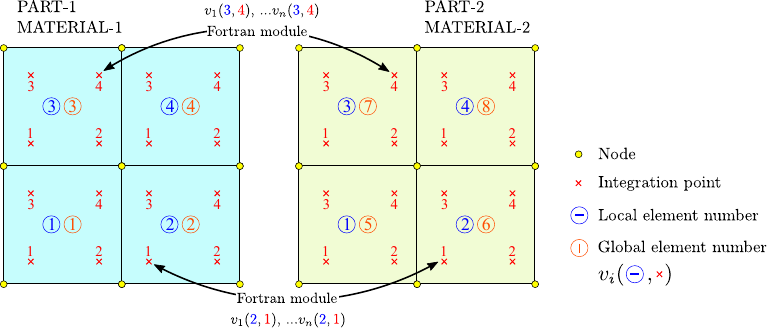}
    \caption{Conjugate pairs of elements for identical geometry and mesh discretization.}
    \label{fig:ConjugateElements}
\end{figure}

The second key innovation of our implementation is the treatment of multiple temperature-like degrees-of-freedom, as required to simultaneously solve several diffusion-like equations (e.g., phase field fracture and hydrogen transport, in the case of hydrogen embrittlement). Thus, we introduce a \emph{twin-part} method, whereby a second part (\texttt{PART-2}) is defined, which duplicates the geometry and mesh of the primary part (\texttt{PART-1}). This is straightforward in Abaqus, as it just implies creating a copy of \texttt{PART-1}, once meshed. While both parts share geometric properties and meshing, they differ in material definitions (\texttt{MATERIAL-1} for \texttt{PART-1} and \texttt{MATERIAL-2} for \texttt{PART-2}) and boundary conditions. Each part includes degrees of freedom for displacement and temperature, and hence the use of conjugate elements enables additional degrees of freedom. For example, the temperature degree of freedom in \texttt{PART-1} represents the phase field ($\phi \equiv T$), while in \texttt{PART-2}, an additional degree of freedom $\xi$ is taken to be analogous to temperature ($\xi \equiv T$). Materials can be differentiated within \texttt{UMAT} and \texttt{UMATHT} subroutines using the \texttt{CMNAME} variable, which allows assignment of distinct materials to each part. Data transfer within the parts is facilitated by the identical local element numbering used by Abaqus when a geometrically identical part with the same mesh is created; the idea of \emph{conjugate elements}, as illustrated in Fig. \ref{fig:ConjugateElements}. In Abaqus, both global and local element numbering systems are used, with local element numbering being specific to each part instance within the model assembly. \\

\begin{figure}[H]
    \centering
    \includegraphics[width=0.58\linewidth]{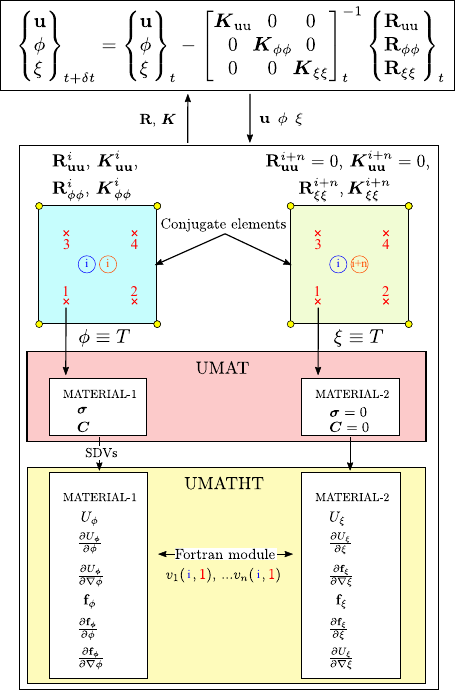}
    \caption{Twin-part procedure to implement coupled problems in Abaqus involving multiple diffusion-type equations. Sketch describing the protocol followed by Abaqus at the integration point level, the element level, and the global equation level. For illustrative purposes, the first part is used to define the balance equations for the displacement field and the phase field, while the second part is used for an additional variable, but this can be changed depending on the solution scheme desired.}
    \label{fig:Coupling}
\end{figure}

The procedure proposed for implementing coupled problems involving multiple diffusion-type equations in Abaqus is described in Fig.  \ref{fig:Coupling}. Abaqus proceeds following the global element number. Thus, for the first element (belonging to \texttt{PART-1}) and the first integration point, the \texttt{UMAT} subroutine is called first. There, the  Cauchy stress $\bm{\sigma}$ and material Jacobian $\bm{C}$ are computed from the strain tensor $\bm{\varepsilon}$ and any other relevant variable, coming from either the inputs of the \texttt{UMAT} (e.g., $T$, representing $\phi$) or from the Fortran module used to communicate between the parts (e.g. $\xi$). The stress tensor and material Jacobian are used by Abaqus to construct the relevant residual and stifness matrix components. Then, within that first integration point, the \texttt{UMATHT} subroutine is called. There, the internal heat energy $U$ and the heat flux vector $\mathbf{f}$ must be defined to construct the residual vector $\mathbf{R}$, along with their variations with respect to temperature $T$ and its gradient $\nabla T$ to form the stiffness matrix $\bm{K}$. This is done in agreement with the definitions provided previously in this section and those given in \ref{Sec:AppDerivativesUMATHT}. Data exchange between the \texttt{UMAT} and \texttt{UMATHT} subroutines can be done in a straightforward manner using state variables (\texttt{SDV}s). \\ 
 
Once Abaqus has looped over all the elements (integration points) in \texttt{PART-1}, it proceeds to \texttt{PART-2}. The \texttt{UMAT} is first called, where $\bm{\sigma}=0$ and $\bm{C}=0$ for the material of the second part (\texttt{MATERIAL-2}). Then, the \texttt{UMATHT} subroutine is called, where one defines $U$, $\mathbf{f}$, and their derivatives for $\xi$, the additional variable of interest. To communicate between the parts, a Fortran module is used, together with the Abaqus utility routine \texttt{GETPARTINFO}, which provides the local element number.\\

\begin{figure}[H]
    \centering
    \includegraphics[width=0.5\linewidth]{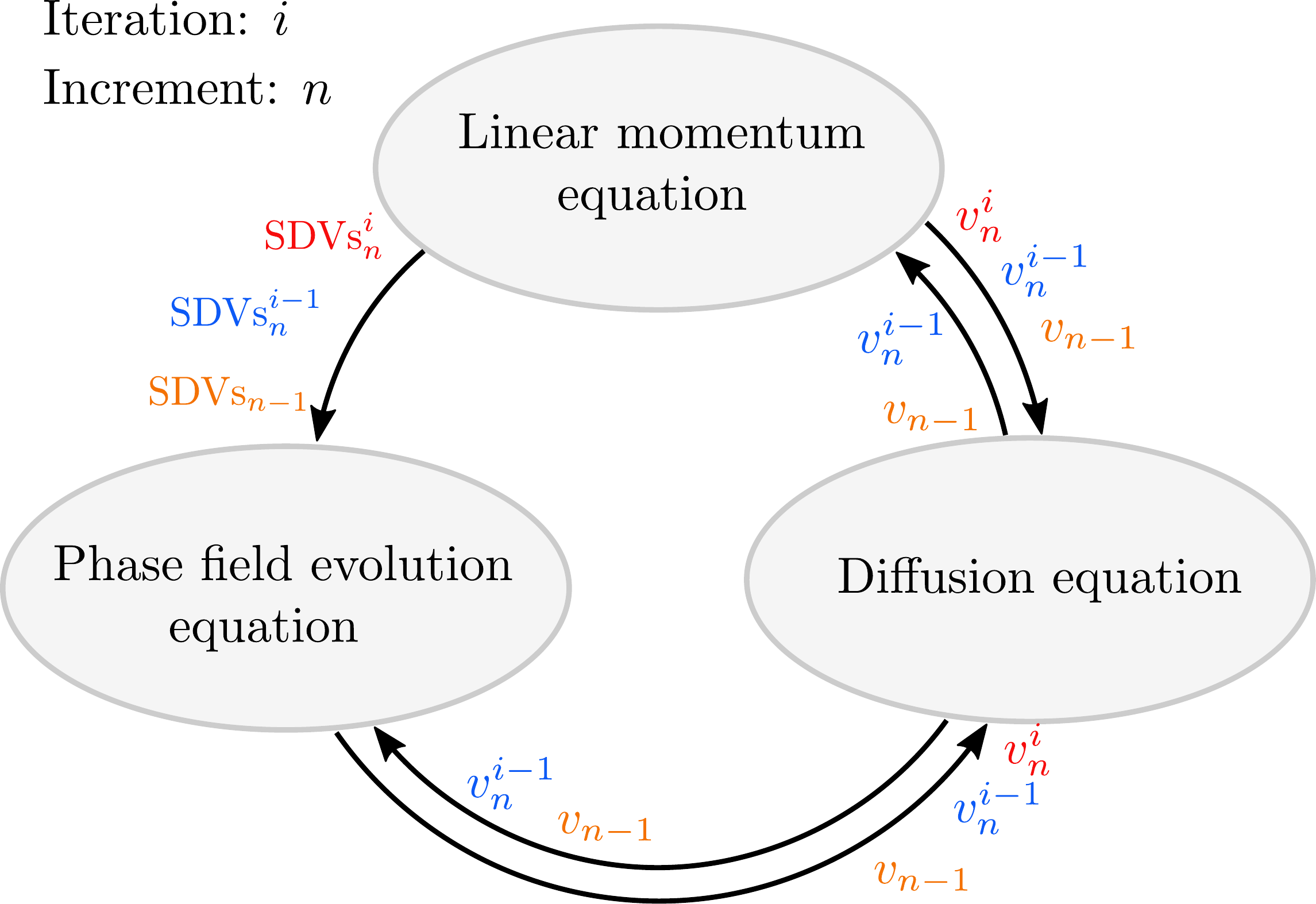}
    \caption{Solution and data exchange schemes available in the current implementation.}
    \label{fig:Scheme}
\end{figure}

In terms of solving the coupled equations, various schemes exist. The monolithic scheme updates all variables simultaneously using the backward Euler method, which is unconditionally stable but can suffer from convergence issues. Conversely, the staggered method, akin to the forward Euler method, updates some primary variables while holding others constant, being more stable but potentially requiring small increments for accuracy. In our approach, the variables solved within one part can be coupled in a monolithic way but the coupling with the third variable must be done in a staggered fashion. Both single-pass and multi-pass staggered approaches are possible, as illustrated in Fig. \ref{fig:Scheme}, where data exchange is shown for an increment $n$ and an iteration $i$. The inability to incorporate fully monolithic approaches, which can be made robust through the use of quasi-Newton schemes \cite{TAFM2020}, is a disadvantage of this approach relative to user element-based implementations. However, the present approach circumvents the coding, validation, and pre- and post-processing issues associated with developing user element subroutines. The present implementation can also be extended to an arbitrary number of fields governed by diffusion-type equations by defining additional \texttt{PARTs} for each new diffusion field — for example, to simultaneously account for hydrogen effects and corrosion. The user subroutines developed are shared freely with the community and are available to download at \url{https://mechmat.web.ox.ac.uk/codes}.

\section{Representative examples}
\label{Sec:Examples}

In this section, we aim to validate and illustrate the capabilities of our multiphysics phase field implementation across different applications. This section includes detailed simulations that replicate established experiments and simulation benchmarks, showcasing the model's versatility and accuracy in various coupled problems. Specifically, we explore case studies involving thermo-mechanical fracture (Section \ref{sec:Quenching}), modelling quenching in ceramic materials; hydraulic fracture (Section \ref{sec:Hydraulic_Fracture}), under conditions that allow validating against an analytical solution for the critical fluid pressure; hydrogen embrittlement (Section \ref{sec:Hydrogen_Embrittlement}), addressing a key validation benchmark; and stress-assisted corrosion (Section \ref{sec:Corrosion-Stress1}), whereby two classic numerical experiments are conducted.

\subsection{Thermo-mechanical fracture: quenching}
\label{sec:Quenching}

To validate our phase field thermal fracture model, we replicate the classic quenching experiment by Jiang et al. \cite{Jiang2012}. The experiment involves immersing a ceramic plate (50 mm x 10 mm), which has been pre-heated up to a temperature $T_0$, into a water bath held at ambient temperature ($T_a = 20 \, ^{\circ}$C). The resulting change in temperature causes thermal fracture, with cracks nucleating at the outer surface and growing in parallel towards the centre of the plate. This problem has been used widely as a benchmark of thermal phase field fracture modelling. The material properties used, taken from the literature \cite{Jiang2012,Sicsic2014,RUAN2023105169}, are listed in Table \ref{tab:Quenching}.

\begin{table}[H]
\caption{Material properties for the quenching case study \cite{Jiang2012,Sicsic2014,RUAN2023105169}.}
    \centering
    \begin{tabular}{lll}
      \hline
      Parameter  &  Value  &  Unit  \\
      \hline  \hline
      Density $\rho$ & 3980 & kg/$\text{m}^3$ \\
      Young's modulus $E$  & 370 & GPa \\
      Poisson's ratio $\nu$  & 0.3 &  \\
      Phase field length scale $\ell$ & 0.1 & mm \\
      Toughness $G_c$ &  42.47 & J/$\text{m}^2$ \\
      Undamaged thermal conductivity $k_0$ & 31 & W/kg K \\
      Heat capacity $c_{T}$ & 880 & J/kg K \\
      Thermal expansion coefficient $\alpha_{T}$ & $7.5 \times 10^{-6}$ & $\text{K}^{-1}$ \\
      \hline
    \end{tabular}
    \label{tab:Quenching}
\end{table}

\begin{figure}[H]
    \centering
    \begin{subfigure}[b]{0.45\textwidth}
         \centering
         \includegraphics[width=\textwidth]{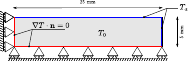}
         \caption{}
     \end{subfigure}\\ \vspace{3mm}
    \begin{subfigure}[b]{0.45\textwidth}
         \centering
         \includegraphics[width=\textwidth]{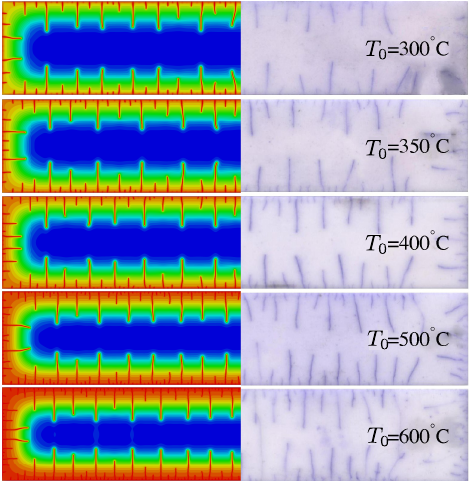}
         \caption{}
         \label{fig:Quenching-phi-a}
     \end{subfigure}
     \begin{subfigure}[b]{0.45\textwidth}
         \centering
         \includegraphics[width=\textwidth]{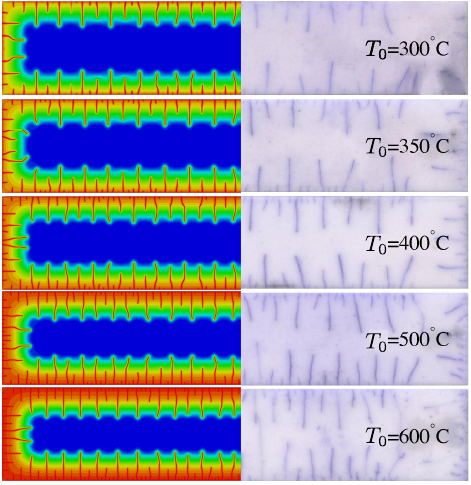}
         \caption{}
         \label{fig:Quenching-phi-b}
     \end{subfigure}\\ \vspace{3mm}
    \begin{subfigure}[b]{0.4\textwidth}
         \centering
         \includegraphics[width=\textwidth]{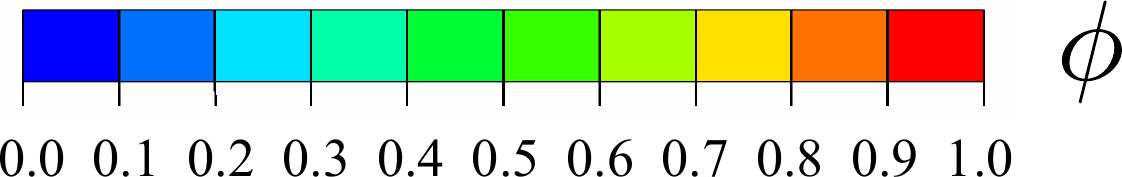}
     \end{subfigure} 
    \caption{Thermo-mechanical fracture (quenching) case study. (a) Geometry and boundary conditions. Comparison between simulation (phase field contours $\phi$) and experimental crack patterns \cite{Jiang2012}. Two scenarios are considered: (b) a non-degraded thermal conductivity ($k_T=k_0$), and (c) a degraded thermal conductivity $k_T=g(\phi) k_0$.}
    \label{fig:Quenching-phi}
\end{figure}

The ceramic plate is discretized using a uniform grid of 4-node plane strain thermally coupled quadrilateral elements (CPE4T in Abaqus) with a characteristic element size of 0.0025 mm. The staggered scheme couples the linear momentum and phase field equations, using a fixed increment size of 0.1 ms over 200 ms of total simulation time. No decomposition of strain energy is considered for the fracture driving force. Due to symmetry, only a quarter of the plate is modelled, as shown in Fig. \ref{fig:Quenching-phi}a. The range of initial temperatures considered is $T_0=\{300, 350, 400, 500, 600\} \, ^{\circ}$C and the boundary condition prescribed in the outer surface of the plate is $T_a = 20 \, ^{\circ}$C.\\

The results obtained are shown in Fig. \ref{fig:Quenching-phi}b,c, in terms of the phase field contour, for the cases where the thermal conductivity is independent of the phase field ($k_T=k_0$) and for the case where it is degraded ($k_T = g(\phi) k_0$). The high-temperature gradient causes uniform damage at the perimeter of the plate, with cracks propagating inwards with uniform spacing. In both experiments and simulations, higher initial temperatures (i.e., higher gradients) result in a higher number of cracks, as expected. The agreement between modelling and experiments is notable, in terms of the number of cracks, their spacing and their extension. Numerical predictions are also in good agreement with the computational literature \cite{Bourdin2014,RUAN2023105169,Tang2016,Chu2017,MANDAL2021113648}. Differences between the degraded and undegraded thermal conductivity calculations are small, but the latter appears to be in closer agreement with experiments (fewer and longer cracks), likely due to anisotropic thermal conductivity post-cracking and potential heat transfer through closed cracks. It is also worth noting that the conventional AT2 phase field model does not include a damage threshold, which can result in a larger degree of distributed damage near the outer surface.

\subsection{Hydraulic fracture}
\label{sec:Hydraulic_Fracture}

The verification of the hydraulic phase field fracture implementation is carried out through two case studies. In the first one, we examine the growth of a pressurized crack located in the centre of a square domain. This analysis aims to validate our estimation of the critical pressure, which represents the water pressure at the time of crack propagation and is denoted as $p_c$. To this end, a comparison with an existing analytical solution is conducted. Subsequently, the second case study is dedicated to simulating an injection-driven fracture scenario in a 3D model, shedding light on the interaction of preexisting cracks. The material parameters listed in Table \ref{tab:ParametersPC} are considered for both case studies. Also, in these analyses, we adopt the no-tension strain energy decomposition \cite{DelPiero1989,Freddi2010} as the fracture driving force, accounting for the anisotropic influence of strain energy decomposition (see \ref{Sec:TSM}). 

\subsubsection{Pressurized crack}

\begin{figure}[H]
    \centering
    \includegraphics{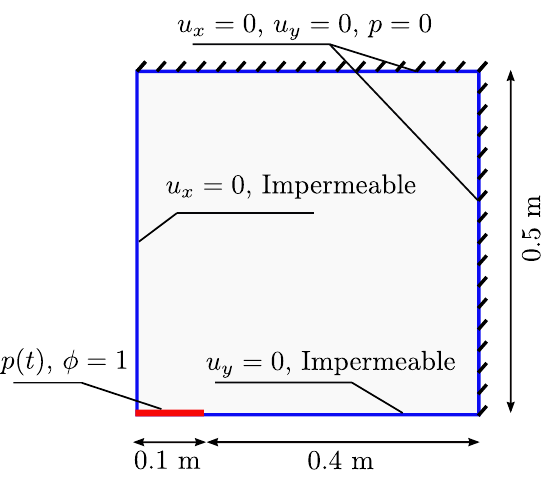}
    \caption{Geometry, dimensions, and boundary conditions of the pressurized crack case study.}
    \label{fig:PC-Config}
\end{figure}

The first case study involves a square domain featuring a centred crack, subjected to a gradually increasing pressure up to \(p=100\) MPa over a period of 2,000 seconds. In practice, this is implemented by defining a temperature Dirichlet boundary condition, exploiting the analogy between pressure and temperature. The geometric configuration and boundary conditions of the model are depicted in Fig. \ref{fig:PC-Config}, with only a quarter of the model simulated to exploit its inherent double symmetry.\\

\begin{table}[H]
\caption{Solid and fluid materials parameters for hydraulic fracture case studies adopted in the analysis of hydraulic fracture.}
    \centering
    \begin{tabular}{lll}
      \hline
      Parameter  &  Value  &  Unit  \\
      \hline  \hline
      Young's modulus $E$  & 210 & GPa \\
      Poisson's ratio $\nu$  & 0.3 &  \\
      Characteristic length scale $\ell$ & 4 & mm \\
      Toughness $G_c$ &  2700 & J/$\text{m}^2$ \\
      Biot's coefficient of reservoir domain $\alpha_{\text{r}}$ & 2 $\times \, 10^{-3}$ & \\
      Porosity in the reservoir domain $n_{\text{pr}}$& 2 $\times \, 10^{-3}$ & \\
      Mass density of the fluid $\rho_{\text{fl}}$& 1000 kg/$\text{m}^3$ & kg/$\text{m}^3$ \\
      Fluid viscosity $\mu_{\text{fl}}$ & 1 $ \times \,10^{-3}$ & Pa$\cdot$s \\
      Fluid compressibility $C_{\text{fl}}$ & 1 $ \times \,10^{-8}$ & $\text{Pa}^{-1}$ \\
        Permeability tensor of reservoir domain $\bm{K}_{\text{r}}$ & 1 $ \times \,10^{-15} \bm{I}$ & $\text{m}^2$ \\
        Permeability tensor of fracture domain $\bm{K}_{\text{f}}$ & 1.333 $\times \, 10^{-6} \bm{I}$ & $\text{m}^2$ \\
        First constants for domain indicator fields $c_1$ & 0.4 & \\
        Second constants for domain indicator fields $c_2$ & 1 & \\
        Transient parameter $b$ & 1.0 & \\
      \hline
    \end{tabular}
    \label{tab:ParametersPC}
\end{table}

The finite element mesh comprised 14,443  eight-node biquadratic displacement, bilinear temperature elements (Abaqus type CPE8T). The mesh was refined in the region where cracking was anticipated, ensuring that the smallest element size was maintained at one-fifth of the characteristic length \(\ell\). The monolithic scheme was employed to solve the coupled deformation-fracture problem, while a multi-pass staggered scheme is adopted to handle the coupling with the fluid equation. 100 increments are used, each spanning 20 seconds.

\begin{figure}[H]
    \centering
    \begin{subfigure}[b]{\textwidth}
         \centering
         \includegraphics[width=0.25\textwidth]{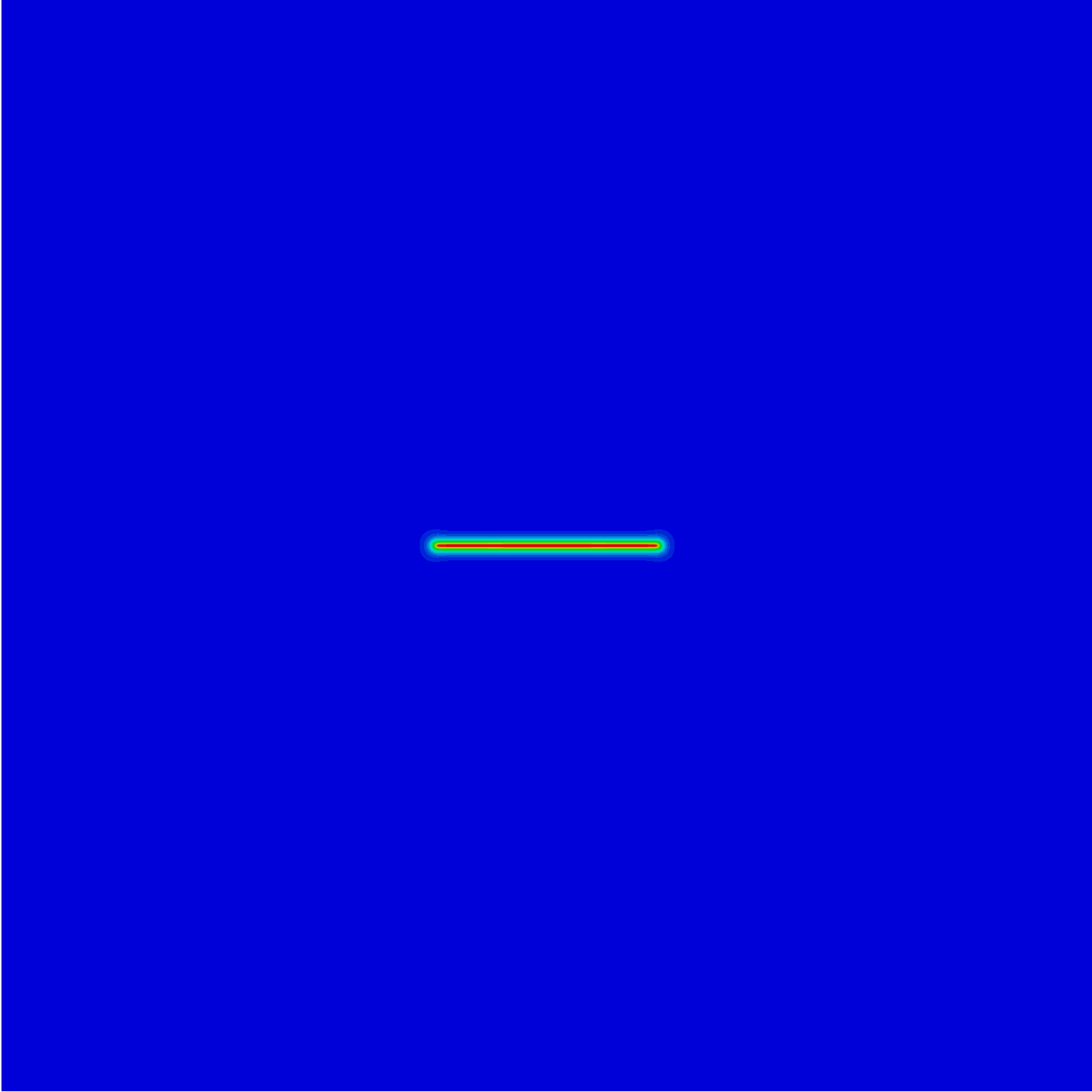}
         \hspace{5mm}
         \includegraphics[width=0.25\textwidth]{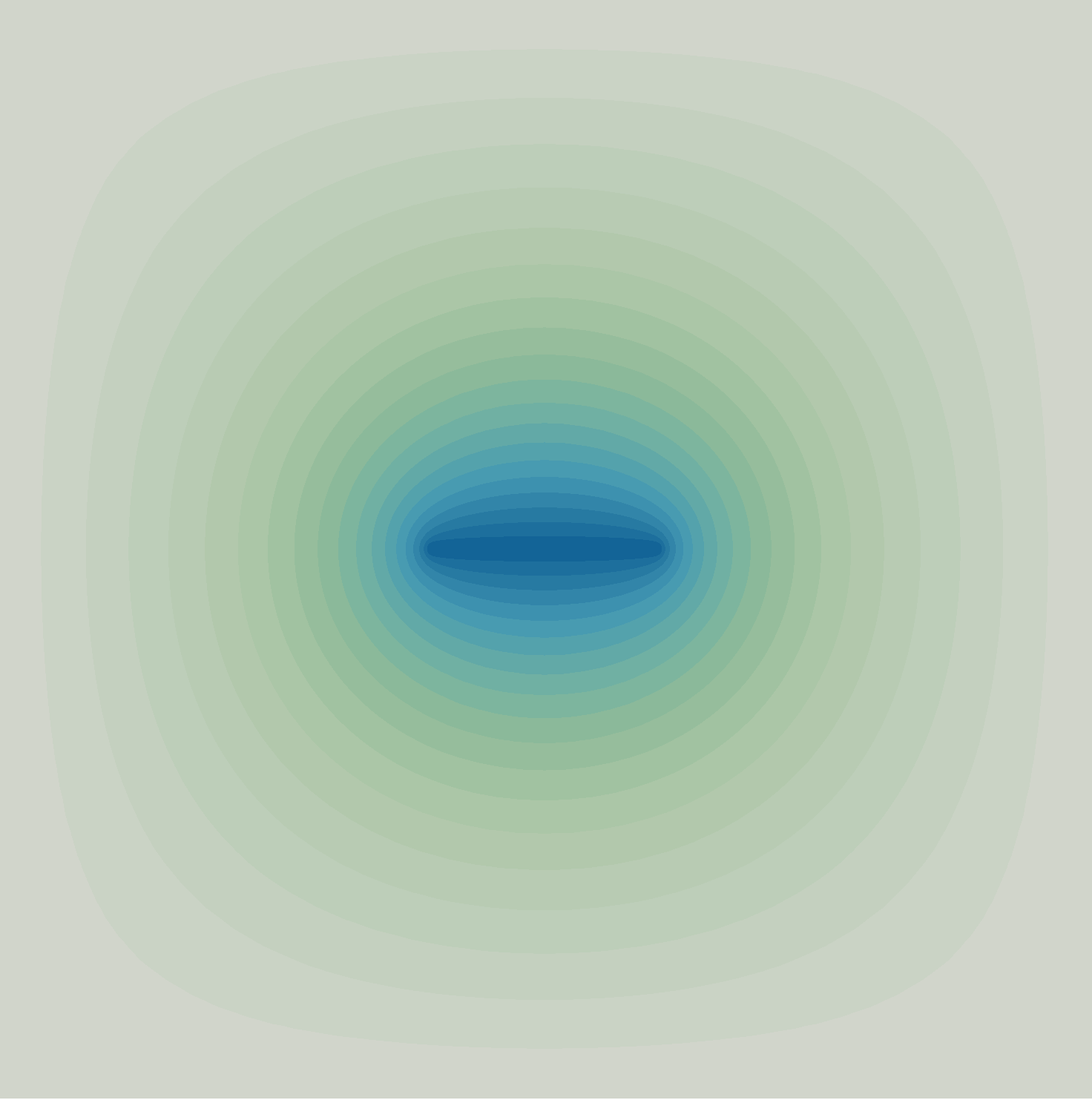}
         \caption{Time $t=1140$ s, $p_{\text{center}}=57$ MPa}
         \label{fig:PC-result-a}
     \end{subfigure} \\ \vspace{5 mm}
     \begin{subfigure}[b]{\textwidth}
         \centering
         \includegraphics[width=0.25\textwidth]{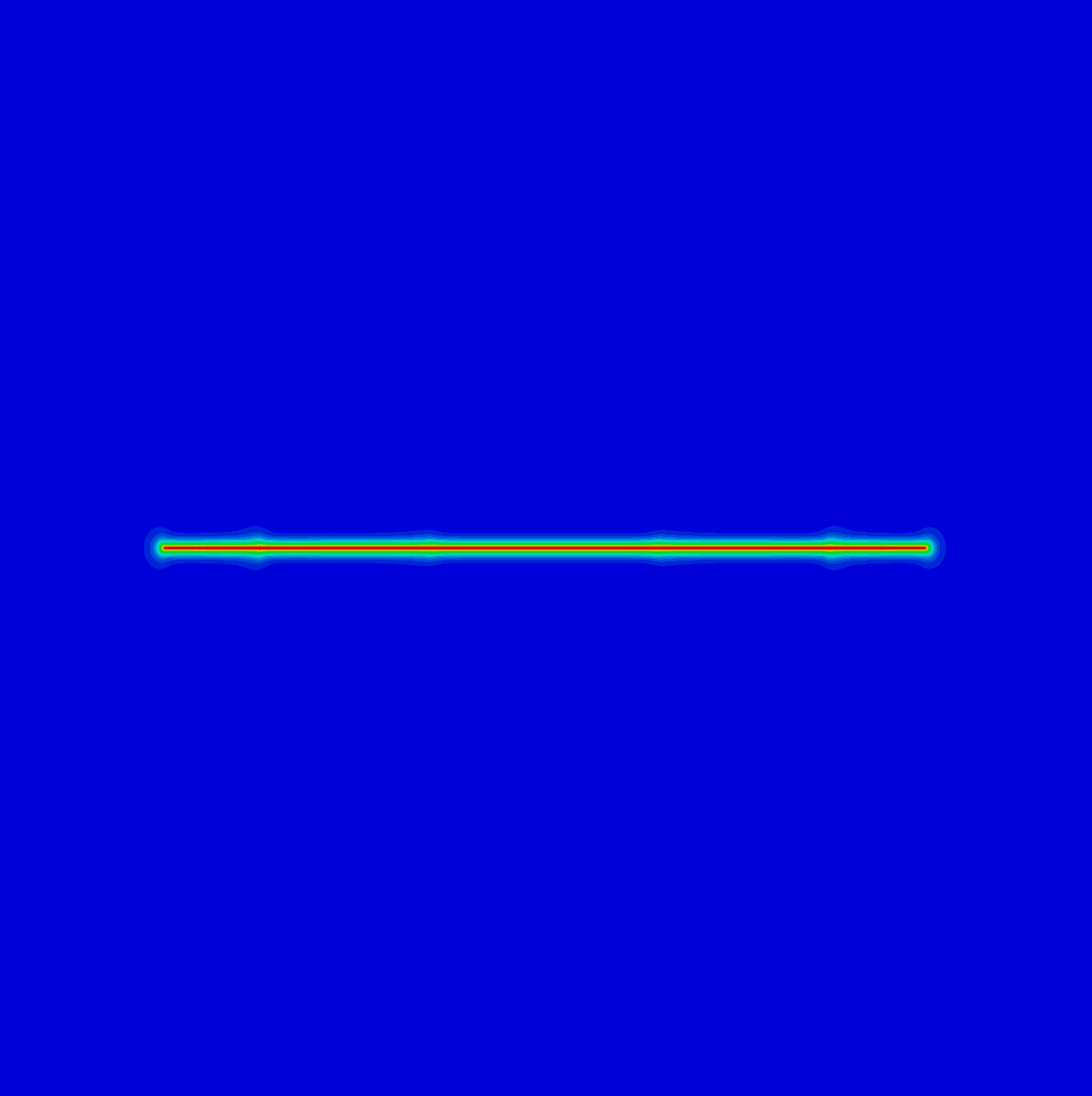}
         \hspace{5 mm}
         \includegraphics[width=0.25\textwidth]{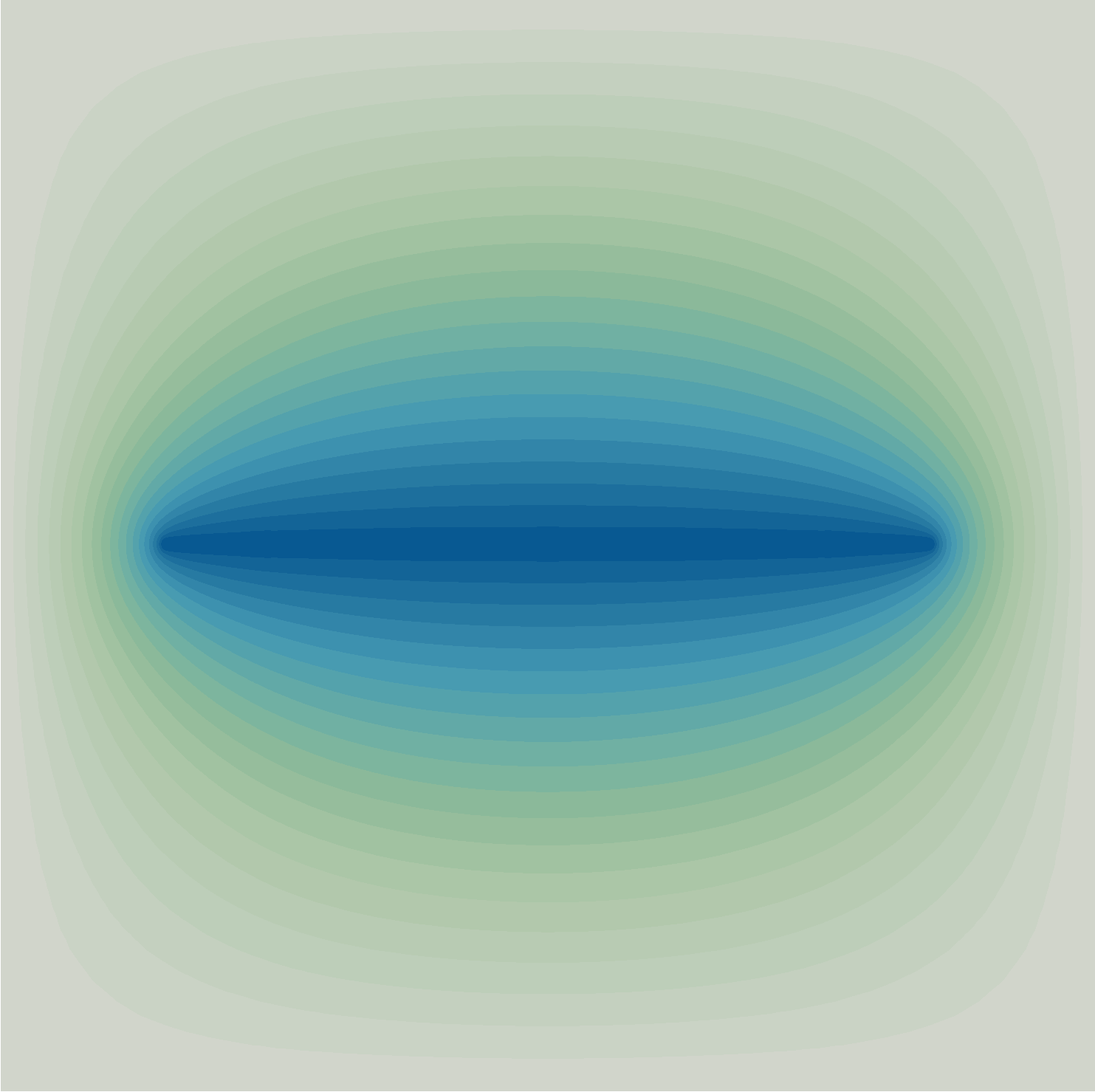}
         \caption{Time $t=1200$ s, $p_{\text{center}}=60$ MPa}
         \label{fig:PC-result-b}
     \end{subfigure} \\ \vspace{5 mm}
     \begin{subfigure}[b]{\textwidth}
         \centering
         \includegraphics[width=0.12\textwidth]{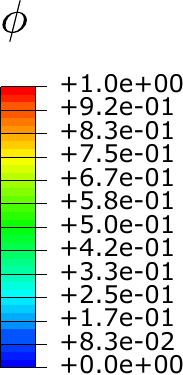}
        \hspace{5 mm}
         \includegraphics[width=.25\textwidth]{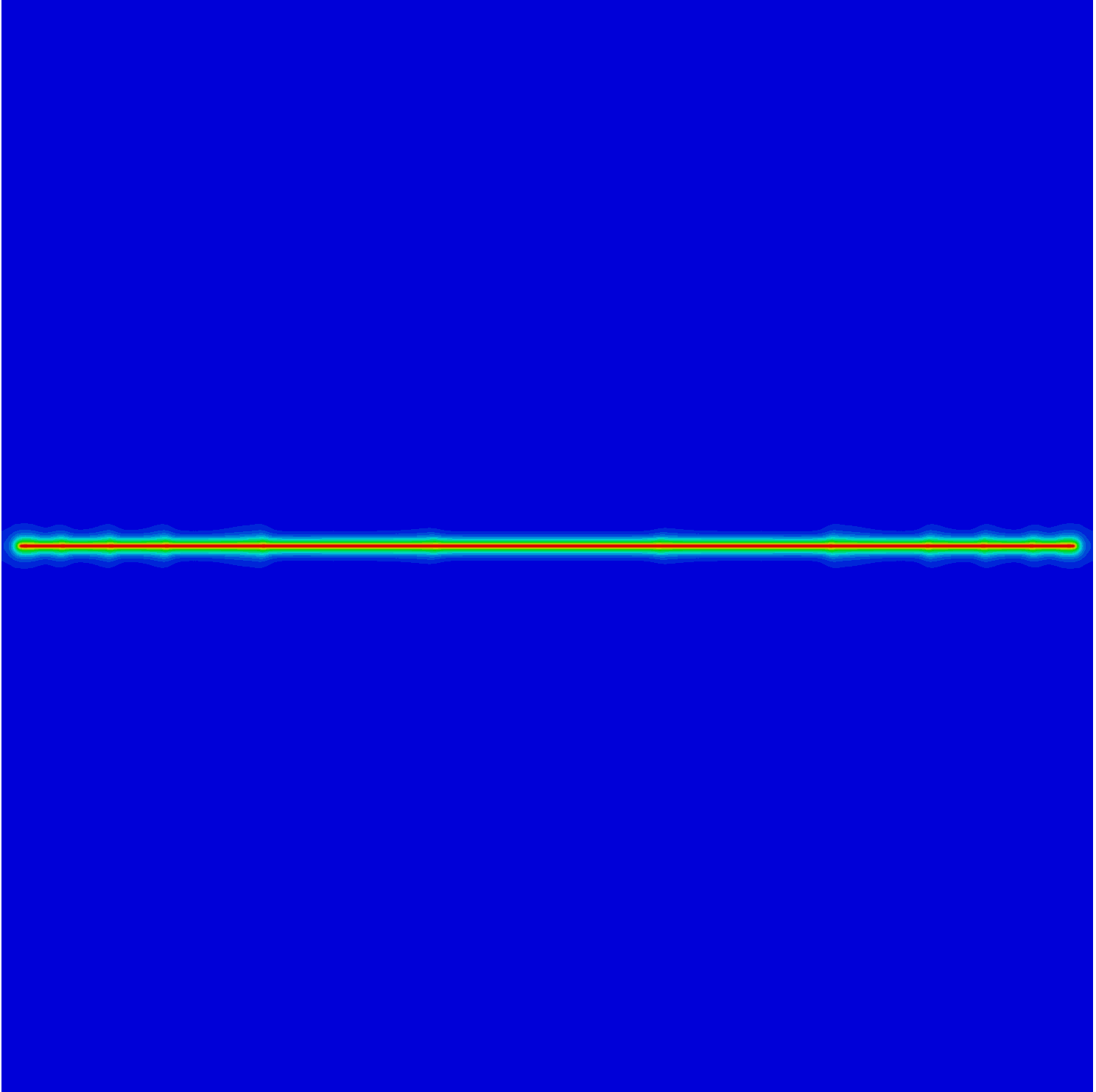}
        \hspace{5 mm}
         \includegraphics[width=.25\textwidth]{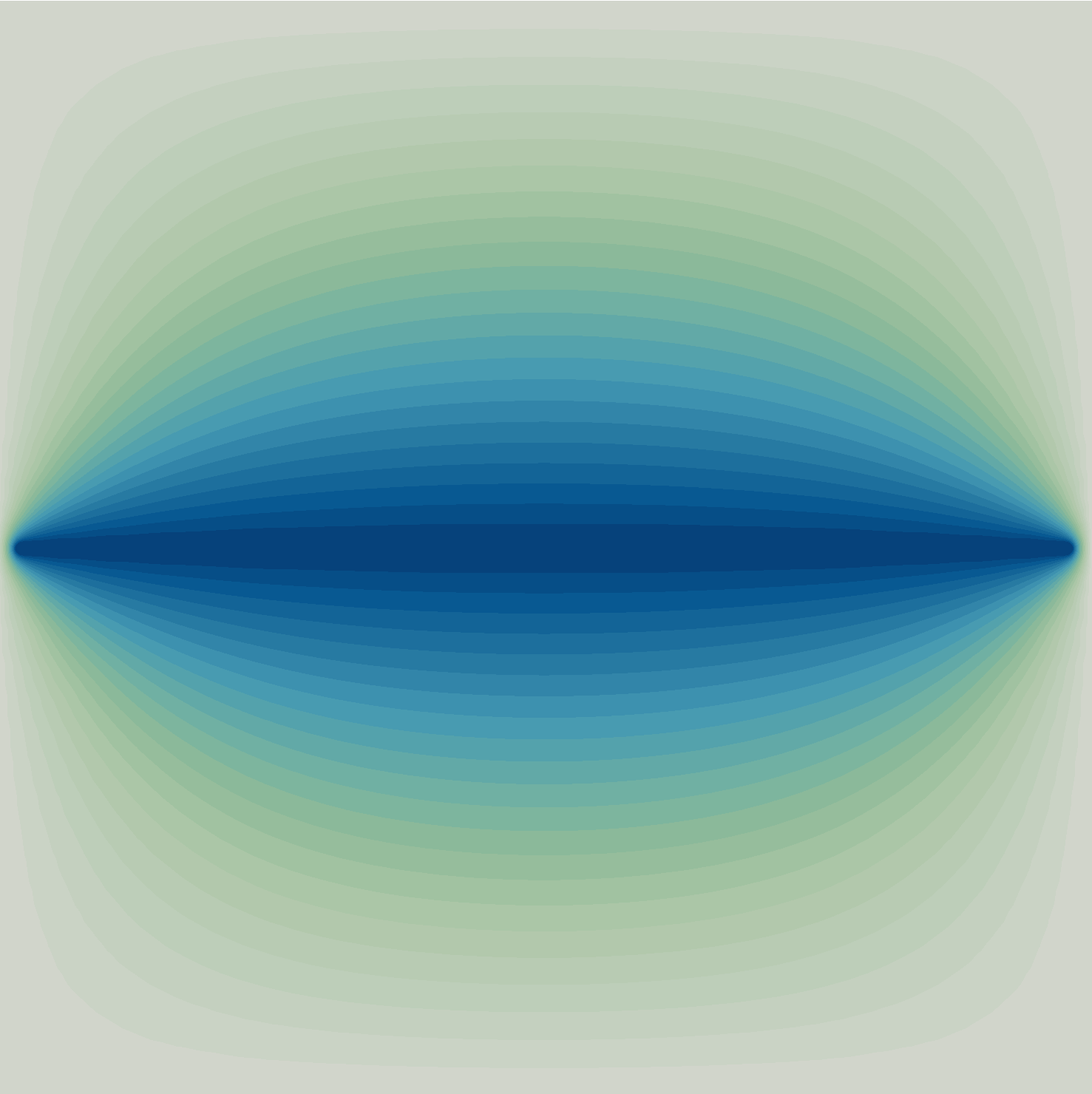}
         \hspace{5 mm}
         \includegraphics[width=0.12\textwidth]{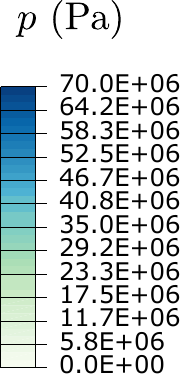}
         \caption{Time $t=1400$ s, $p_{\text{center}}=70$ MPa}
         \label{fig:PC-result-c}
     \end{subfigure}
    \caption{Contours of phase field $\phi$ (left) and fluid pressure $p$ (right) for a pressurized crack at different times: (a) $t=1140$ s, $p_{\text{center}}=57$ MPa, (b) $t=1200$ s, $p_{\text{center}}=60$ MPa, and (c) $t=1400$ s, $p_{\text{center}}=70$ MPa.}
    \label{fig:PC-result}
\end{figure}

The evolution of the crack and the pressure distribution are shown in Fig. \ref{fig:PC-result} for three selected time intervals: 1140, 1200 and 1400 s. The initiation of crack growth occurs at a time of 1140 s, as shown in Fig. \ref{fig:PC-result-a}, with a central domain pressure $p_{\text{center}}$ of $57$ MPa. Subsequenlty, the crack extends along the mode I crack trajectory until reaching the edge of the domain, with the associated increase in pressure along the cracked domain being appropriately captured by the model, see Figs. \ref{fig:PC-result}b and \ref{fig:PC-result}c. 

\begin{figure}[H]
    \centering
    \includegraphics[width=0.5\linewidth]{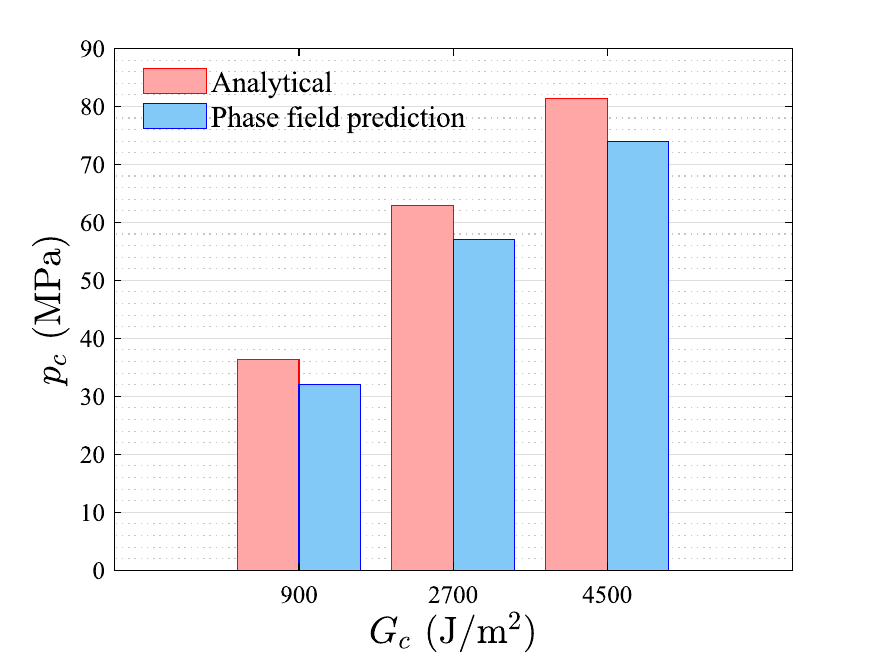}
    \caption{Comparison of critical fluid pressure $p_c$ predictions, as obtained from the analytical solution given in Eq. \eqref{eq:analytical} and through the present phase field model.}
    \label{fig:Critical-Pressure}
\end{figure}

We proceed to compare the critical pressure $p_c$ estimates of the model with the analytical solution by Yoshioka and Bourdin \cite{YOSHIOKA2016137} (see also Ref. \cite{Zhou2018c}). The dependency of $p_c$ on the initial crack length $a_0$, material toughness $G_c$ and elastic properties is given by,
\begin{equation}\label{eq:analytical}
p_c=\left(\frac{4 E' G_c}{\pi a_0}\right)^{\frac{1}{2}},
\end{equation}

\noindent where $E' = E/{(1-\nu^2)}$ is the plane strain Young's modulus, with $\nu$ being Poisson's ratio. The comparison between the critical pressure estimates from Eq. (\ref{eq:analytical}) and those from the present phase field-based numerical framework are shown in Fig. \ref{fig:Critical-Pressure}. Results are obtained for various choices of the material toughness $G_c$. Overall, the analytical and numerical results are in very good agreement, with the numerical results slightly underpredicting the fluid pressure at the time at crack propagation.

\subsubsection{3D cracking due to an injected fluid}

\begin{figure}[H]
    \centering
    \begin{subfigure}[b]{0.35\textwidth}
    \centering
        \includegraphics[width=\linewidth]{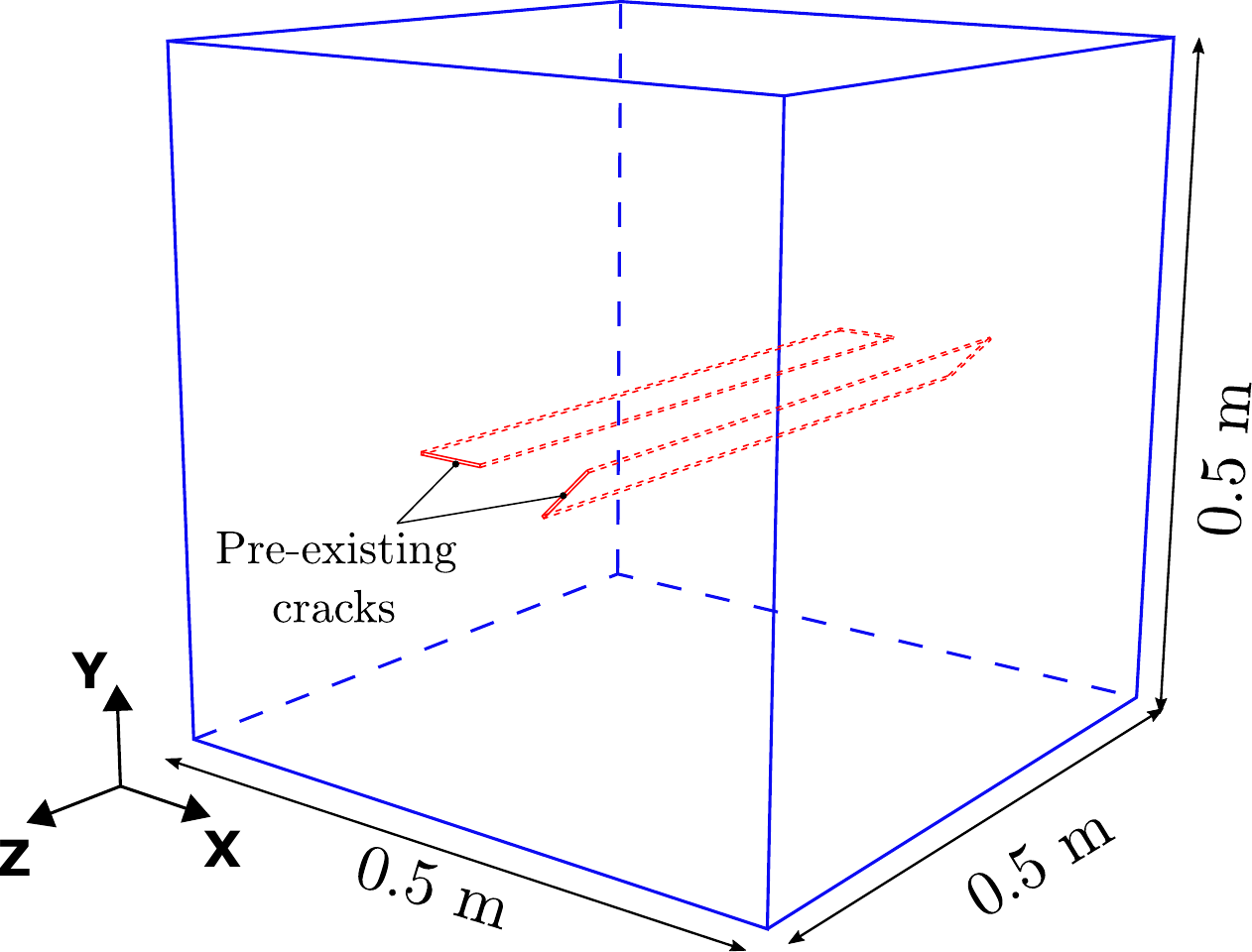}
        \caption{}
    \end{subfigure} \hspace{7mm}
    \begin{subfigure}[b]{0.3\textwidth}
    \centering
        \includegraphics[width=\linewidth]{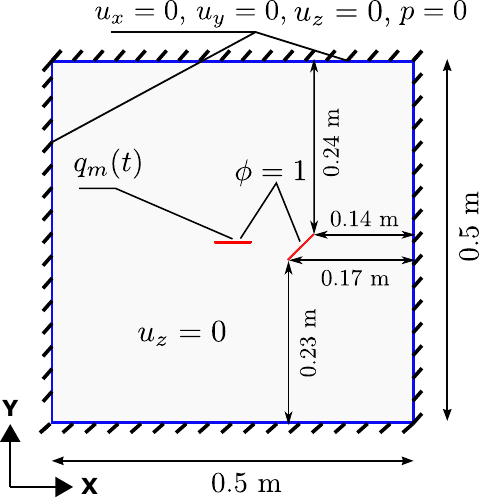}
        \caption{}
    \end{subfigure}
    \caption{Boundary value problem: Cubic domain with two preexisting cracks, (a) 3D geometry, and (b) boundary conditions shown in XY plane.}
    \label{fig:INJ-Inter-Config}
\end{figure}

The second case study investigates the interaction of pre-existing cracks in a three-dimensional configuration. A cubic domain with a characteristic length of 0.5 m is considered, featuring two cracks in the $XY$ plane, prolonged along the $Z$ direction (Fig. \ref{fig:INJ-Inter-Config}a), each with a length of 0.05 m. One crack is positioned horizontally at the centre of the domain in the XY plane, while the other crack is inclined and located away from the centre, as illustrated in Fig. \ref{fig:INJ-Inter-Config}b. The domain is discretized using approximately 268,000 8-node thermally coupled brick, trilinear displacement and temperature elements (C3D8T), with the characteristic finite element length being 2 mm. Different to the previous case study, crack growth is here driven by fluid injection. Specifically, a fluid source of $q_m=4000 \, \text{kg/(m}^3 \cdot \text{s)}$ is applied and held constant over a time of 300 seconds. This is achieved by defining a body heat flux ($r$ in Eq. \eqref{Eq:GHEAT1}) on the elements located in the central crack. A staggered solution scheme is employed in this model with a time increment of 1 second.

\begin{figure}[H]
    \centering
    \begin{subfigure}[b]{0.12\textwidth}
         \centering
         \includegraphics[width=\textwidth]{Legend-V-phi.pdf}
         \vspace{5mm}
     \end{subfigure}  \hspace{1mm}
    \begin{subfigure}[b]{0.35\textwidth}
         \centering
         \includegraphics[width=\textwidth]{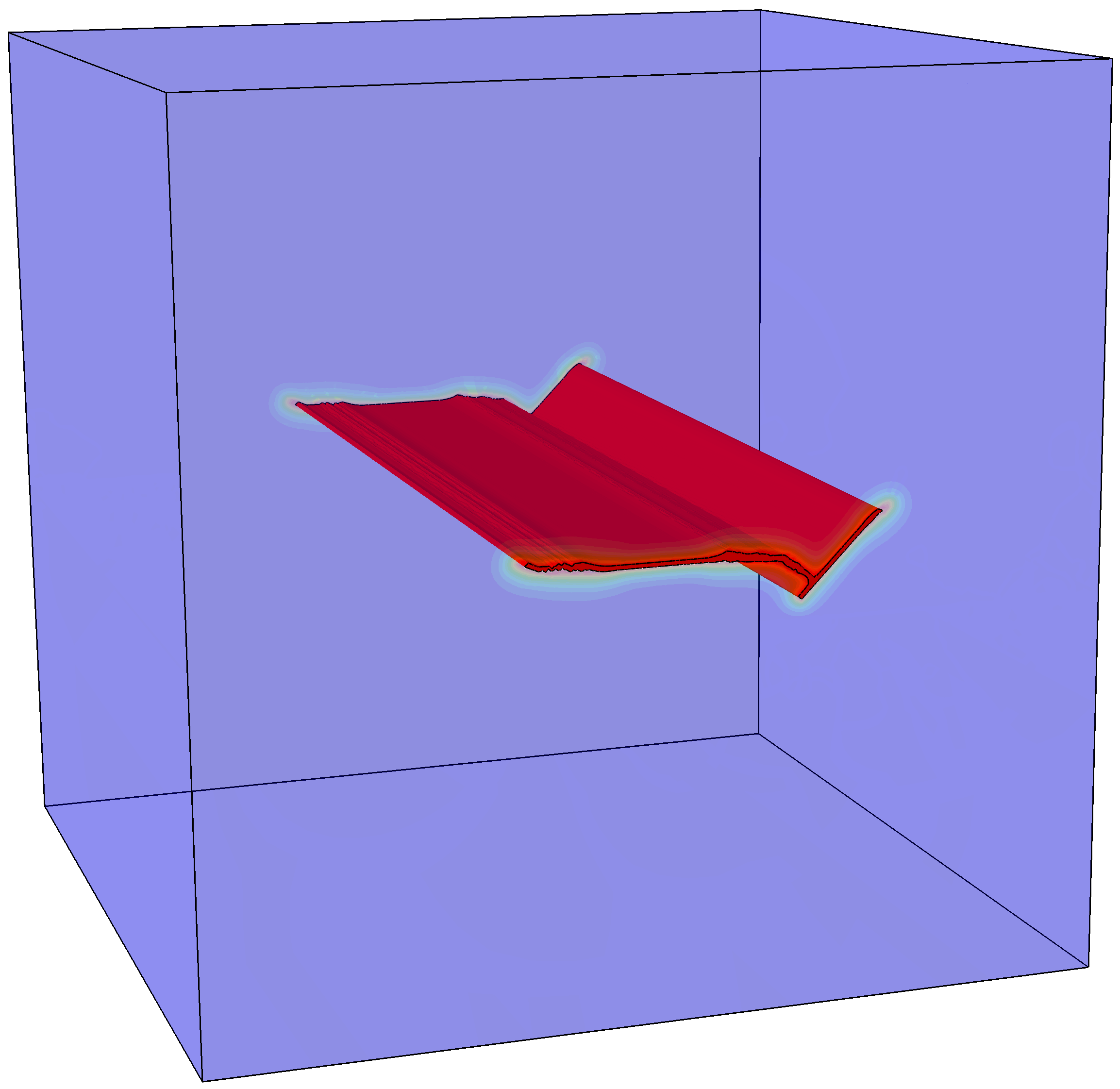}
         \caption{}
     \end{subfigure} \hspace{5mm}
     \begin{subfigure}[b]{0.35\textwidth}
         \centering
         \includegraphics[width=\textwidth]{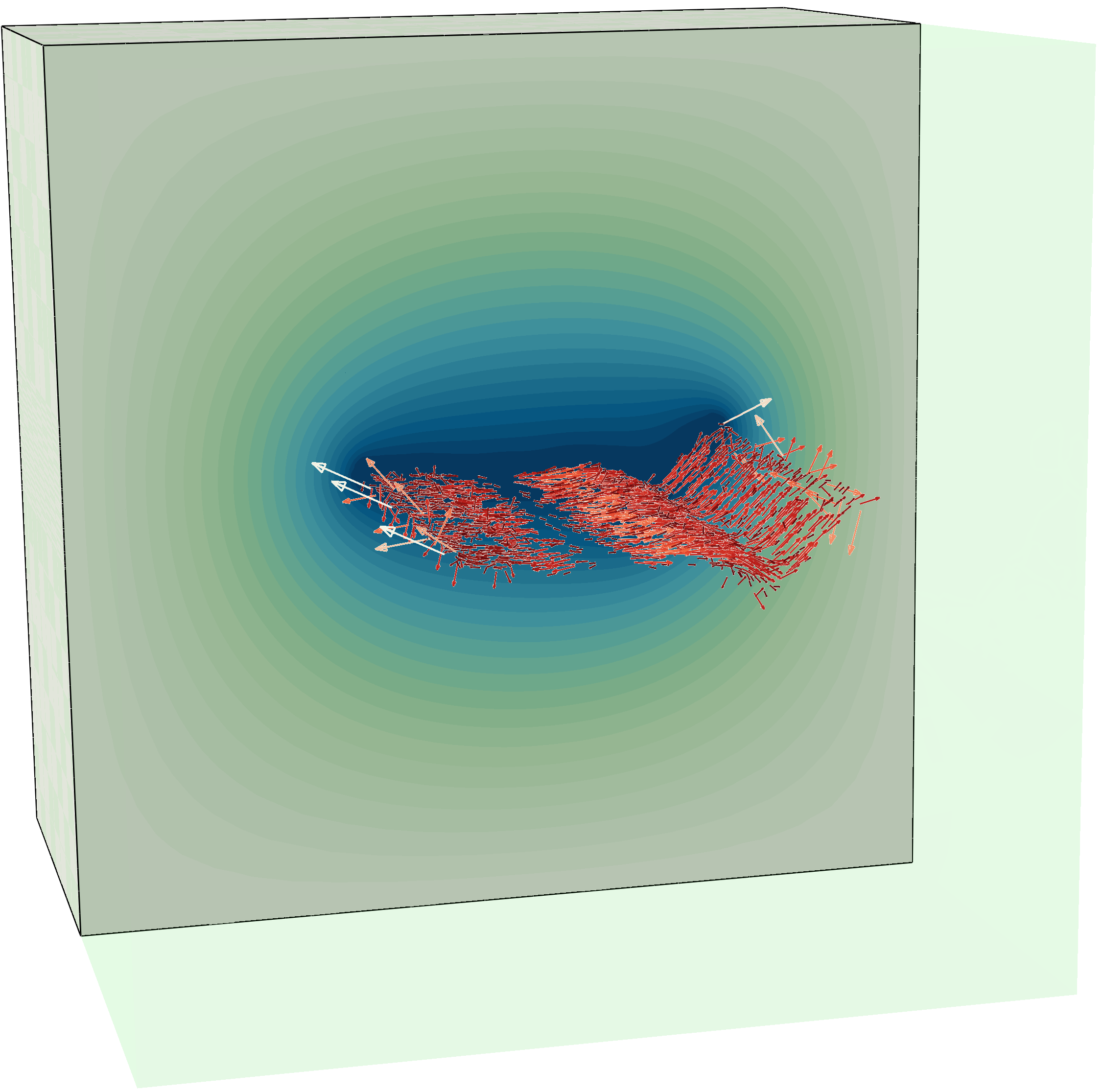}
         \caption{}
     \end{subfigure}\hspace{2mm}
      \begin{subfigure}[b]{0.08\textwidth}
         \centering
         \includegraphics[width=\textwidth]{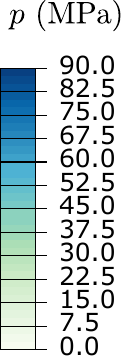}
         \vspace{5mm}
     \end{subfigure}
    \begin{subfigure}[b]{0.7\textwidth}
         \centering
         \includegraphics[width=\textwidth]{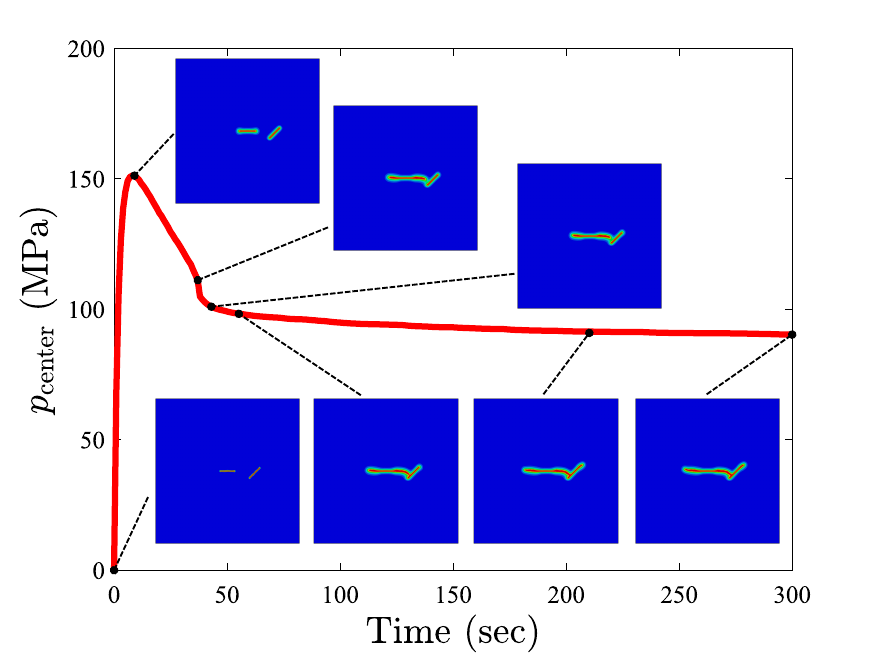}
         \caption{}
     \end{subfigure} 
    \caption{Cracking due to an injected fluid in a cube: (a) phase field contours depicting the crack trajectory, (b) fluid flux vector and fluid pressure contour $p$ in the $XY$ plane at $Z=-0.25$ m, and (c) fluid pressure at the center of the domain $p_{\text{center}}$ versus time, also showing phase field contours at selected times.}
    \label{fig:INJ-Inter-PT}
\end{figure}

The results obtained are shown in Fig. \ref{fig:INJ-Inter-PT}. The crack path is shown using the phase field contour in Fig. \ref{fig:INJ-Inter-PT}a for time of $t=300$ s. The pressure contour at that time, the steady state situation, is given in Fig. \ref{fig:INJ-Inter-PT}b. The fluid flux vector, computed based on Darcy's law ($\mathbf{q}=-\rho_{\text{fl}} \frac{\bm{K}_{\text{fl}}}{\mu_{\text{fl}}} \nabla p$), is also superimposed on the figure. This quantity is equivalent to the heat flux vector $\mathbf{f}$ in the thermal analogy presented in Eq. (\ref{Eq:dMassContent6}). The time evolution of the pressure at the centre of the domain is given in Fig. \ref{fig:INJ-Inter-PT}c. Insets of the phase field contour are also included to depict crack evolution over time. The result shows that the pressure increases upon water injection until the initiation of crack growth. At this point, the maximum water pressure exceeds 150 MPa. As crack propagation occurs, the pressure starts to decrease, with a sharper decline when the two cracks approach each other. Over time, both cracks propagate and coalesce. However, after reaching a steady state, which is characterised by a constant pressure at the centre, no further crack propagation is observed.

\subsection{Hydrogen embrittlement}
\label{sec:Hydrogen_Embrittlement}

\begin{figure}[H]
    \centering
    \includegraphics[width=0.5\linewidth]{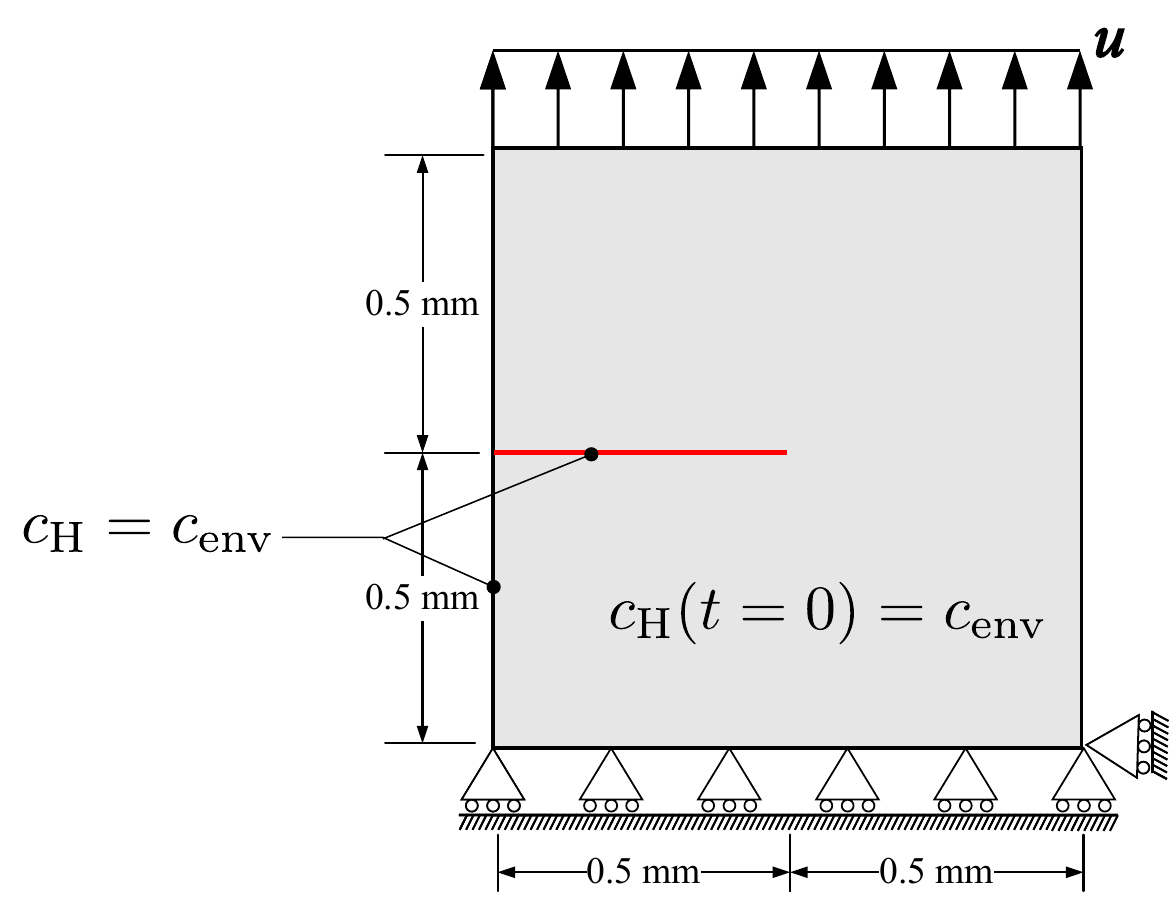}
    \caption{Geometry and boundary conditions for the notched square plate under tension exposed to hydrogen.}
    \label{fig:SQ-Config}
\end{figure}

The ability of the present framework to simulate hydrogen-assisted fractures is here demonstrated by benchmarking against literature results of the classic edge-cracked square plate boundary value problem, first studied by Martínez-Pañeda et al. \cite{MARTINEZPANEDA2018742}. The geometry and loading configuration are given in Fig. \ref{fig:SQ-Config}. The specimen is initially saturated with a constant hydrogen concentration $c_{\mathrm{H}}$, equivalent to the environmental concentration $c_{\text{env}}$. Subsequently, a displacement is applied to the top of the plate over $10^7$ seconds in 2,000 increments. The mechanical and hydrogen transport parameters are listed in Table \ref{tab:Hydrogen}.

\begin{table}[H]
\caption{Material properties for hydrogen embrittlement.}
    \centering
    \begin{tabular}{lll}
      \hline
      Parameter  &  Value  &  Unit  \\
\hline  \hline
Young's modulus $E$  & 210 & GPa \\
Poisson's ratio $\nu$  & 0.3 &  \\
Phase field length scale $\ell$ & 0.0075 & mm \\
Toughness $G_c$ & 2.7 & $\text{kJ}/\text{m}^2$ \\
Hydrogen diffusion coefficient $D_{\mathrm{H}}$ & 0.0127 & $\text{mm}^2/\text{s}$ \\
Trap binding energy $\Delta g_b^0$ & 30 & $\text{kJ}/\text{mol}$ \\
Hydrogen damage coefficient $\chi_{\mathrm{H}}$ & 0.89 & \\
Temperature $T_\text{k}$ & 300 & K\\
\hline
    \end{tabular}
    \label{tab:Hydrogen}
\end{table}

The domain is discretised into approximately 36,000 8-node plane strain thermally coupled quadrilateral elements with biquadratic displacement and bilinear temperature discretisation (denoted as CPE8T in Abaqus). The mesh around the predicted crack path was refined, with the element size being five times smaller than the characteristic phase field length $\ell$. As shown in Eq. \eqref{Eq:HydrogenAna}, one must compute the gradient of hydrostatic stress $\nabla \sigma_h$. This is achieved here by extrapolating the integration point values of $\sigma_h$ to the nodes using appropriate shape functions. To build these shape functions, at the beginning of the analysis we store all the relevant information (node numbers and coordinates, and element connectivity), using a \texttt{UEXTERNALDB} subroutine (as shown in the codes provided). Then, the strain-displacement matrices (so-called $\bm{B}$-matrices) are used to estimate $\nabla \sigma_h$ at the integration points. A staggered solution scheme was employed. 

\begin{figure}[H]
    \centering
    \includegraphics[width=.7\linewidth]{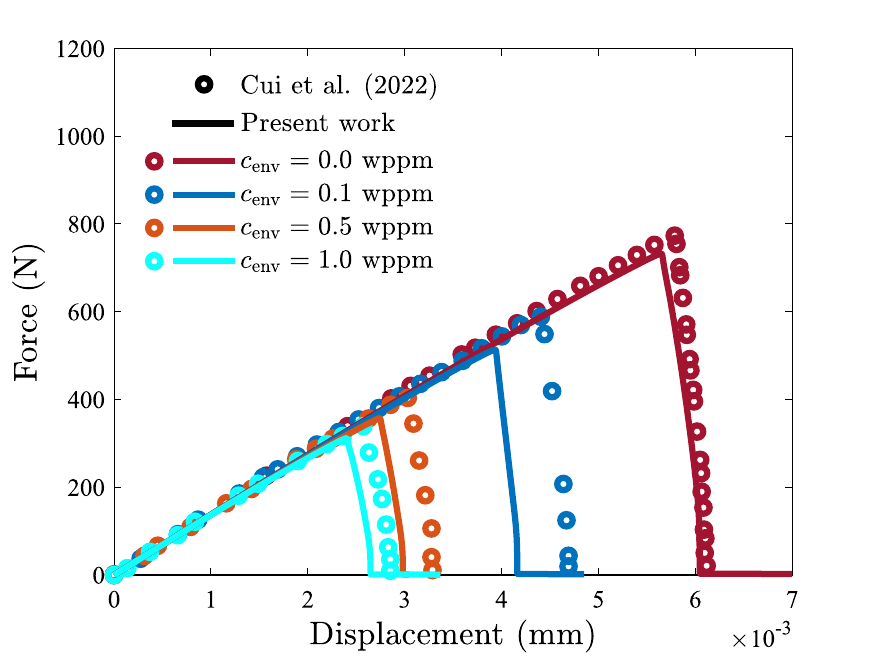}
    \caption{Hydrogen embrittlement of a square plate under tension. Load versus applied displacement predictions for various hydrogen contents, as obtained with the present implementation and as reported in the work by Cui et al. \cite{cui2022generalised}.}
    \label{fig:SQ-LD}
\end{figure}

The comparison of the load versus displacement responses obtained with our numerical framework and those from Cui et al. \cite{cui2022generalised} are given in Fig. \ref{fig:SQ-LD}. The agreement is satisfactory but not excellent. Differences could be due to the loading rate (not reported in Ref. \cite{cui2022generalised}), the different approach employed to estimate the gradient of the hydrostatic stress $\sigma_h$, the lack of a penalty boundary condition on our simulation, or the use of fully integrated elements (as opposed to reduced integration elements in Ref. \cite{cui2022generalised}). For both cases, it can be seen that increasing hydrogen content results in a drop in the peak load and a reduction in the failure displacement. A sharper softening is observed in our simulation, suggesting a more accurate representation of material behaviour (unstable crack growth is expected). 

\begin{figure}[H]
    \centering
    \begin{subfigure}[b]{\textwidth}
         \centering
         \includegraphics[width=0.25\textwidth]{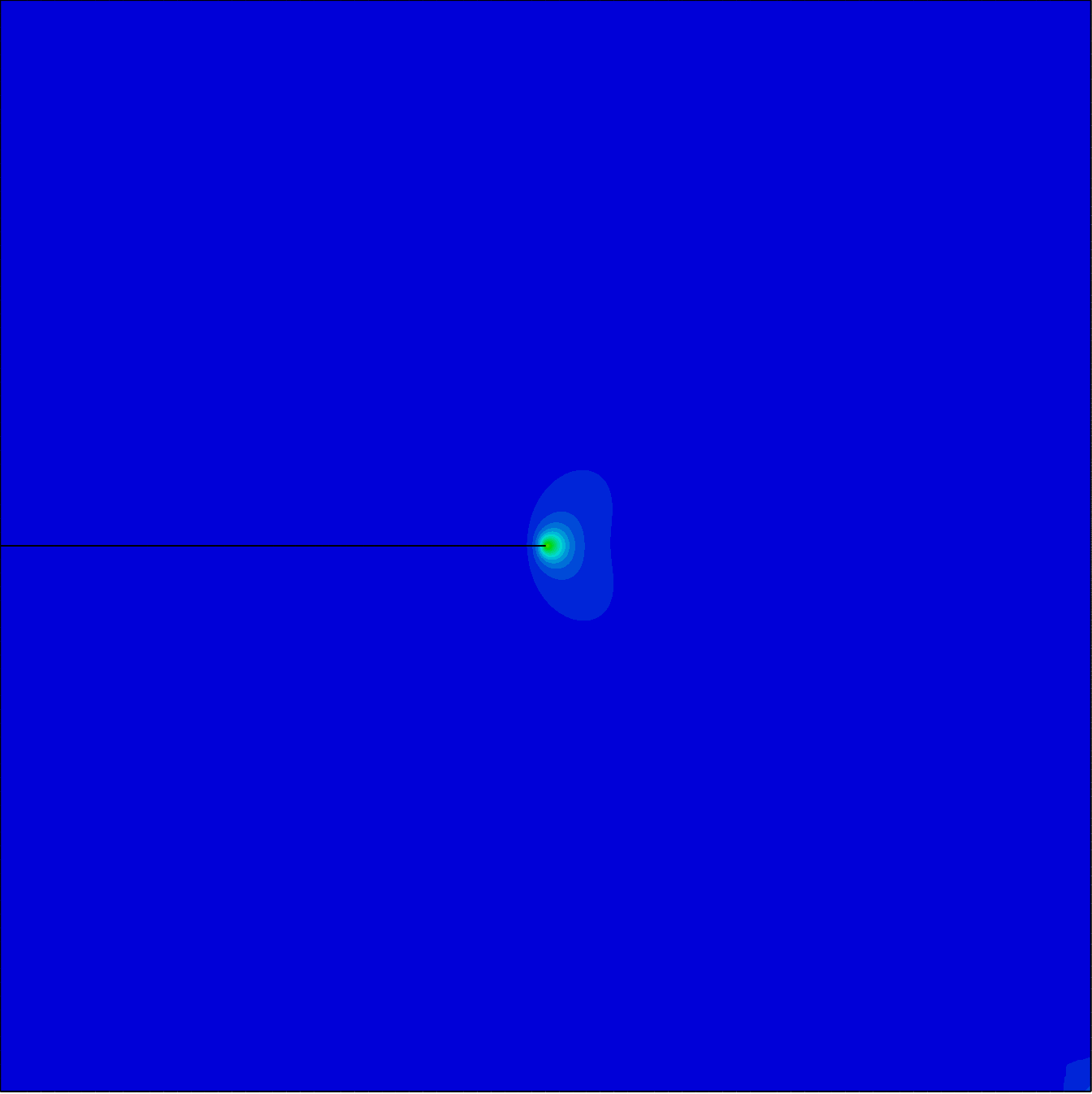}
         \hspace{5 mm}
         \includegraphics[width=0.25\textwidth]{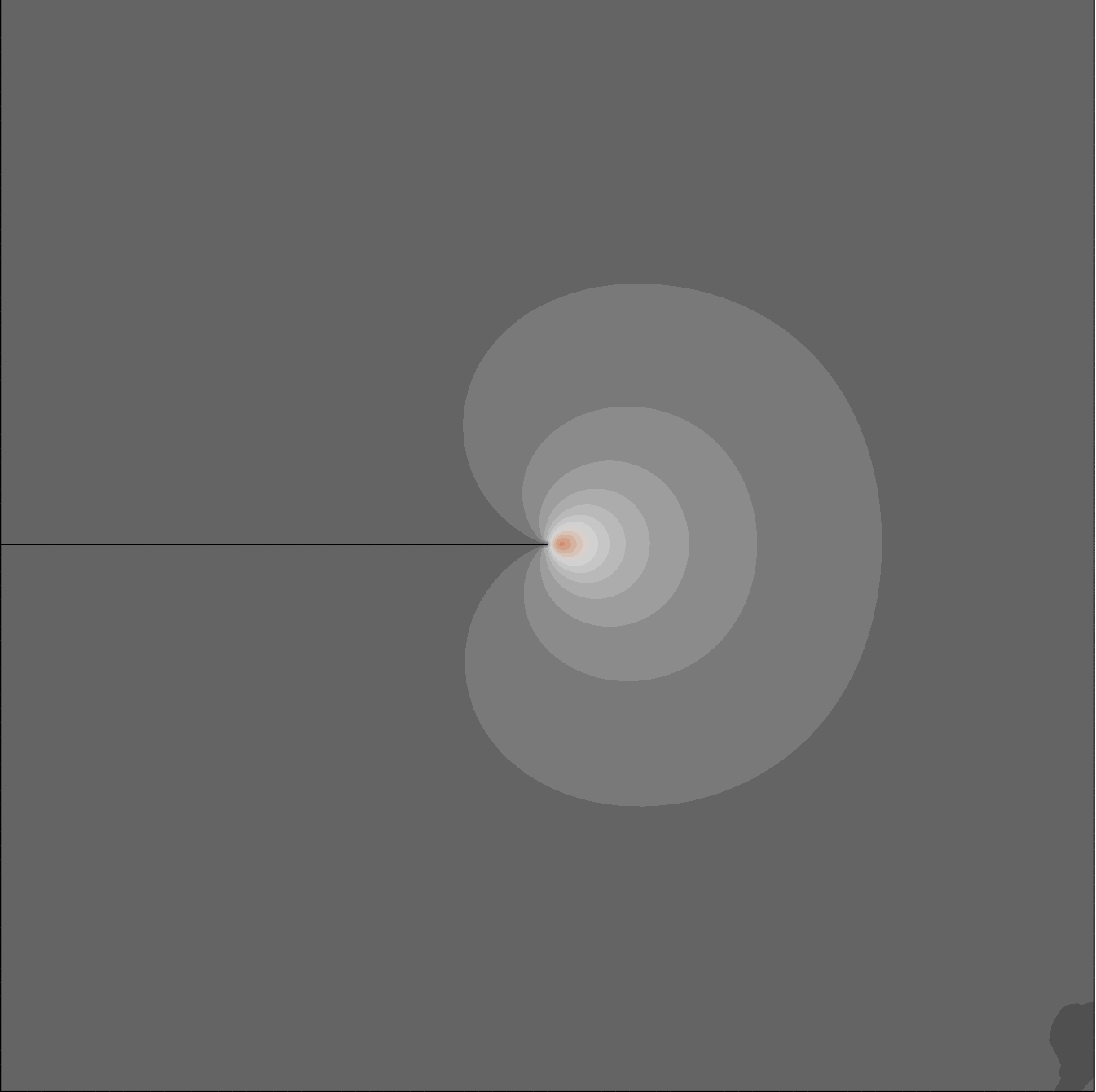}
         \caption{$u=0.00273$ mm}
         \label{fig:Hydrogen-countor-phi-ch-a}
     \end{subfigure} \\ \vspace{5 mm}
     \begin{subfigure}[b]{\textwidth}
         \centering
         \includegraphics[width=0.25\textwidth]{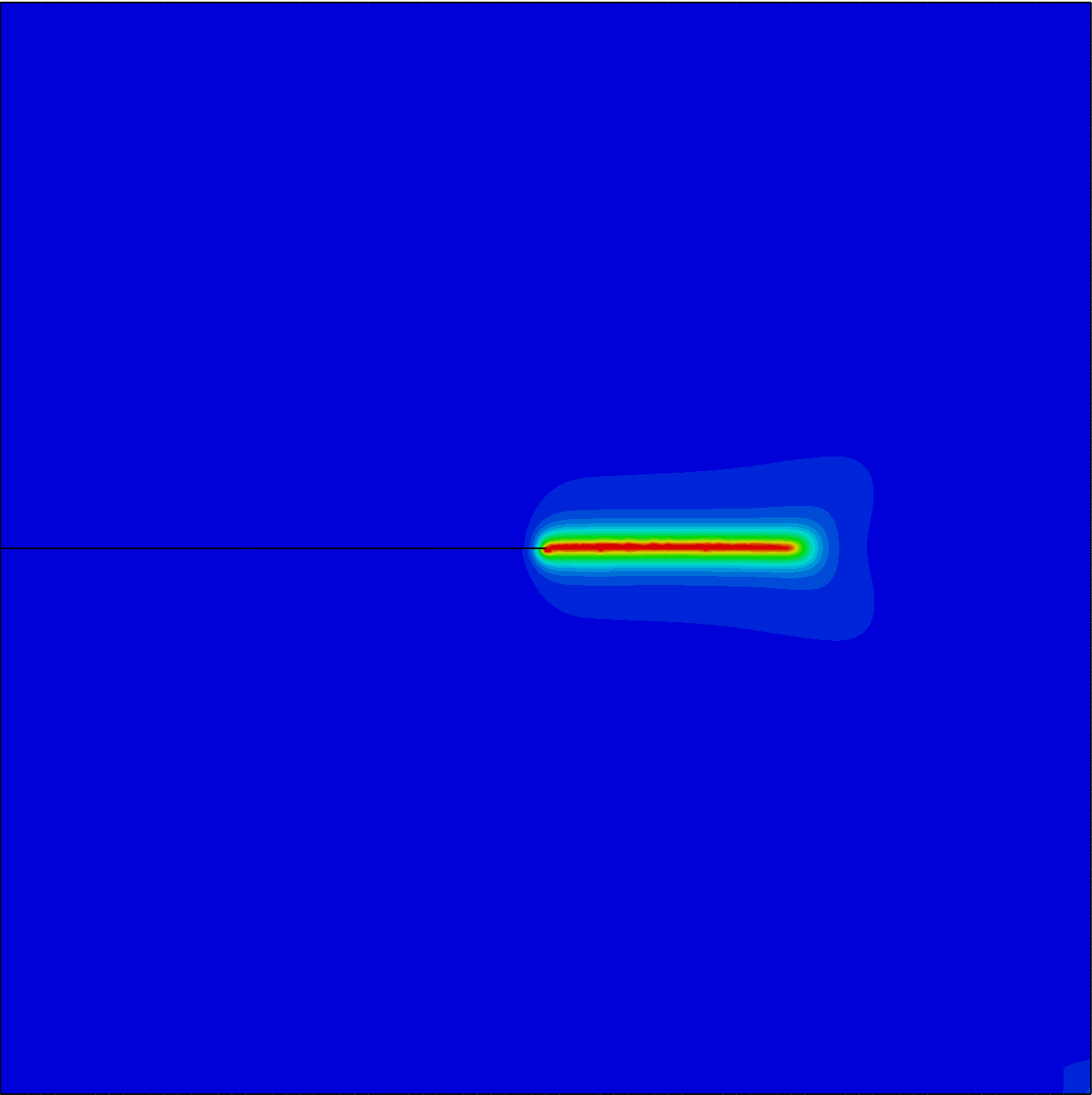}
         \hspace{5 mm}
         \includegraphics[width=0.25\textwidth]{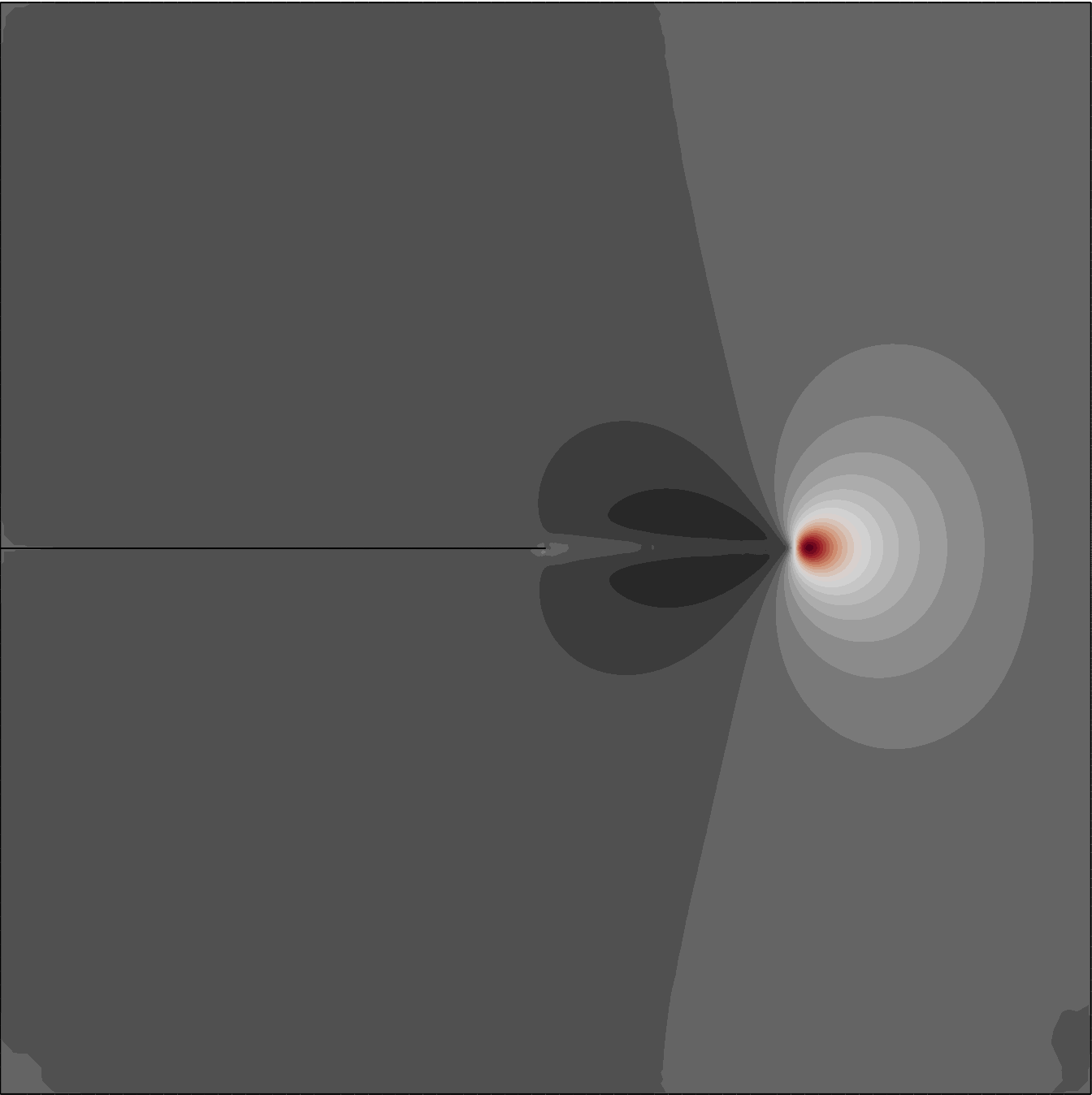}
         \caption{$u=0.00287$ mm}
         \label{fig:Hydrogen-countor-phi-ch-b}
     \end{subfigure} \\ \vspace{5 mm}
     \begin{subfigure}[b]{\textwidth}
        \centering
        \includegraphics[width=0.12\textwidth]{Legend-V-phi.pdf}
        \hspace{5 mm}
         \includegraphics[width=.25\textwidth]{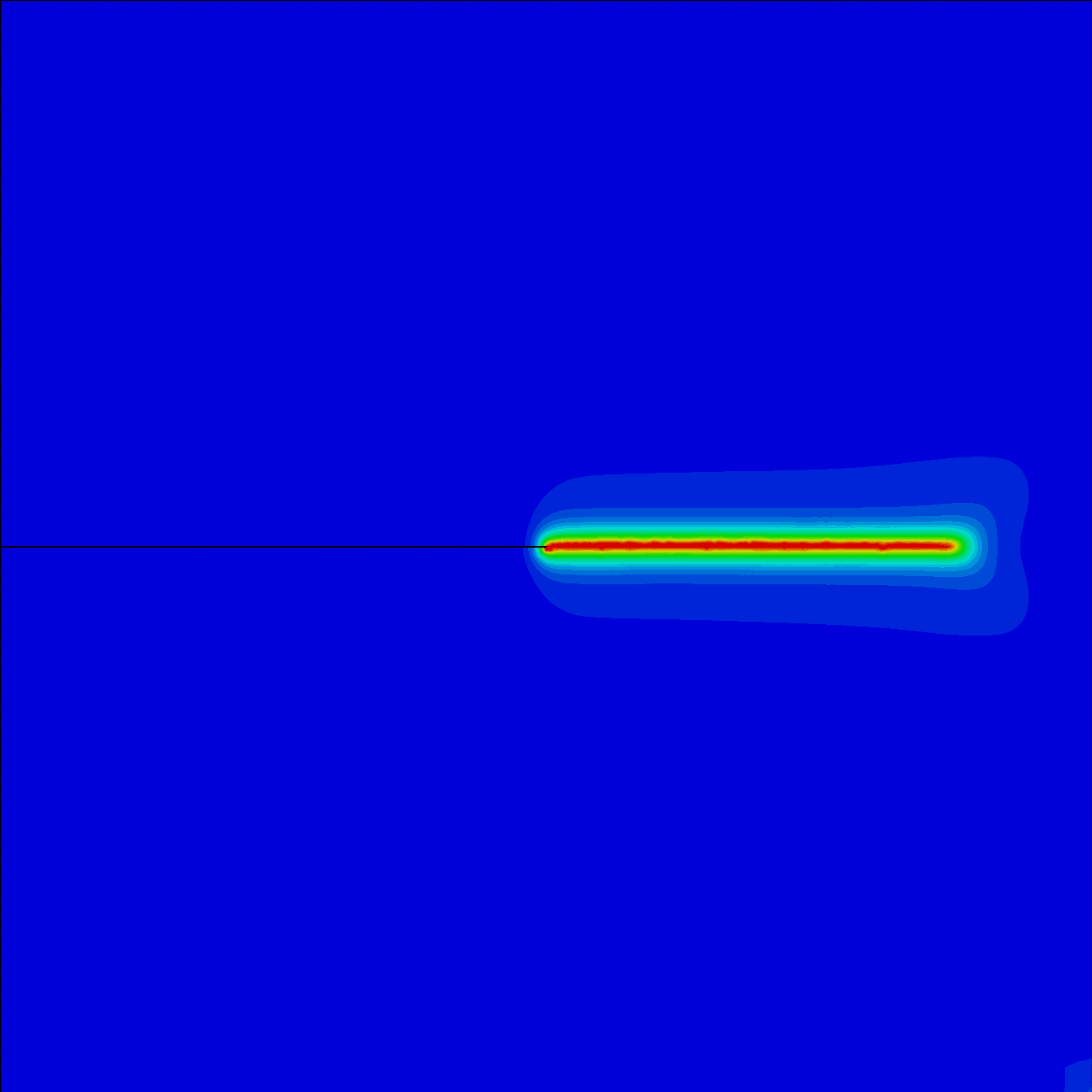}
         \hspace{5 mm}
         \includegraphics[width=.25\textwidth]{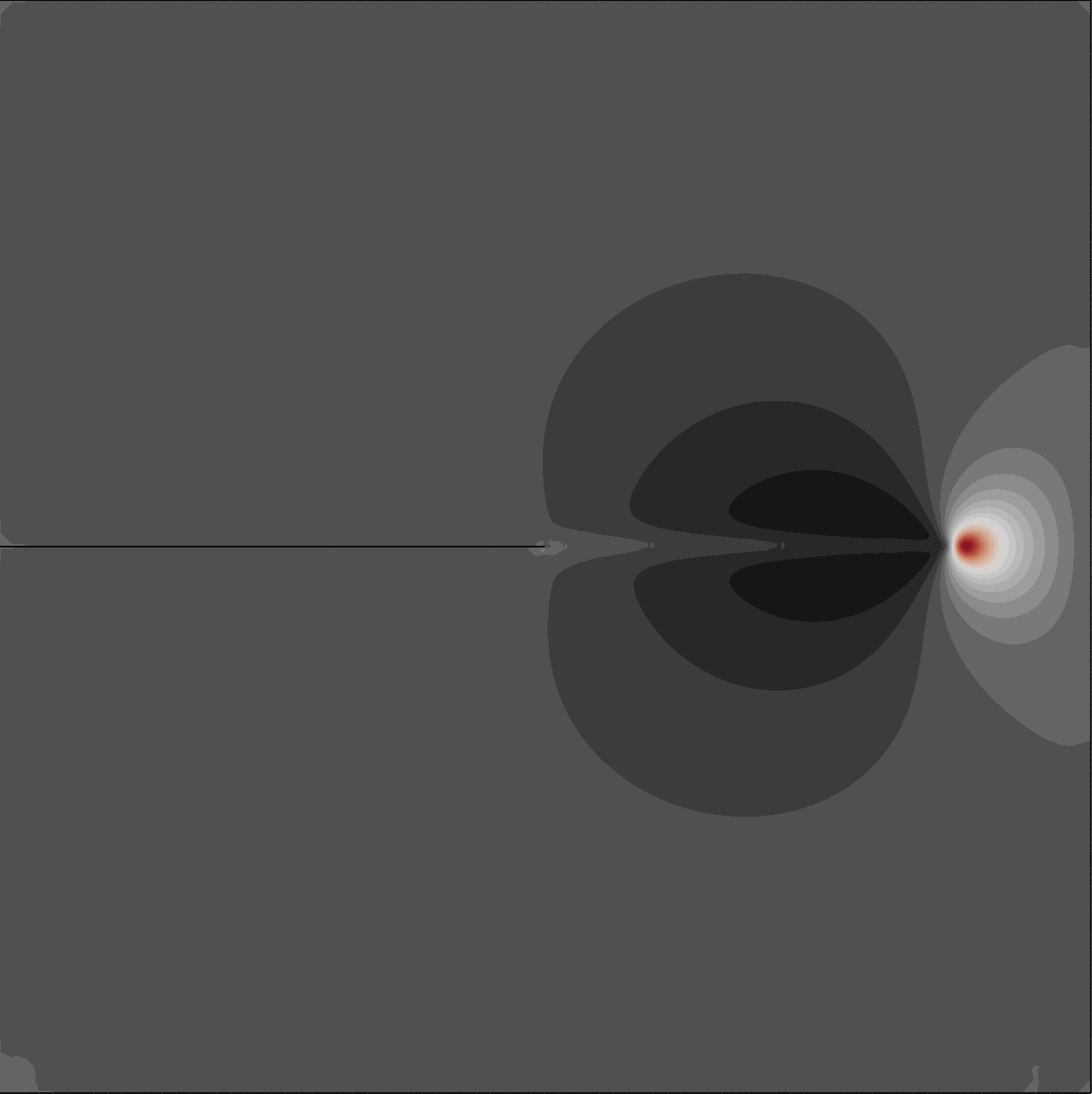}
         \hspace{5 mm}
         \includegraphics[width=0.135\textwidth]{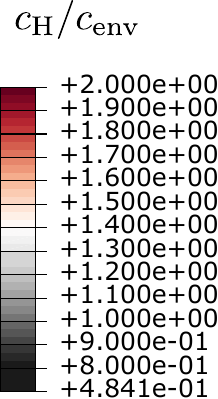}
         \caption{$u=0.00294$ mm}
         \label{fig:Hydrogen-countor-phi-ch-c}
     \end{subfigure} 
    \caption{Phase field contours (left) and normalized hydrogen concentration (right) for $c_{\mathrm{env}}=0.5$ wppm in the square plate under tension at different displacements $u$: (a) $u=0.00273$ mm, (b) $u=0.00287$ mm, and (c) $u=0.00294$ mm.}
    \label{fig:Hydrogen-countor-phi-ch}
\end{figure}

The evolution of the phase field contour and of the normalised hydrogen concentration are illustrated in Fig. \ref{fig:Hydrogen-countor-phi-ch}, for selected time steps and the case of $c_{\text{env}} = 0.5$ wppm. It can be seen that the model captures the accumulation of hydrogen near the crack tip, due to the role that hydrostatic stresses play in driving hydrogen diffusion. The hydrogen concentration contours are smooth and follow the crack tip as it propagates, as a result of the low load rate considered. The good agreement attained with experimentally-validated models (see, e.g., Refs. \cite{IJP2021,mandal2024computational}) demonstrates that the present framework is also capable of predicting hydrogen-assisted failures in laboratory and practical conditions.

\begin{figure}[H]
    \centering
    \includegraphics[width=.7\linewidth]{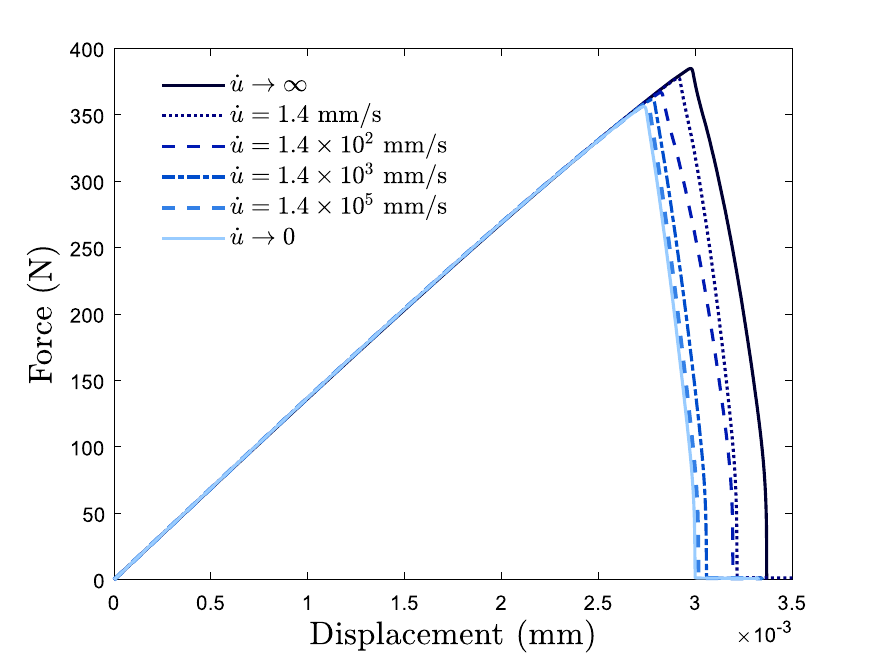}
    \caption{Hydrogen embrittlement of a square plate under tension: effect of the loading rate. Load versus applied displacement predictions for various loading rates (including the two limit cases of very fast and very slow tests) for the case of $c_{\mathrm{env}} = 0.5$ wppm.}
    \label{fig:SQ-LD1}
\end{figure}

Finally, calculations are conducted under selected loading rates, to showcase the model's ability to capture the qualitative trend observed in experiments and bring insight into the differences observed with the predictions by Cui et al. \cite{cui2022generalised}. For slow loading rates, there is less time for the hydrogen to accumulate in the fracture process zone and therefore, a higher resistance to fracture is expected. The results obtained are given in Fig.~\ref{fig:SQ-LD1}, for the case of $c_{\mathrm{env}} = 0.5$ wppm. The two limit cases of $\dot{u} \to 0$ (where $c_H$ follows the steady state solution) and $\dot{u} \to 0$ (where the hydrogen concentration is uniform and equal to the initial hydrogen concentration), together with four selected intermediate cases. The results show that the model is able to capture the sensitivity to loading rate and reveal the variation in critical load that can be attained with changes in loading rate.

\subsection{Stress corrosion cracking}
\label{sec:Corrosion-Stress1}

The last case study showcases the ability of the present framework to simulate stress-assisted corrosion. This is achieved by considering two boundary value problems involving localised corrosion phenomena: pitting and (anodic dissolution-driven) stress corrosion cracking. The first example examines a plate with a semi-circular pit in the absence of mechanical load, to validate the phase field corrosion implementation, while the second involves a plate with a semi-elliptical pit subjected to a remote displacement. Our results are compared with findings from the literature. In both case studies, a monolithic scheme is used to couple displacement with the phase field evolution equation, while the metallic ion transport equation is coupled via a multi-pass staggered method.

\subsubsection{Growth of a semi-circular pit}

\begin{figure}[H]
    \centering
    \begin{subfigure}[b]{0.55\textwidth}
         \centering
         \includegraphics[width=\textwidth]{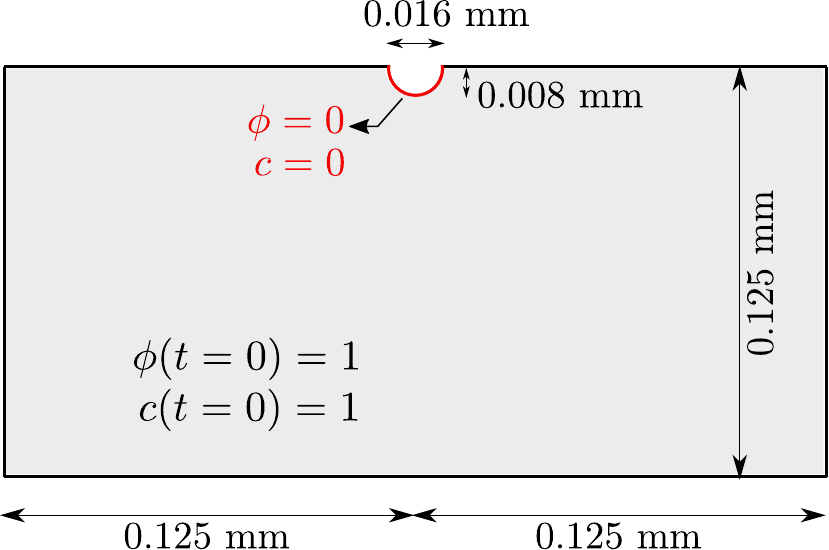}
     \end{subfigure}
    \caption{Geometry and boundary conditions for the semi-circular pit example.}
    \label{fig:Semi-Circular-pit-Config}
\end{figure}

We validate pit growth in the absence of stress using a rectangular plate with an initial semi-circular pit. This setup was previously simulated by Duddu \cite{Duddu2014} using the level set method and by Mai et al. \cite{Mai2016}, Gao et al. \cite{Gao2020}, and Cui et al. \cite{cui2022generalised} using the phase field method. The geometry and boundary conditions are shown in Fig. \ref{fig:Semi-Circular-pit-Config}, and the material parameters considered are listed in Table \ref{tab:Corrosion}. To achieve diffusion-controlled pit growth, a large value of \(L_0\) is considered. The minimum element size is set to 0.001 mm, and approximately 30,000 8-node reduced integration elements that provide biquadratic discretisation for the displacement field and bilinear discretisation for the temperature field are adopted (denoted as CPE8RT in Abaqus). No film rupture or repassivation effects are considered in this case study (i.e., $k=0$).\\

\begin{table}[H]
\caption{Material parameters for the stress-assisted corrosion case study.}
    \centering
    \begin{tabular}{lll}
      \hline 
      Parameter  &  Value  &  Unit  \\
      \hline \hline 
      Gradient energy coefficient $\kappa$ & $51 \times 10^{-6}$  & J/m \\
      Height of the double-well potential $\omega$ & $35.3 \times 10^{6}$ & J/m$^3$ \\
      Temperature $T$ & 300 & K \\
      Diffusion coefficient of metal ion $D_m$ & $8.5 \times 10^{-4}$ & $\mathrm{mm}^2$/s \\
      Interface kinetics coefficient $L_0$ & $2 \times 10^6$ & $\mathrm{mm}^2$/(N s) \\
      Free energy density curvature $A$ & 53.5 & N / $\mathrm{mm}^2$ \\
      Average concentration of metal $c_{\text{solid}}$ & 143 & mol/L \\
      Average saturation concentration $c_{\text{sat}}$ & 5.1 & mol/L \\
      \hline
    \end{tabular}
    \label{tab:Corrosion}
\end{table}

\begin{figure}[H]
    \centering
    \begin{subfigure}[b]{0.4\textwidth}
         \centering
         \includegraphics[width=\textwidth]{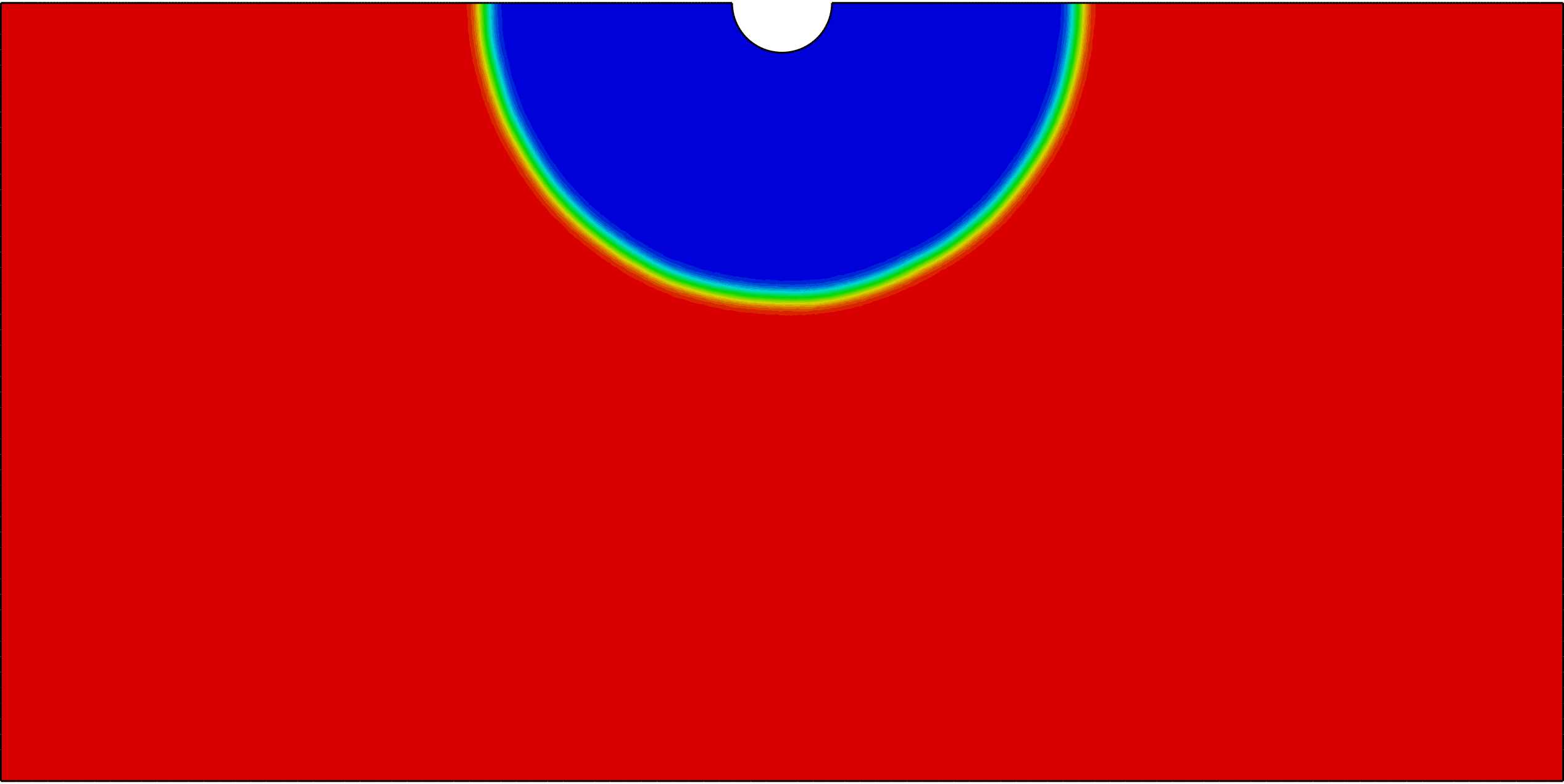}
         \caption{Present work}
         \label{fig:Semi-Cir-phi-a}
     \end{subfigure}  \hspace{2mm}
     \begin{subfigure}[b]{0.4\textwidth} 
         \centering 
         \includegraphics[width=\textwidth]{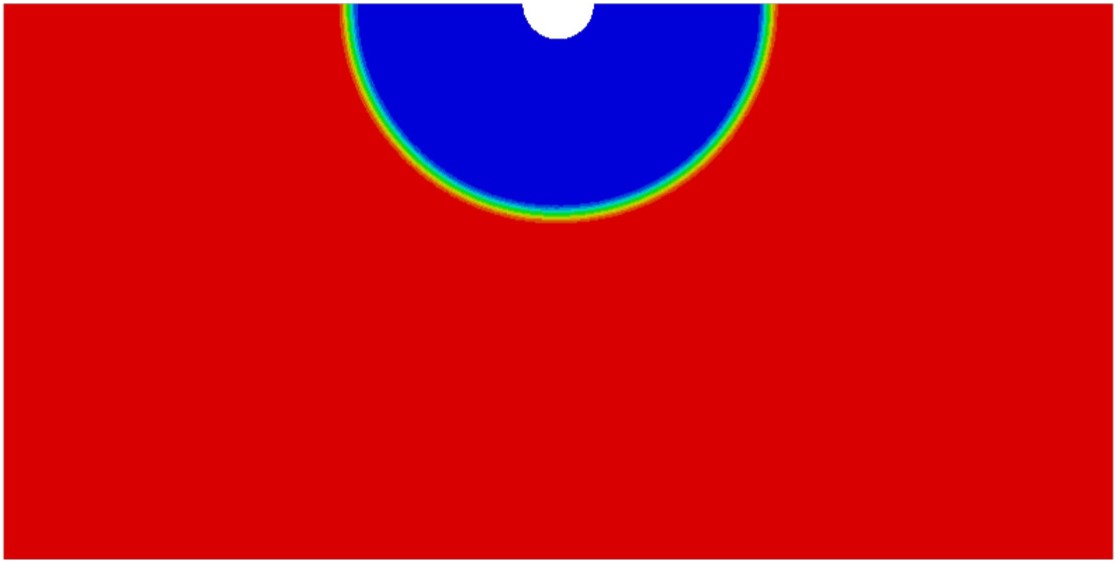}
         \caption{Cui et al. \cite{cui2022generalised}}
         \label{fig:Semi-Cir-phi-b}
     \end{subfigure} \hspace{2mm}
     \begin{subfigure}[b]{0.13\textwidth} 
         \centering
         \includegraphics[width=\textwidth]{Legend-V-phi.pdf}
         \vspace{0mm}
     \end{subfigure} 
     \begin{subfigure}[b]{0.5\textwidth} 
         \centering 
         \includegraphics[width=\textwidth]{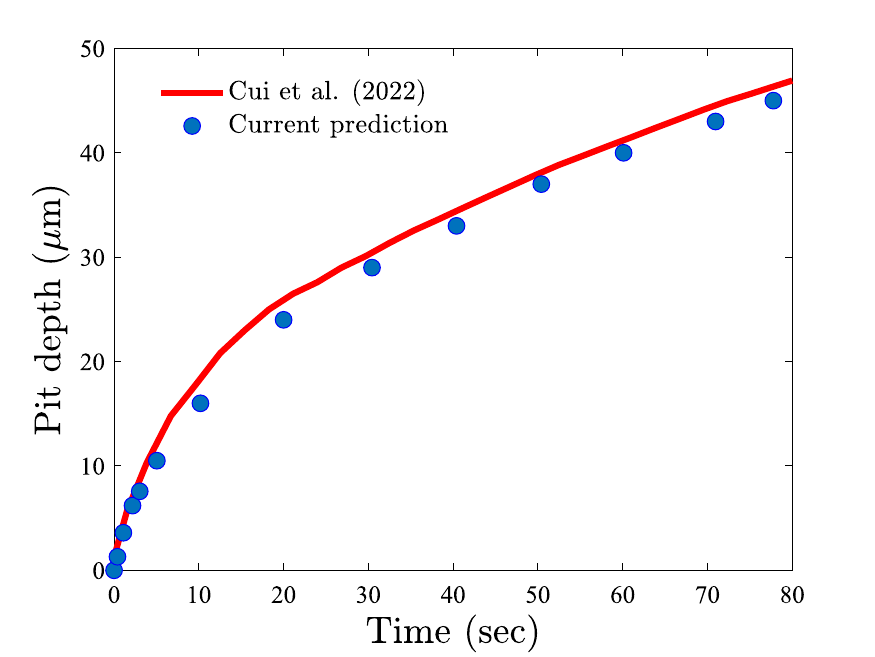}
         \label{fig:Semi-Cir-phi-c}
     \end{subfigure}
    \caption{Pit growth for a semi-circular pit. Phase field contours at a time of $t=50$ s, as obtained with (a) our current implementation, and (b) by Cui et al. \cite{cui2022generalised}. A quantitative comparison of pit depth versus time is given in (c).}
    \label{fig:Semi-Cir-phi}
\end{figure}

The results obtained are shown in Figs. \ref{fig:Semi-Cir-phi}a and \ref{fig:Semi-Cir-phi}b, depicting pit growth by means of phase field contours. The results are shown for a time of 50 seconds for our implementation and that by Cui et al. \cite{cui2022generalised}, respectively. An excellent agreement is observed. Moreover, predictions match the experimental results by Ernst and Newman \cite{Ernst2002}, showing that a semi-circular pit remains semi-circular in the absence of stress during corrosion. The evolution of pit depth over time is quantitatively compared with the preditions by Cui et al. \cite{cui2022generalised} in Fig. \ref{fig:Semi-Cir-phi}c, showing a very good agreement. 

\subsubsection{Growth of a semi-elliptical pit under stress}

\begin{figure}[H]
    \centering
    \begin{subfigure}[b]{0.55\textwidth}
         \centering
         \includegraphics[width=\textwidth]{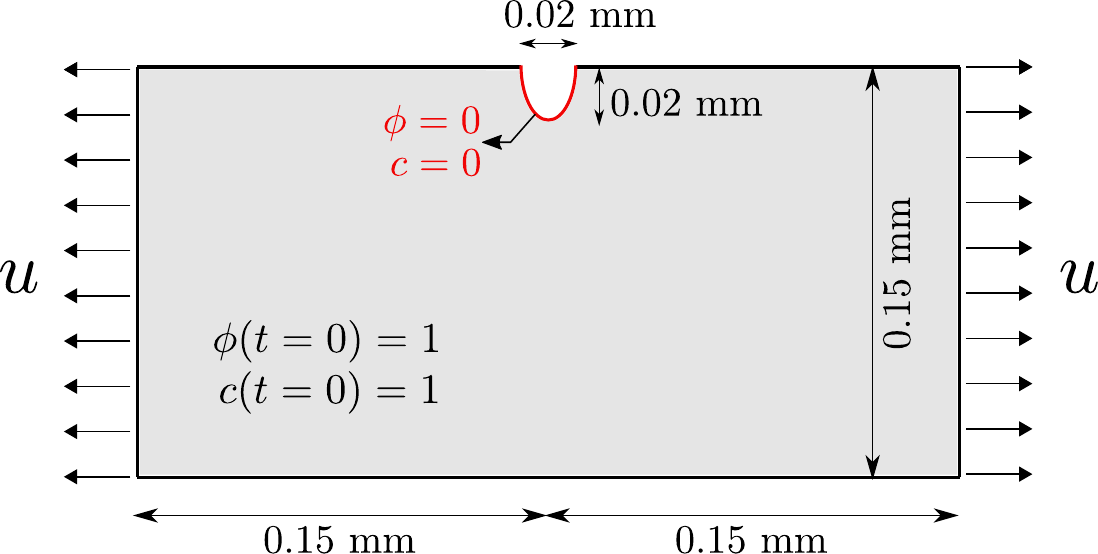}
     \end{subfigure}
    \caption{Geometry and boundary conditions of the semi-elliptical pit.}
    \label{fig:SCC-Config}
\end{figure}

The ability to couple corrosion and mechanical phenomena is investigated by predicting the evolution of a pre-existing semi-elliptical pit in a stainless steel plate subjected to a remote displacement. This case study was first simulated by Cui et al. \cite{JMPS2021} and reproduced by others since. The geometry and boundary conditions are illustrated in Fig. \ref{fig:SCC-Config}. A uniform displacement of $u=0.0002$ mm is applied linearly over 1 second to both the right and left sides and then maintained constant throughout the simulation. Over the pit domain (red line in Fig. \ref{fig:SCC-Config}), we enforce $\phi=0$ and $c=0$, while the remaining regions are initialised with $\phi (t=0) = 1$ and $c (t=0) = 1$. Unless otherwise specified, the material parameters adopted correspond to those used in the previous case study and provided in Table \ref{tab:Corrosion}. A stability parameter of $k=5 \times 10^{-4}$ is considered, with a time interval before decay begins set at $t_0=10$ s and a critical strain for film rupture of $\varepsilon_f=3 \times 10^{-3}$. The initial interface kinetics coefficient is given as $L_0=0.001$ mm$^2$/ (N s). The mechanical parameters include a Young’s modulus of $E=190$ GPa, a Poisson’s ratio of $\nu=0.3$, an initial yield stress of $\sigma_y=520$ MPa, and a strain hardening exponent of $N=0.067$. The computational domain is discretised using over 13,000 8-node elements with reduced integration (CPE8RT). The characteristic element size in the central region is set to 0.005 mm.

\begin{figure}[H]
    \centering
    \begin{subfigure}[b]{0.09\textwidth} 
         \centering
         \includegraphics[width=\textwidth]{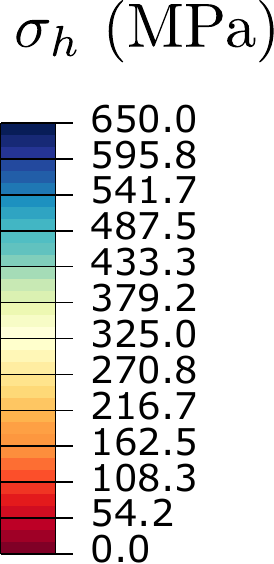}
         \vspace{0mm}
     \end{subfigure}
    \begin{subfigure}[b]{0.2\textwidth}
         \centering
         \includegraphics[width=\textwidth]{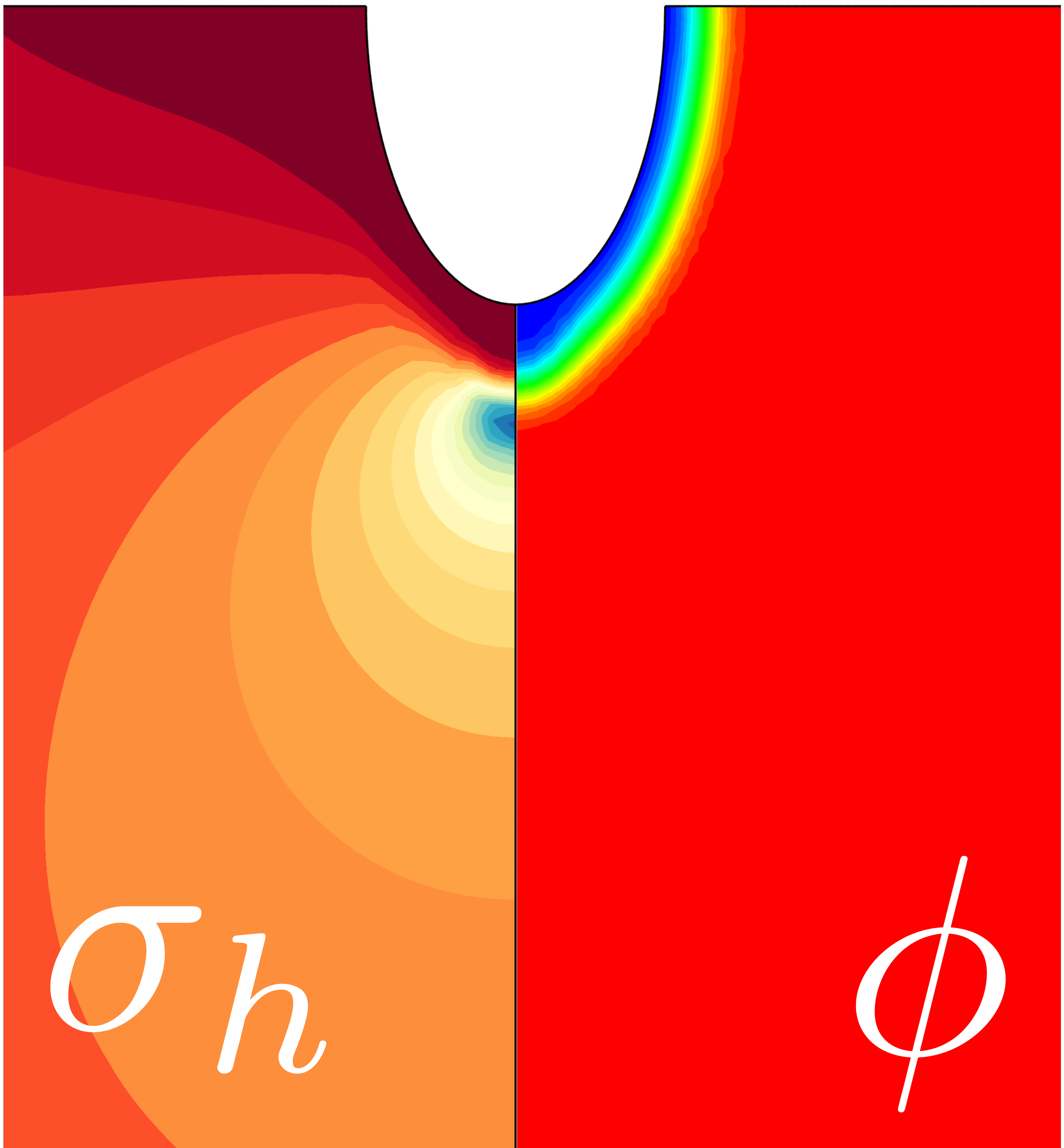}
         \caption{Time=100 sec}
         \label{fig:SCC-phi-a}
     \end{subfigure} \hspace{2mm}
     \begin{subfigure}[b]{0.2\textwidth} 
         \centering 
         \includegraphics[width=\textwidth]{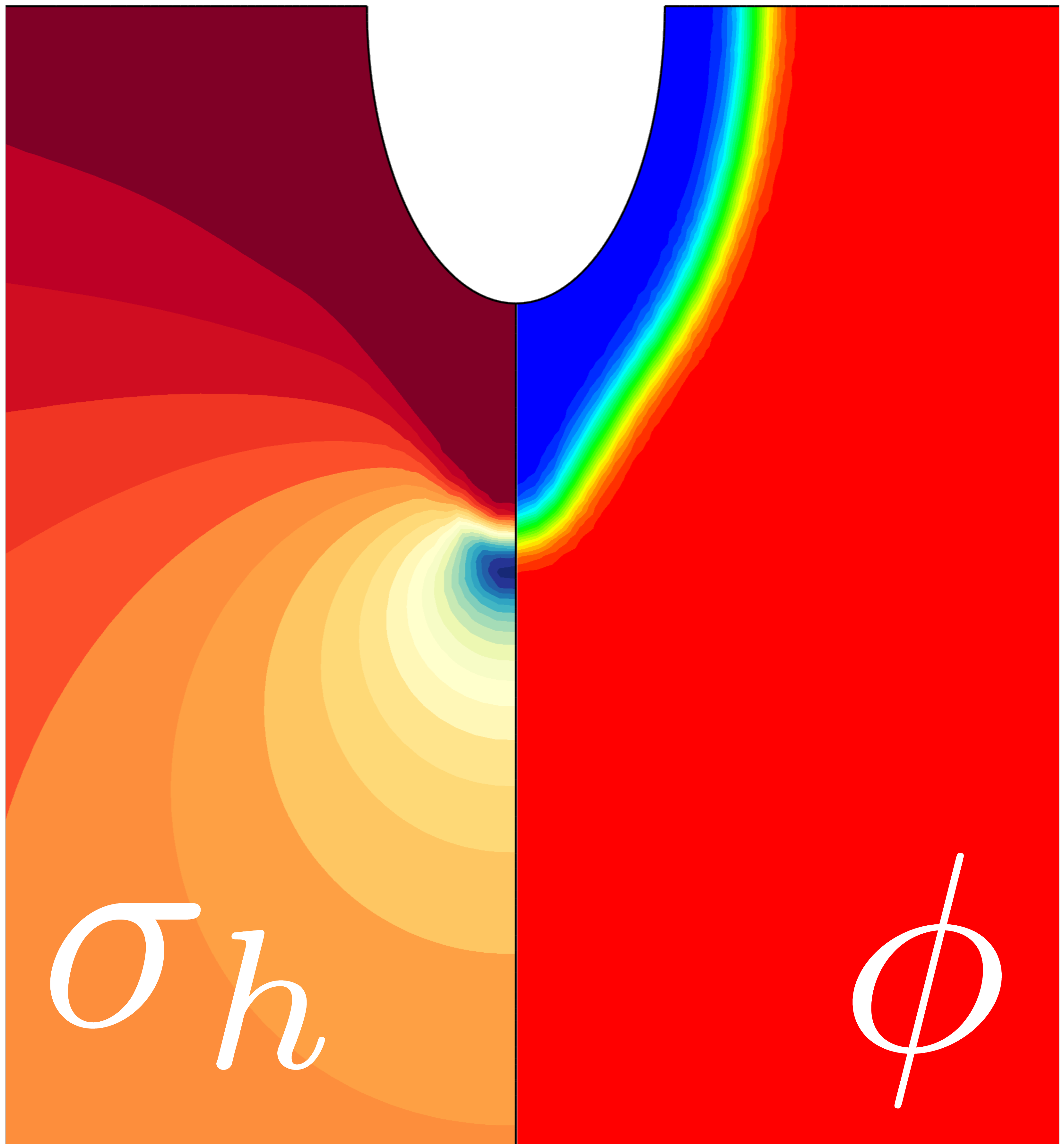}
         \caption{Time=300 sec}
         \label{fig:SCC-phi-b}
     \end{subfigure} \hspace{2.5mm}
     \begin{subfigure}[b]{0.2\textwidth}
         \centering
         \includegraphics[width=\textwidth]{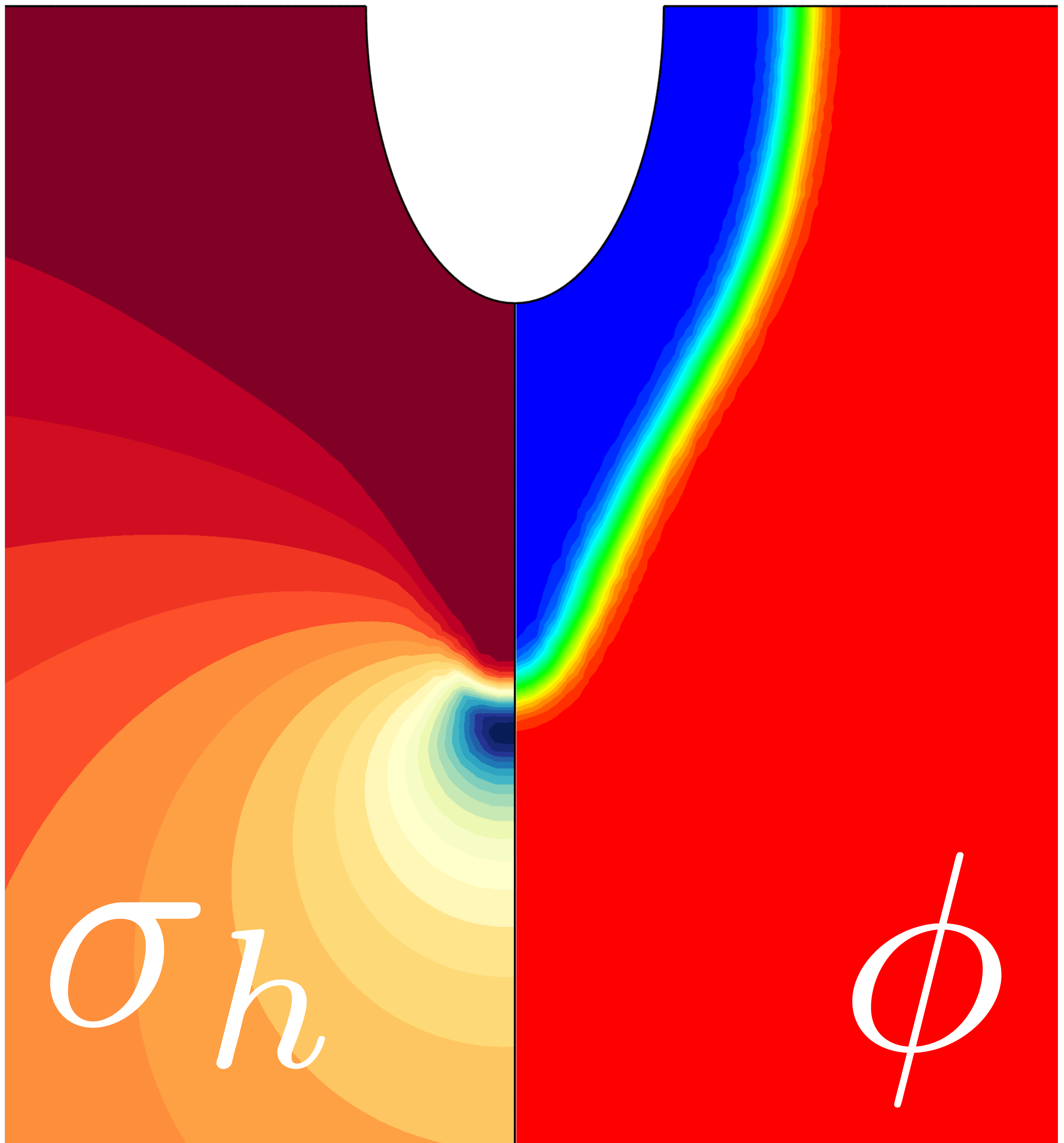}
         \caption{Time=500 sec}
         \label{fig:SCC-phi-a}
     \end{subfigure} \hspace{2mm}
     \begin{subfigure}[b]{0.09\textwidth} 
         \centering
         \includegraphics[width=\textwidth]{Legend-V-phi.pdf}
         \vspace{0mm}
     \end{subfigure} \\ \hspace{35mm} \\
     \begin{subfigure}[b]{0.45\textwidth} 
         \centering 
         \includegraphics[width=\textwidth]{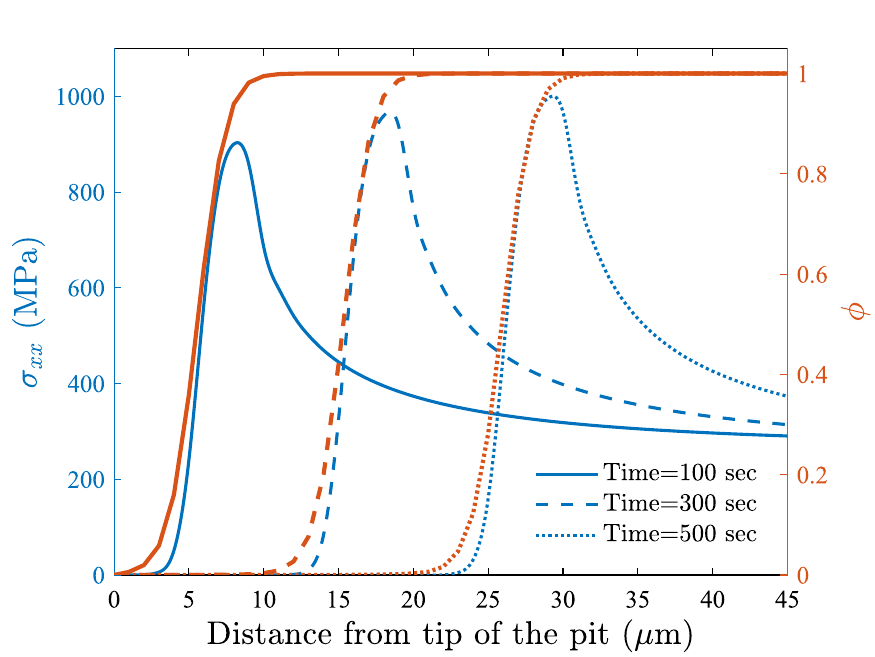}
         \caption{}
         \label{}
     \end{subfigure}
     \begin{subfigure}[b]{0.45\textwidth} 
         \centering 
         \includegraphics[width=\textwidth]{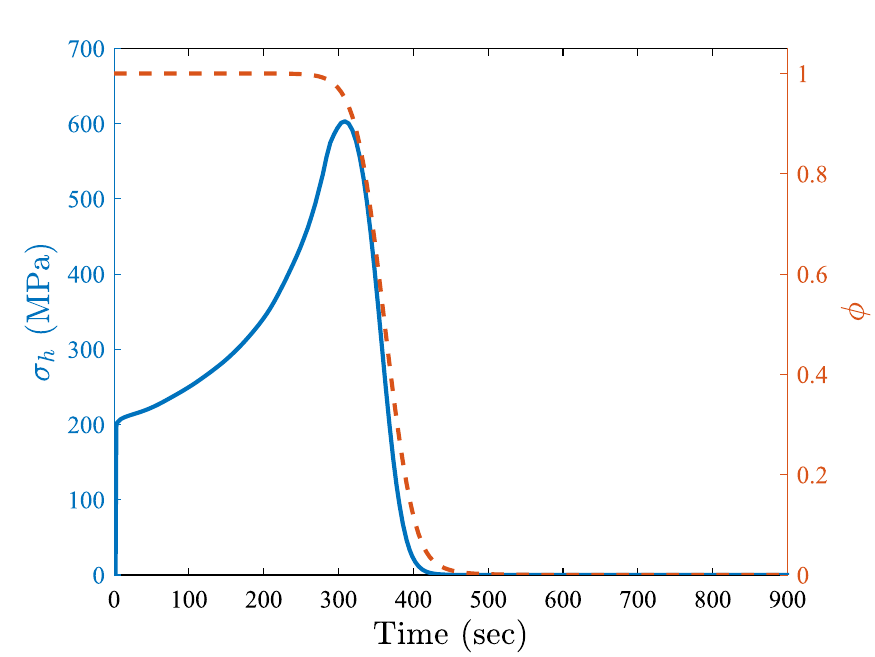}
         \caption{}
         \label{}
     \end{subfigure}
    \caption{Stress-assisted corrosion originating from a semi-elliptical pit. Contours of hydrostatic stress $\sigma_h$ (left) and phase field variable $\phi$ (right) are shown for: (a) $t = 100$ s, (b) $t = 300$ s, and (c) $t = 500$ s. The distributions of normal stress $\sigma_{xx}$ and phase field variable $\phi$ along the vertical line from the pit tip at $t = \{100, 300, 500\}$ s are presented in (d). The temporal evolution of hydrostatic stress $\sigma_h$ and phase field variable $\phi$ at a vertical distance of 20~$\mu$m from the pit tip is shown in (e).}
    \label{fig:SCC-evol}
\end{figure}

The results of the present study are illustrated in Fig. \ref{fig:SCC-evol}. Upon applying the prescribed boundary conditions, the phase field corrosion process initiates and evolves over time. As corrosion progresses, the material undergoes degradation, leading to a reduction in its stiffness, as evidenced by the contours of hydrostatic stress $\sigma_h$ and the phase field variable $\phi$ at different time steps shown in Fig. \ref{fig:SCC-evol}a-c. The role of the evolving corrosion pit in acting as a stress concentrator is also shown. Fig. \ref{fig:SCC-evol}d presents the distributions of normal stress $\sigma_{xx}$ and the phase field variable $\phi$ along a vertical line extending from the initial pit tip at times $t = \{100, 300, 500\}$ s. These results show how the stress vanishes in regions where $\phi = 0$, representing fully corroded material. Fig. \ref{fig:SCC-evol}e shows the temporal evolution of hydrostatic stress and the phase field variable at a point located 20~$\mu$m below the initial pit tip. As observed, hydrostatic stress initially increases during the first second due to the application of displacement. Subsequently, as the pit grows and the corrosion front approaches this point, the hydrostatic stress continues to rise. However, once the solid material at this location transitions to the electrolyte phase (i.e., as $\phi$ approaches 0), the loss of stiffness leads to a decrease in hydrostatic stress.

\begin{figure}[H]
    \centering
    \begin{subfigure}[b]{0.35\textwidth}
         \centering
         \includegraphics[width=\textwidth]{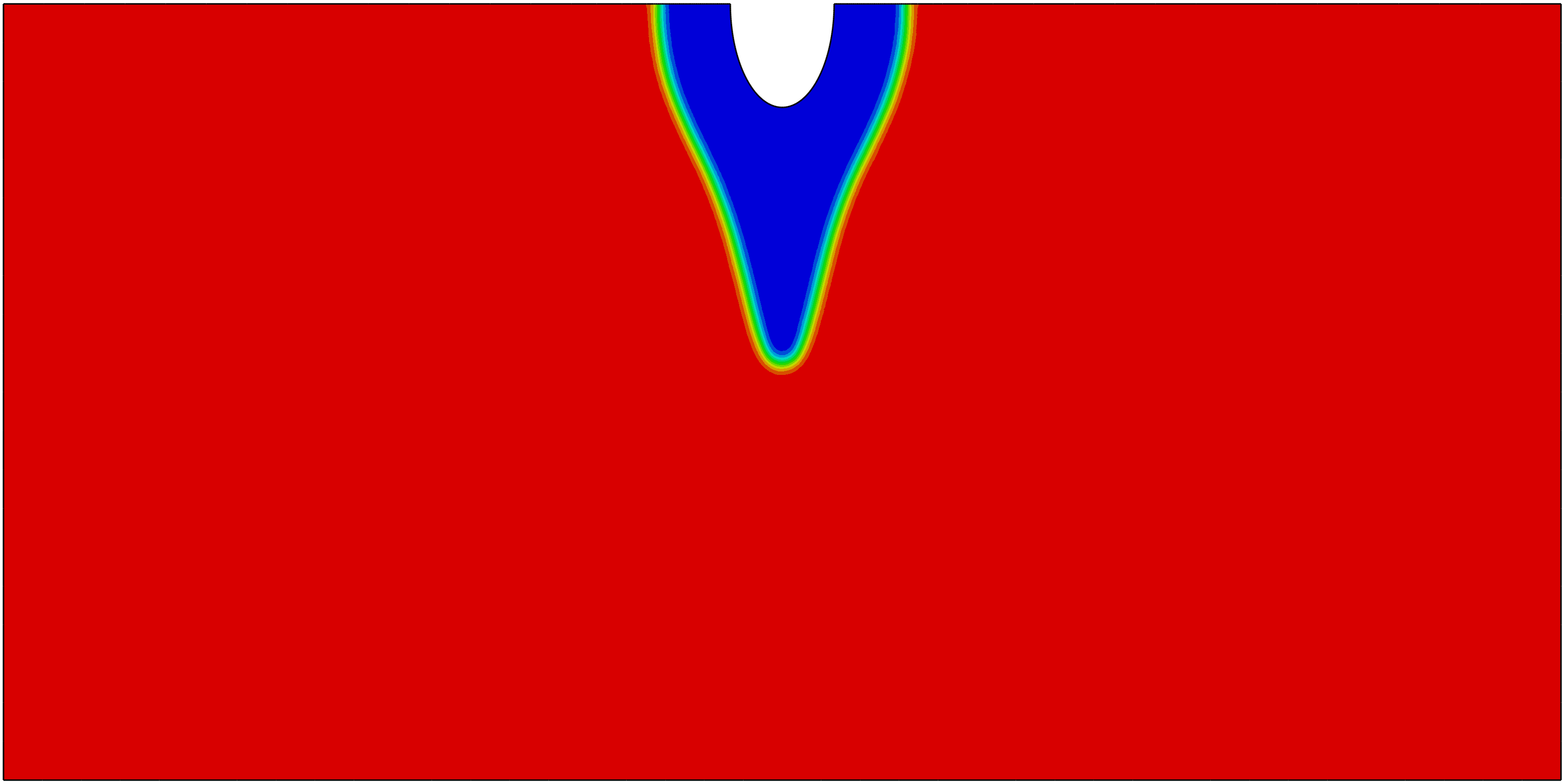}
         \caption{Present work}
         \label{fig:SCC-phi-a}
     \end{subfigure} \hspace{2mm}
     \begin{subfigure}[b]{0.35\textwidth} 
         \centering 
         \includegraphics[width=\textwidth]{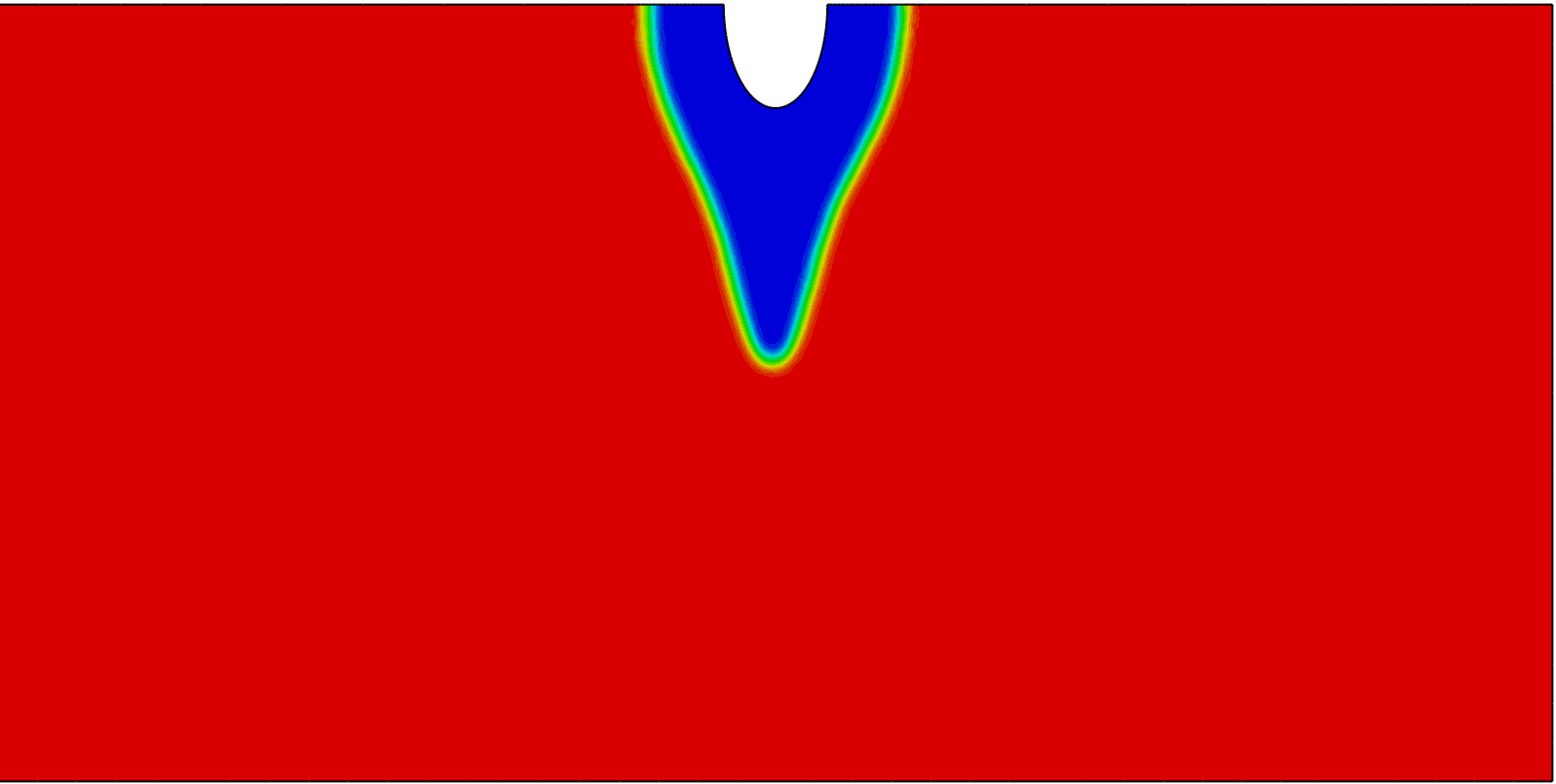}
         \caption{Cui et al. \cite{JMPS2021}}
         \label{fig:SCC-phi-b}
     \end{subfigure} \hspace{2.5mm}
     \begin{subfigure}[b]{0.09\textwidth} 
         \centering
         \includegraphics[width=\textwidth]{Legend-V-phi.pdf}
         \vspace{0mm}
     \end{subfigure} \\ \hspace{35mm} \\
     \begin{subfigure}[b]{0.5\textwidth} 
         \centering 
         \includegraphics[width=\textwidth]{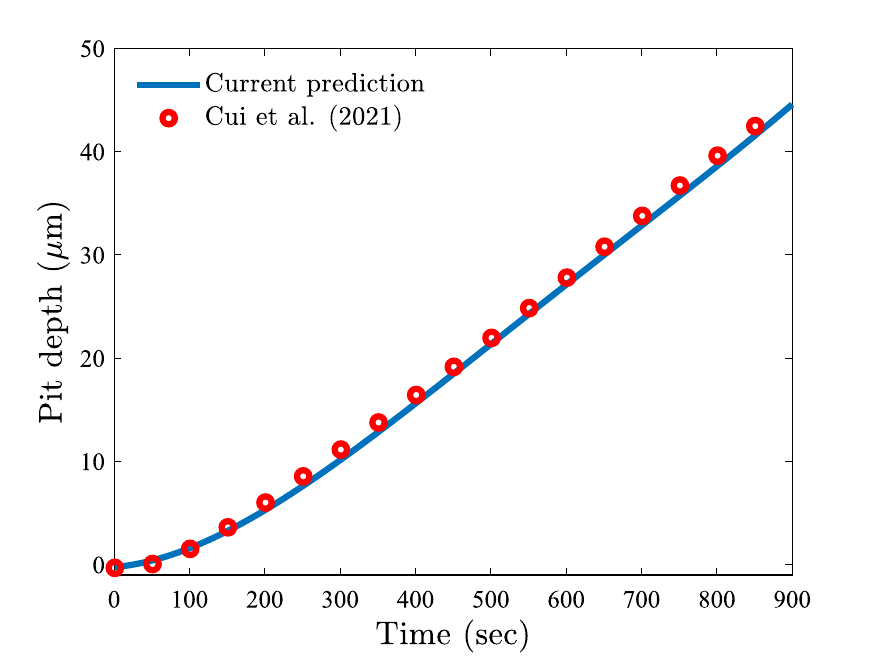}
         \caption{}
         \label{}
     \end{subfigure}
    \caption{Stress-assisted corrosion from a semi-elliptical pit. Phase field contours at a time of $t=900$ s for: (a) the present implementation, and (b) as obtained with the UEL code by Cui et al. \cite{JMPS2021}. A quantitative comparison of pit depth versus time is given in (c).}
    \label{fig:SCC-phi1}
\end{figure}

The results are compared with those obtained from the UEL code provided by Cui et al. \cite{JMPS2021}. A comparison of the phase field contours from our current simulation (Fig. \ref{fig:SCC-phi1}a) with those from the UEL code by Cui et al. \cite{JMPS2021} (Fig. \ref{fig:SCC-phi1}b) demonstrates an excellent agreement. This is corroborated further by the quantitative comparison of pit depth versus time shown in Fig. \ref{fig:SCC-phi1}c. It is important to note that the reference results by Cui et al. \cite{JMPS2021} have been computed using the UEL code provided, which corresponds to the formulation used in Ref. \cite{cui2022generalised}. Slight differences exist due to the arrangement of the weak form, as discussed in \ref{sec:Arang}.

\section{Conclusions}
\label{Sec:Conclusions}

We have presented a generalised formulation to handle coupled phase field fracture and corrosion problems under the same theoretical structure. The versatility of the heat transfer equation is then exploited to provide a novel computational framework capable of handling coupled phase field problems. This significantly facilitates the numerical implementation, which can be done at the integration point level in commercial finite element packages. This is demonstrated in Abaqus but the approach is universal and could be adopted in other platforms. The applications explored span hydraulic fracture, thermal fracture, hydrogen embrittlement and stress-assisted corrosion. 2D and 3D problems are considered. The associated user material subroutines are made freely available to showcase the simplicity of the approach. The results obtained with the novel computational framework presented are well-aligned with experimental and computational results from the literature involving the quenching of ceramic plates, the propagation of pressure and fluid injection-driven cracks in porous media, the growth of cracks assisted by hydrogen in metallic materials, and the corrosion of metals and their interplay with mechanical fields. 

\section*{Acknowledgments}
\label{Sec:Acknowledge of funding}

\noindent Y. Navidtehrani and C. Betegón acknowledge financial support from the Ministry of Science, Innovation, and Universities of Spain through grant MCINN-22-TED2021-130306B-100. Additionally, C. Betegón acknowledges financial support from the Ministry of Science, Innovation, and Universities of Spain under grant MCINN-23-PID2022-1420150B-100. E. Martínez-Pañeda was supported by a UKRI Future Leaders Fellowship (grant MR/V024124/1).

\appendix

\section{Using a UMATHT subroutine to implement diffusion-type equations}
\label{Sec:AppDerivativesUMATHT}

Diffusion-type equations can be readily implemented into commercial finite element packages exploiting the analogy with heat transfer, as discussed in Section \ref{sec:ThermalAnalogy}. In Abaqus, this is achieved through a user-defined thermal material behavior subroutine (\texttt{UMATHT}). One must define relevant variables, including the internal thermal energy per unit mass $U$, the variation of internal thermal energy per unit mass with respect to temperature $\partial U / \partial T$, the variation of internal thermal energy per unit mass with respect to the spatial gradients of temperature $\partial U / \partial (\nabla T)$, the heat flux vector $\mathbf{f}$, the variation of the heat flux vector with respect to temperature $\partial \mathbf{f} / \partial T$, and the variation of the heat flux vector with respect to the spatial gradients of temperature $\partial \mathbf{f} / \partial (\nabla T)$. Table \ref{tab:Unified-UMATHT} provides the mapping of those variables to the relevant variables of each of the governing equations considered in this work (heat transfer, hydrogen transport, fluid flow, corrosion, transport of metallic ions, phase field corrosion, phase field fracture).

\begin{landscape}
\begin{table}[H]
\small
\caption{Unified table of quantities to be defined in a \texttt{UMATHT} subroutine for various equations and their thermal analogies.}
\begin{tabular}{|c |c |c |c |c |c |c |c|}
\hline
Equation & $U$ & DUDT & DUDG & FLUX & DFDT & DFDG \\
\hline
\hline
\makecell{General heat transfer \\ Eq. \eqref{Eq:GHEAT1}} & $U_{t} + \dot{U} \delta t$ & $\frac{\partial U}{\partial T}$ & $\frac{\partial U}{\partial (\nabla T)}$ & $\mathbf{f}$ & $\frac{\partial \mathbf{f}}{\partial T}$ & $\frac{\partial \mathbf{f}}{\partial (\nabla T)}$ \\
\hline
\makecell{General phase field \\ Eq. \eqref{eq;thermal-general-corrosion}} & $\begin{array} {c} U_{t} - \eta \delta \phi - w'(\phi) \delta t - \\ g'(\phi)( f_{b1}(\beta) - f_{b2}(\beta) ) \delta t \end{array}$ & $\begin{array} {c} - \eta -  w^{\prime\prime}(\phi) \delta t - \\ g^{\prime\prime}(\phi) ( f_{b1}(\beta) - f_{b2}(\beta) ) \delta t \end{array}$ & $0$ & $\kappa \nabla \phi$ & $0$ & $\kappa  \bm{I}$ \\
\hline
\makecell{Phase field fracture \\ Eq. \eqref{Eq:phasefieldfracture}} & $U_{t} + \left(\frac{\phi}{\ell^2} + g^{\prime}(\phi) \frac{\mathcal{H}}{G_c \ell}\right) \delta t$ & $\left(\frac{1}{\ell^2} + g^{\prime\prime}(\phi) \frac{\mathcal{H}}{G_c \ell}\right) \delta t$ & $0$ & $-\nabla \phi$ & $0$ & $-\bm{I}$ \\
\hline
\makecell{Metal ion transport \\ Eq. \eqref{Eq:Coros}} & $U_{t} + \delta c$ & $1$ & $0$ & $\begin{array} {c} -D_m \nabla c + \\ D_m \nabla( h(\phi)(c_{\mathrm{Se}} - c_{\mathrm{Le}})- c_{\mathrm{Le}}) \end{array}$ & $0$ & $-D_m\bm{I}$ \\
\hline
\makecell{Phase field corrosion \\ Eq. \eqref{Eq:phasefieldCoro}} & $U_{t} -\frac{1}{ L} \delta {\phi} - \frac{\partial \psi^{\mathrm{ch}}}{\partial \phi} \delta t$ & $-\frac{1}{L} - \frac{\partial^2 \psi^{\mathrm{ch}}}{\partial \phi^2} \delta t$ & $0$ & $\kappa \nabla \phi$ & $0$ & $\kappa \bm{I}$ \\
\hline
\makecell{Fluid flow \\ Eq. \eqref{Eq:dMassContent6}}  & $U_{t} +  \left(S \delta p + \alpha_p \chi_{\text{r}} \delta \varepsilon_{\text{vol}}\right)$ & $S$ & $0$ & $-\rho_{\mathrm{fl}} \frac{\bm{K}_{\text{fl}}}{\mu_{\text{fl}}} \nabla p$ & $0$ & $-\rho_{\mathrm{fl}} \frac{\bm{K}_{\text{fl}}}{\mu_{\text{fl}}} \bm{I}$ \\
\hline
\makecell{Hydrogen transport \\ Eq. \eqref{Eq:HydrogenAna}}  & $U_{t} + \delta c_{\mathrm{H}}$ & $1$ & $0$ & $ \begin{array} {c}-D_{\mathrm{H}} \nabla c_{\mathrm{H}} + \frac{D_{\mathrm{H}}}{R T_{\mathrm{k}}} c_{\mathrm{H}} V_{\mathrm{H}} \nabla \sigma_h \end{array}$ & $\frac{D_{\mathrm{H}}}{R T_{\mathrm{k}}} c_{\mathrm{H}} V_{\mathrm{H}} \nabla \sigma_h$ & $-D_{\mathrm{H}} \bm{I}$ \\
\hline
\makecell{Heat transfert \\ Eq. \eqref{Eq:SHeatEQ}} & $U_{t} +  c_{T} \delta T$ & $c_{T}$ & $0$ & $-\color{black}k_{T} \nabla T$  &  $0$ & $k_{T} \bm{I}$ \\
\hline
\end{tabular}
\label{tab:Unified-UMATHT}
\end{table}
\end{landscape}

\section{Finite element discretisation of multiphysics phase field models}
\label{App:FEdetails}

While in this work the numerical implementation is conducted at the integration point level, details of a more general implementation are also provided for completeness. In a three-field problem involving mechanical deformation, phase transformations and a diffusion-type equation one typically solves for the displacement vector $\mathbf{u}$, the phase field order parameter $\phi$ and the diffusion field $\xi$ as primary (nodal) variables. Their associated balance equations are provided in Eqs. (\ref{Eq:StrMulPhys11}-\ref{Eq:StrMulPhys3}). Then, using test functions $\delta \mathbf{u}$, $\delta \phi$, and $\delta \xi$, the integral form can be written as:
\begin{equation}
\begin{aligned}
    &\int_{\Omega} \left[\nabla \cdot \bm{\sigma} + \mathbf{b} \right] \delta \mathbf{u} \, \mathrm{d} V = 0, \\
    &\int_{\Omega} \left[\kappa \nabla^2 \phi - w'(\phi) - g'(\phi) \big( f_{\text{b1}} - f_{\text{b2}} \big) - \eta \dot{ \phi} \right] \delta \phi \, \mathrm{d} V = 0, \\
    &\int_{\Omega} \left[\rho \dot{U_{\xi}} + \nabla \cdot \mathbf{f_{\xi}} - q_{\xi} \right] \delta \xi \, \mathrm{d} V = 0.
\end{aligned}
\end{equation}

\noindent Using the divergence theorem, the weak form of each equation becomes:
\begin{equation}
\begin{aligned}
    &\int_{\Omega} \left[- \bm{\sigma} : \nabla^{\text{sym}} \delta \mathbf{u} + \mathbf{b} \delta \mathbf{u} \right] \, \mathrm{d} V + \int_{\partial \Omega} \delta \mathbf{u} \bm{\sigma} \cdot \mathbf{n} \, \mathrm{d} S = 0, \\
    &\int_{\Omega} \left[\left( w'(\phi) + g'(\phi) \big( f_{\text{b1}} + f_{\text{b2}} \big) + \eta \dot{ \phi}\right) \delta \phi + \kappa \nabla \phi \cdot \nabla \delta \phi \right] \, \mathrm{d} V - \int_{\partial \Omega} \delta \phi \kappa \nabla \phi \cdot \mathbf{n} \, \mathrm{d} S = 0, \\
    &\int_{\Omega} \left[\left(\rho \dot{U}_{\xi} - q_{\xi} \right) \delta \xi - \mathbf{f}_{\xi} \cdot \nabla \delta \xi \right] \, \mathrm{d} V + \int_{\partial \Omega} \delta \xi \mathbf{f} \cdot \mathbf{n} \, \mathrm{d} S = 0.
\end{aligned}
\end{equation}

\noindent With the following boundary conditions applying on $\partial \Omega$:
\begin{equation}
\begin{aligned}
    &\mathbf{T} = \bm{\sigma} \cdot \mathbf{n}, \\
    &\nabla \phi \cdot \mathbf{n} = 0, \\
    &f_{\xi} = \mathbf{f}_{\xi} \cdot \mathbf{n},
\end{aligned}
\end{equation}

\noindent where $\mathbf{T}$ is the surface traction, $f_{\xi}$ the diffusion field flux.\\

To derive the finite element discretised form, we approximate the variables $\mathbf{u}$, $\phi$, and $\xi$ using shape functions $N_i$ (shape function of node $i$) and nodal values of each field as follows:
\begin{equation}\label{Eq:ShapeFun}
\mathbf{u} = \sum_i^n \bm{N}_i \mathbf{u}_i, \quad \phi = \sum_i^n N_i \phi_i, \quad \xi = \sum_i^n N_i \xi_i.
\end{equation}

The gradients of each field are found by taking derivatives of the shape functions with respect to space, creating the $\bm{B}$-matrices:
\begin{equation}\label{Eq:BMatrices}
\boldsymbol{\varepsilon} = \sum_i^n \bm{B}_i^u \boldsymbol{u}_i, \quad \nabla \phi = \sum_i^n \bm{B}_i \phi_i, \quad \nabla \xi = \sum_i^n \bm{B}_i \xi_i.
\end{equation}

Using Eqs. (\ref{Eq:ShapeFun}) and (\ref{Eq:BMatrices}), the residual for each equation becomes:
\begin{equation}
\begin{aligned}
&\mathbf{R}_i^\mathbf{u} = \int_{\Omega} \left[(\bm{B}_i^\mathbf{u})^T \bm{\sigma} - (\bm{N}_i)^T \mathbf{b}\right] \, \mathrm{d} V - \int_{\partial \Omega} (\bm{N}_i)^T \mathbf{T} \, \mathrm{d} S, \\
&R_i^\phi = \int_{\Omega} \left[\left(\eta \frac{\phi^{t+\delta t} - \phi^{t}}{\delta t} + w^{\prime}(\phi) + g'(\phi) \big( f_{\text{b1}} + f_{\text{b2}} \big)\right) N_i - \kappa \bm{B}_i^T \nabla \phi \right] \, \mathrm{d} V, \\
&R_i^\xi = \int_{\Omega} \left[\left(\rho \frac{U_{\xi}^{t+\delta t} - U_{\xi}^{t}}{\delta t} - q_{\xi}\right) N_i - \bm{B}_i^T \mathbf{f}_{\xi}\right] \, \mathrm{d} V - \int_{\partial \Omega} N_i f_{\xi} \, \mathrm{d} S.
\end{aligned}
\end{equation}

Finally, by taking the variation of each residual with respect to the field variable, we obtain the stiffness matrix:
\begin{equation}
\begin{aligned}
&\bm{K}_{ij}^\mathbf{u} = \frac{\partial \mathbf{R}_i^\mathbf{u}}{\partial \mathbf{u}_j} = \int_{\Omega} (\bm{B}_i^\mathbf{u})^T \bm{C} \bm{B}_j^\mathbf{u} \, \mathrm{d} V, \\
&\bm{K}_{ij}^{\phi} = \frac{\partial {R}_i^{\phi}}{\partial \phi_j} = 
 \int_{\Omega} \big(\eta + w^{\prime\prime}(\phi) + g^{\prime\prime}(\phi) ( f_{\text{b1}} + f_{\text{b2}})\big) N_i N_j \, \mathrm{d} V - \int_{\Omega} (\bm{B}_i)^T \bm{B}_j \, \mathrm{d} V, \\
&\bm{K}_{ij}^{\xi} = \frac{\partial {R}_i^{\xi}}{\partial \xi_j} = 
\frac{1}{\delta t} \int_{\Omega} N_i \rho \frac{\partial U_{\xi}}{\partial \xi} N_j \, \mathrm{d} V + \frac{1}{\delta t} \int_{\Omega} N_i \rho \frac{\partial U_{\xi}}{\partial \nabla \xi} \cdot \bm{B}_j \, \mathrm{d} V \\
& \hspace{10mm} - \int_{\Omega} (\bm{B}_i)^T \cdot \frac{\partial \mathbf{f}_{\xi}}{\partial \xi} N_j \, \mathrm{d} V - \int_{\Omega} (\bm{B}_i)^T \cdot \frac{\partial \mathbf{f}_{\xi}}{\partial \nabla \xi} \cdot \bm{B}_j \, \mathrm{d} V \\
& \hspace{10mm} - \int_{\Omega} N_i \frac{\partial r}{\partial \xi} N_j \, \mathrm{d} V - \int_{\partial \Omega} N_i \frac{\partial f_{\xi}}{\partial \xi} N_j \, \mathrm{d} S.
\end{aligned}
\end{equation}

The off-diagonal stiffness matrices are considered zero to reduce computational effort, allowing the inversion process to handle lower-dimensional stiffness matrices more efficiently.

\section{On the arrangement of the phase field equation}
\label{sec:Arang}

We show here how the arrangement of the strong form of the phase field equation (or any type of diffusion equation) can lead to different weak forms (and thus results) if the mobility coefficient varies in space ($L(\mathbf{x})$). This is for example the case when the mobility coefficient is enhanced to capture the role of mechanics in corrosion, as shown in Eq. (\ref{eq:Lmech}). We consider three distinct cases, which should be equivalent for a constant (homogeneous) mobility coefficient $L$:
\begin{equation}\label{Eq:StrongDif}
\begin{aligned}
& \frac{1}{L(\mathbf{x})} \dot{\phi} + \nabla \cdot \nabla \phi = 0, \\
& \dot{\phi} + L(\mathbf{x}) \nabla \cdot \nabla \phi = 0, \\
& \dot{\phi} + \nabla \cdot (L(\mathbf{x}) \cdot \nabla \phi) = 0,
\end{aligned}
\end{equation}

\noindent The weak form of each equation can be readily obtained by introducing the test function $\delta \phi$,
\begin{equation}
\begin{aligned}
& \int_{\Omega} \left( \frac{1}{ L(x)} \dot{\phi} + \nabla \cdot \nabla \phi \right) \delta \phi \, \mathrm{d} V = 0, \\
& \int_{\Omega} \left( \dot{\phi} +  L(x) \color{black} \nabla \cdot \nabla \phi \right) \delta \phi \, \mathrm{d} V = 0, \\
& \int_{\Omega} \left( \dot{\phi} + \nabla \cdot ( L(x) \color{black} \cdot \nabla \phi) \right) \delta \phi \, \mathrm{d} V = 0.
\end{aligned}
\end{equation}

Utilising the divergence theorem, the weak forms can also be formulated as,
\begin{equation}
\begin{aligned}
& \int_{\Omega} \left( \frac{1}{ L(x) \color{black} } \dot{\phi} \delta \phi - \nabla \delta \phi \cdot \nabla \phi \right) \, \mathrm{d} V + \int_{\partial \Omega} \nabla \phi \cdot \mathbf{n} \, \delta \phi \, \mathrm{d} S = 0, \\
& \int_{\Omega} \left( \dot{\phi} \delta \phi -  L(x) \color{black}  \nabla \delta \phi \cdot \nabla \phi - \nabla  L(x) \color{black}  \cdot \nabla \phi \, \delta \phi \right) \, \mathrm{d} V + \int_{\partial \Omega} \nabla \phi \cdot \mathbf{n} \, \delta \phi \, \mathrm{d} S = 0, \\
& \int_{\Omega} \left( \dot{\phi} \delta \phi -  L(x) \color{black}  \nabla \delta \phi \cdot \nabla \phi \right) \, \mathrm{d} V + \int_{\partial \Omega}  L(x) \color{black}  \nabla \phi \cdot \mathbf{n} \, \delta \phi \, \mathrm{d} S = 0.
\end{aligned}
\end{equation}

These derived weak forms yield distinct results. In the first equation, the parameter $L(\mathbf{x})$ acts as a regularization parameter influencing the system's equilibrium. The second equation incorporates the gradient of this parameter into its weak form, while in the third equation, $L(\mathbf{x})$ is a parameter for the gradient term of the field variable. Eq. \eqref{Eq:StrongDif}a is the approach typically adopted \cite{cui2022generalised}. Both Eq. \eqref{Eq:StrongDif}a and Eq. \eqref{Eq:StrongDif}c can be implemented using a \texttt{UMATHT} subroutine.\\

\section{Tangential stiffness matrix for the no-tension split}
\label{Sec:TSM}

In the hydraulic fracture analyses, we chose to adopt the no-tension strain energy decomposition \cite{DelPiero1989,Freddi2010} for both the phase field and the balance of linear momentum equations (i.e., not using the hybrid approach but retaining variational consistency). This requires defining a suitable tangential stiffness tensor, which is the aim of this Appendix. Let us begin by defining the total strain energy density as:
\begin{equation}
    \psi \left( \bm{\varepsilon}, \phi \right) = g \left( \phi \right) \psi_1^{\mathrm{M}} \left( \bm{\varepsilon} \right) + \left( 1 - g \left( \phi \right) \right) \psi_2^{\mathrm{M}} \left( \bm{\varepsilon} \right) \, ,
\end{equation}

\noindent with the tangential stiffness tensor being defined as:
\begin{equation}\label{Eq:Stiffness}
\bm{C} = g \left( \phi \right) \frac{\partial^2 \psi_1^{\mathrm{M}}}{\partial \boldsymbol\varepsilon^2} + \left( 1 - g \left( \phi \right) \right) \frac{\partial^2 \psi_2^{\mathrm{M}}}{\partial \boldsymbol\varepsilon^2} = g \left( \phi \right) \bm{C}_1^{\mathrm{M}} + \left( 1 - g \left( \phi \right) \right) \bm{C}_2^{\mathrm{M}} \, .
\end{equation}

\noindent Here, $\bm{C}_1^{\mathrm{M}}$ and $\bm{C}_2^{\mathrm{M}}$ are the tangential moduli of the first and second materials, respectively.

For the no-tension model, the strain energy can be expressed as:
\begin{equation}
\begin{aligned}
\psi= & \frac{E \nu}{2(1+\nu)(1-2 \nu)}\left\{\left[\epsilon_1-(1-\sqrt{g}) \epsilon_1^t\right]+\left[\epsilon_2-(1-\sqrt{g}) \epsilon_2^t\right]+\left[\epsilon_3-(1-\sqrt{g}) \epsilon_3^t\right]\right\}^2 \\
& + \frac{E}{2(1+\nu)}\left\{\left[\epsilon_1-(1-\sqrt{g}) \epsilon_1^t\right]^2+\left[\epsilon_2-(1-\sqrt{g}) \epsilon_2^t\right]^2+\left[\epsilon_3-(1-\sqrt{g}) \epsilon_3^t\right]^2\right\} \, ,
\end{aligned}
\end{equation}

\noindent where $\epsilon_i$ and $\epsilon_i^t$ are the principal strain and the principal tensile strain, respectively, with $i$ indexing from the minimum to the maximum principal strain ($\epsilon_3 \geq \epsilon_2 \geq \epsilon_1$). The principal tensile strains are defined based on the strain state:
\begin{equation}
\begin{aligned}
\text{if} & \quad \epsilon_1>0 & & \quad \Rightarrow \quad \epsilon_1^t = \epsilon_1, \quad \epsilon_2^t = \epsilon_2, \quad \epsilon_3^t = \epsilon_3, \\
\text{elseif} & \quad \epsilon_2 + \nu \epsilon_1 > 0 & & \quad \Rightarrow \quad \epsilon_1^t = 0, \quad \epsilon_2^t = \epsilon_2 + \nu \epsilon_1, \quad \epsilon_3^t = \epsilon_3 + \nu \epsilon_1, \\
\text{elseif} & \quad (1-\nu) \epsilon_3 + \nu (\epsilon_1 + \epsilon_2) > 0 & & \quad \Rightarrow \quad \epsilon_1^t = 0, \quad \epsilon_2^t = 0, \quad \epsilon_3^t = \epsilon_3 + \frac{\nu}{1-\nu} (\epsilon_1 + \epsilon_2), \\
& & & \quad \text{else} \quad \Rightarrow \quad \epsilon_1^t = 0, \quad \epsilon_2^t = 0, \quad \epsilon_3^t = 0.
\end{aligned}
\end{equation}

Based on this definition, the strain energy of the second phase can be expressed as:
\begin{equation}
\psi_2^{\text{M}}(\boldsymbol{\varepsilon})=
\begin{cases}
0 & \epsilon_1>0 \\
\frac{E}{2} \epsilon_1^2 & \epsilon_2+\nu \epsilon_1>0 \\
\frac{E}{2\left(1-v^2\right)}\left(\epsilon_1^2+\epsilon_2^2+2 v \epsilon_1 \epsilon_2\right) & (1-\nu)\epsilon_3+\nu (\epsilon_1+\epsilon_2)>0 \\
\frac{E v}{2(1+\nu)(1-2 v)}\left(\epsilon_1+\epsilon_2+\epsilon_3\right)^2+\frac{E}{2(1+\nu)}\left(\epsilon_1^2+\epsilon_2^2+\epsilon_3^2\right) & \text{else}
\end{cases}
\end{equation}

For a fully damaged material, the material Jacobian in the principal direction, $(\bm{C}_2^{\mathrm{M}})'$, is defined as:
\begin{equation}
(C_2^{\mathrm{M}})^{'}_{ijkl}=\frac{\partial \psi_s}{\partial \epsilon_{ij} \partial \epsilon_{kl}}=
\begin{cases}
0 & \epsilon_1>0, \\
\delta_{i1} \delta_{j1} \delta_{k1} \delta_{l1} \, E &  \epsilon_2 + \nu \epsilon_1 > 0, \\
\delta_{ij} \delta_{kl} (1-\delta_{i3}) \big(\delta_{ik}+(1-\delta_{k3}) \, \nu \big) \, \frac{E}{1 - \nu^2} & (1-\nu)\epsilon_3 + \nu (\epsilon_1 + \epsilon_2) > 0, \\
\bm{C}_0 ,
\end{cases}
\end{equation}

\noindent where $\delta_{ij}$  is the Kronecker delta. The tangential matrix of the first phase is equal to the elastic stiffness matrix so $\bm{C}_1^{\mathrm{M}}=\bm{C}_0$. The tangential stiffness matrix in the principal direction, $\bm{C}^{'}$, can be written as:
\begin{equation}
\bm{C}^{'} = g \left( \phi \right)(\bm{C}_0) + \left( 1 - g \left( \phi \right) \right)(\bm{C}_2^{\mathrm{M}})'.
\end{equation}

The Jacobian matrix in the original direction, $\bm{C}$, is obtained by rotating $\bm{C}^{'}$ to the original orientation using:
\begin{equation}\label{Eq:Rotation}
C_{qrst} = a_{qi} a_{rj} a_{sk} a_{tl} C_{ijkl}^{'} \, ,
\end{equation}

\noindent where $\bm{a}$ is the transpose of the direction cosines matrix for the principal directions, defined as:
\begin{equation}
\bm{a}^{'} = [\boldsymbol{v}_1, \boldsymbol{v}_2, \boldsymbol{v}_3] \, ,
\end{equation}

\noindent where $\boldsymbol{v}_1$, $\boldsymbol{v}_2$, and $\boldsymbol{v}_3$ are the principal vectors of the strain tensor, satisfying:
\begin{equation}
(\boldsymbol{\varepsilon} - \epsilon_{ii} \bm{I}) \cdot \boldsymbol{v}_i = 0,
\end{equation}

with $i=1,2,3$ and $\bm{I}$ being the identity matrix.


\begin{thebibliography}{10}
\expandafter\ifx\csname url\endcsname\relax
  \def\url#1{\texttt{#1}}\fi
\expandafter\ifx\csname urlprefix\endcsname\relax\def\urlprefix{URL }\fi
\expandafter\ifx\csname href\endcsname\relax
  \def\href#1#2{#2} \def\path#1{#1}\fi

\bibitem{cahn1958free}
J.~W. Cahn, J.~E. Hilliard, Free energy of a nonuniform system. i. interfacial
  free energy, The Journal of chemical physics 28~(2) (1958) 258--267.

\bibitem{wheeler1992phase}
A.~A. Wheeler, W.~J. Boettinger, G.~B. McFadden, Phase-field model for
  isothermal phase transitions in binary alloys, Physical Review A 45~(10)
  (1992) 7424.

\bibitem{Chen2002}
L.~Q. Chen, {Phase-field models for microstructure evolution}, Annual Review of
  Materials Science 32 (2002) 113--140.

\bibitem{zhao2022phase}
Y.~Zhao, R.~Wang, E.~Mart{\'\i}nez-Pa{\~n}eda, A phase field
  electro-chemo-mechanical formulation for predicting void evolution at the
  li--electrolyte interface in all-solid-state batteries, Journal of the
  Mechanics and Physics of Solids 167 (2022) 104999.

\bibitem{MOKBEL2018823}
D.~Mokbel, H.~Abels, S.~Aland, A phase-field model for fluid–structure
  interaction, Journal of Computational Physics 372 (2018) 823--840.

\bibitem{Bourdin2000}
B.~Bourdin, G.~A. Francfort, J.-J. Marigo, {Numerical experiments in revisited
  brittle fracture}, Journal of the Mechanics and Physics of Solids 48~(4)
  (2000) 797--826.

\bibitem{Bourdin2008}
B.~Bourdin, G.~A. Francfort, J.~J. Marigo, {The variational approach to
  fracture}, Springer Netherlands, 2008.

\bibitem{Karma2001}
A.~Karma, D.~A. Kessler, H.~Levine, {Phase-field model of mode III dynamic
  fracture}, Physical Review Letters 87~(4) (2001) 45501--1--45501--4.

\bibitem{Mai2016}
W.~Mai, S.~Soghrati, R.~G. Buchheit, {A phase field model for simulating the
  pitting corrosion}, Corrosion Science 110 (2016) 157--166.

\bibitem{JMPS2021}
C.~Cui, R.~Ma, E.~Mart{\'{i}}nez-Pa{\~{n}}eda, {A phase field formulation for
  dissolution-driven stress corrosion cracking}, Journal of the Mechanics and
  Physics of Solids 147 (2021) 104254.

\bibitem{Quintanas-Corominas2019}
A.~Quintanas-Corominas, J.~Reinoso, E.~Casoni, A.~Turon, J.~A. Mayugo, {A phase
  field approach to simulate intralaminar and translaminar fracture in long
  fiber composite materials}, Composite Structures 220 (2019) 899--911.

\bibitem{CST2021}
W.~Tan, E.~Mart{\'{i}}nez-Pa{\~{n}}eda, {Phase field predictions of microscopic
  fracture and R-curve behaviour of fibre-reinforced composites}, Composites
  Science and Technology 202 (2021) 108539.

\bibitem{CMAME2021}
M.~Simoes, E.~Mart{\'{i}}nez-Pa{\~{n}}eda, {Phase field modelling of fracture
  and fatigue in Shape Memory Alloys}, Computer Methods in Applied Mechanics
  and Engineering 373 (2021) 113504.

\bibitem{lotfolahpour2023phase}
A.~Lotfolahpour, W.~Huber, M.~A. Zaeem, A phase-field model for interactive
  evolution of phase transformation and cracking in superelastic shape memory
  ceramics, Computational Materials Science 216 (2023) 111844.

\bibitem{Wu2021}
J.-Y. Wu, Y.~Huang, V.~P. Nguyen, {Three-dimensional phase-field modeling of
  mode I + II / III failure in solids}, Computer Methods in Applied Mechanics
  and Engineering 373 (2021) 113537.

\bibitem{shishvan2021mechanism}
S.~S. Shishvan, S.~Assadpour-asl, E.~Martinez-Paneda, A mechanism-based
  gradient damage model for metallic fracture, Engineering Fracture Mechanics
  255 (2021) 107927.

\bibitem{Sun2021}
X.~Sun, R.~Duddu, Hirshikesh, {A poro-damage phase field model for
  hydrofracturing of glacier crevasses}, Extreme Mechanics Letters 45 (2021)
  101277.

\bibitem{Clayton2022}
T.~Clayton, R.~Duddu, M.~Siegert, E.~Mart{\'{i}}nez-Pa{\~{n}}eda, {A
  stress-based poro-damage phase field model for hydrofracturing of creeping
  glaciers and ice shelves}, Engineering Fracture Mechanics 272 (2022) 108693.

\bibitem{Schuler2020}
L.~Schuler, A.~G. Ilgen, P.~Newell, {Chemo-mechanical phase-field modeling of
  dissolution-assisted fracture}, Computer Methods in Applied Mechanics and
  Engineering 362 (2020) 112838.

\bibitem{navidtehrani2024damage}
Y.~Navidtehrani, R.~Duddu, E.~Mart{\'\i}nez-Pa{\~n}eda, Damage mechanics
  challenge: Predictions based on the phase field fracture model, Engineering
  Fracture Mechanics 301 (2024) 110046.

\bibitem{AHMADIAN2024110417}
H.~Ahmadian, M.~R. Mehraban, M.~R. Ayatollahi, Y.~Navidtehrani, B.~Bahrami,
  Phase-field approach for fracture prediction of brittle cracked components,
  Engineering Fracture Mechanics 309 (2024) 110417.

\bibitem{narayan2019gradient}
S.~Narayan, L.~Anand, A gradient-damage theory for fracture of quasi-brittle
  materials, Journal of the Mechanics and Physics of Solids 129 (2019)
  119--146.

\bibitem{korec2024predicting}
E.~Korec, L.~Mingazzi, F.~Freddi, E.~Mart{\'\i}nez-Pa{\~n}eda, Predicting the
  impact of water transport on carbonation-induced corrosion in variably
  saturated reinforced concrete, Materials and Structures 57~(4) (2024) 91.

\bibitem{CPB2019}
Hirshikesh, S.~Natarajan, R.~K. Annabattula, E.~Mart{\'{i}}nez-Pa{\~{n}}eda,
  {Phase field modelling of crack propagation in functionally graded
  materials}, Composites Part B: Engineering 169 (2019) 239--248.

\bibitem{Kumar2021}
P.~K. A.~V. Kumar, A.~Dean, J.~Reinoso, P.~Lenarda, M.~Paggi, {Phase field
  modeling of fracture in Functionally Graded Materials: G -convergence and
  mechanical insight on the effect of grading}, Thin-Walled Structures 159
  (2021) 107234.

\bibitem{Bourdin2012}
B.~Bourdin, C.~Chukwudozie, K.~Yoshioka, {A variational approach to the
  numerical simulation of hydraulic fracturing}, Proceedings - SPE Annual
  Technical Conference and Exhibition 2 (2012) 1442--1452.

\bibitem{Heider2021}
Y.~Heider, A review on phase-field modeling of hydraulic fracturing,
  Engineering Fracture Mechanics 253 (2021) 107881.

\bibitem{Boyce2022}
A.~M. Boyce, E.~Mart{\'{i}}nez-Pa{\~{n}}eda, A.~Wade, Y.~S. Zhang, J.~J.
  Bailey, T.~M. Heenan, P.~R. {Brett, Dan J. L., Shearing}, {Cracking
  predictions of lithium-ion battery electrodes by X-ray computed tomography
  and modelling}, Journal of Power Sources 526 (2022) 231119.

\bibitem{Ai2022}
W.~Ai, B.~Wu, E.~Mart{\'{i}}nez-Pa{\~{n}}eda, {A multi-physics phase field
  formulation for modelling fatigue cracking in lithium-ion battery electrode
  particles}, Journal of Power Sources 544 (2022) 231805.

\bibitem{zhang2019phase}
G.~Zhang, T.~F. Guo, Z.~Zhou, S.~Tang, X.~Guo, A phase-field model for fracture
  in water-containing soft solids, Engineering Fracture Mechanics 212 (2019)
  180--196.

\bibitem{zheng2022phase}
S.~Zheng, R.~Huang, R.~Lin, Z.~Liu, A phase field solution for modelling
  hyperelastic material and hydrogel fracture in abaqus, Engineering Fracture
  Mechanics 276 (2022) 108894.

\bibitem{korec2023phase}
E.~Korec, M.~Jir{\'a}sek, H.~S. Wong, E.~Mart{\'\i}nez-Pa{\~n}eda, A
  phase-field chemo-mechanical model for corrosion-induced cracking in
  reinforced concrete, Construction and Building Materials 393 (2023) 131964.

\bibitem{fang2023multi}
X.~Fang, Z.~Pan, R.~Ma, et~al., A multi-phase-field framework for non-uniform
  corrosion and corrosion-induced concrete cracking, Computer Methods in
  Applied Mechanics and Engineering 414 (2023) 116196.

\bibitem{Bourdin2014}
B.~Bourdin, J.~J. Marigo, C.~Maurini, P.~Sicsic, {Morphogenesis and propagation
  of complex cracks induced by thermal shocks}, Physical Review Letters 112~(1)
  (2014) 1--5.

\bibitem{Miehe2015c}
C.~Miehe, L.~M. Sch{\"{a}}nzel, H.~Ulmer, {Phase field modeling of fracture in
  multi-physics problems. Part I. Balance of crack surface and failure criteria
  for brittle crack propagation in thermo-elastic solids}, Computer Methods in
  Applied Mechanics and Engineering 294 (2015) 449--485.

\bibitem{MARTINEZPANEDA2018742}
E.~Martínez-Pañeda, A.~Golahmar, C.~F. Niordson, A phase field formulation
  for hydrogen assisted cracking, Computer Methods in Applied Mechanics and
  Engineering 342 (2018) 742--761.

\bibitem{CUI2024315}
C.~Cui, P.~Bortot, M.~Ortolani, E.~Martínez-Pañeda, Computational predictions
  of hydrogen-assisted fatigue crack growth, International Journal of Hydrogen
  Energy 72 (2024) 315--325.

\bibitem{abdollahi2015phase}
A.~Abdollahi, I.~Arias, Phase-field modeling of fracture in ferroelectric
  materials, Archives of Computational Methods in Engineering 22 (2015)
  153--181.

\bibitem{quinteros2023electromechanical}
L.~Quinteros, E.~Garc{\'\i}a-Mac{\'\i}as, E.~Mart{\'\i}nez-Pa{\~n}eda,
  Electromechanical phase-field fracture modelling of piezoresistive cnt-based
  composites, Computer Methods in Applied Mechanics and Engineering 407 (2023)
  115941.

\bibitem{martinez2024phase}
E.~Mart{\'\i}nez-Pa{\~n}eda, Phase-field simulations opening new horizons in
  corrosion research, MRS Bulletin 49 (2024) 603--612.

\bibitem{cui2022generalised}
C.~Cui, R.~Ma, E.~Mart{\'\i}nez-Pa{\~n}eda, A generalised, multi-phase-field
  theory for dissolution-driven stress corrosion cracking and hydrogen
  embrittlement, Journal of the Mechanics and Physics of Solids 166 (2022)
  104951.

\bibitem{allen1979microscopic}
S.~M. Allen, J.~W. Cahn, A microscopic theory for antiphase boundary motion and
  its application to antiphase domain coarsening, Acta metallurgica 27~(6)
  (1979) 1085--1095.

\bibitem{goldenfeld2018lectures}
N.~Goldenfeld, Lectures on phase transitions and the renormalization group, CRC
  Press, 2018.

\bibitem{Amor2009}
H.~Amor, J.~J. Marigo, C.~Maurini, {Regularized formulation of the variational
  brittle fracture with unilateral contact: Numerical experiments}, Journal of
  the Mechanics and Physics of Solids 57~(8) (2009) 1209--1229.

\bibitem{Miehe2010a}
C.~Miehe, M.~Hofacker, F.~Welschinger, {A phase field model for
  rate-independent crack propagation: Robust algorithmic implementation based
  on operator splits}, Computer Methods in Applied Mechanics and Engineering
  199~(45-48) (2010) 2765--2778.

\bibitem{Freddi2010}
F.~Freddi, G.~Royer-Carfagni, {Regularized variational theories of fracture: A
  unified approach}, Journal of the Mechanics and Physics of Solids 58~(8)
  (2010) 1154--1174.

\bibitem{Lorenzis2021}
L.~De~Lorenzis, C.~Maurini, Nucleation under multi-axial loading in variational
  phase-field models of brittle fracture, International Journal of Fracture
  237~(1-2) (2022) 61--81.

\bibitem{Navidtehrani2022}
Y.~Navidtehrani, C.~Beteg{\'{o}}n, E.~Mart{\'{i}}nez-Pa{\~{n}}eda, {A general
  framework for decomposing the phase field fracture driving force,
  particularised to a Drucker–Prager failure surface}, Theoretical and
  Applied Fracture Mechanics 121 (2022) 103555.

\bibitem{cui2023electro}
C.~Cui, R.~Ma, E.~Mart{\'\i}nez-Pa{\~n}eda, Electro-chemo-mechanical phase
  field modeling of localized corrosion: theory and comsol implementation,
  Engineering with Computers 39~(6) (2023) 3877--3894.

\bibitem{PETROVA2020102605}
V.~Petrova, S.~Schmauder, A theoretical model for the study of thermal fracture
  of functionally graded thermal barrier coatings with a system of edge and
  internal cracks, Theoretical and Applied Fracture Mechanics 108 (2020)
  102605.

\bibitem{LI2021793}
W.~Li, K.~Shirvan, Multiphysics phase-field modeling of quasi-static cracking
  in urania ceramic nuclear fuel, Ceramics International 47~(1) (2021)
  793--810.

\bibitem{RUAN2024105756}
H.~Ruan, X.-L. Peng, Y.~Yang, D.~Gross, B.-X. Xu, Phase-field ductile fracture
  simulations of thermal cracking in additive manufacturing, Journal of the
  Mechanics and Physics of Solids 191 (2024) 105756.

\bibitem{10.1115/1.3231067}
D.~A. Mendelsohn, A review of hydraulic fracture modeling—part i: General
  concepts, 2d models, motivation for 3d modeling, Journal of Energy Resources
  Technology 106~(3) (1984) 369--376.

\bibitem{chen2021review}
B.~Chen, B.~R. Barboza, Y.~Sun, J.~Bai, H.~R. Thomas, M.~Dutko, M.~Cottrell,
  C.~Li, A review of hydraulic fracturing simulation, Archives of Computational
  Methods in Engineering (2021) 1--58.

\bibitem{Lee2016}
S.~Lee, M.~F. Wheeler, T.~Wick, {Pressure and fluid-driven fracture propagation
  in porous media using an adaptive finite element phase field model}, Comput.
  Methods Appl. Mech. Engrg 305 (2016) 111--132.

\bibitem{Gangloff2012}
R.~P. Gangloff, B.~P. Somerday, {Gaseous Hydrogen Embrittlement of Materials in
  Energy Technologies}, Woodhead Publishing Limited, Cambridge, 2012.

\bibitem{CHEN2024}
Y.-S. Chen, C.~Huang, P.-Y. Liu, H.-W. Yen, R.~Niu, P.~Burr, K.~L. Moore,
  E.~Martínez-Pañeda, A.~Atrens, J.~M. Cairney, Hydrogen trapping and
  embrittlement in metals – a review, International Journal of Hydrogen
  Energy (2024).

\bibitem{CMAME2018}
E.~Mart{\'{i}}nez-Pa{\~{n}}eda, A.~Golahmar, C.~F. Niordson, {A phase field
  formulation for hydrogen assisted cracking}, Computer Methods in Applied
  Mechanics and Engineering 342 (2018) 742--761.

\bibitem{JMPS2020}
P.~K. Kristensen, C.~F. Niordson, E.~Mart{\'{i}}nez-Pa{\~{n}}eda, {A phase
  field model for elastic-gradient-plastic solids undergoing hydrogen
  embrittlement}, Journal of the Mechanics and Physics of Solids 143 (2020)
  104093.

\bibitem{CupertinoMalheiros2024}
L.~Cupertino-Malheiros, T.~K. Mandal, F.~Thébault, E.~Martínez-Pañeda, On
  the suitability of single-edge notch tension (sent) testing for assessing
  hydrogen-assisted cracking susceptibility, Engineering Failure Analysis 162
  (2024) 108360.

\bibitem{Parkins1987}
R.~N. Parkins, {Factors Influencing Stress Corrosion Crack Growth Kinetics.},
  Corrosion 43~(3) (1987) 130--139.

\bibitem{Navidtehrani2021a}
Y.~Navidtehrani, C.~Beteg{\'{o}}n, E.~Mart{\'{i}}nez-Pa{\~{n}}eda, {A Unified
  Abaqus Implementation of the Phase Field Fracture Method Using Only a User
  Material Subroutine}, Materials 14 (2021) 1913.

\bibitem{Navidtehrani2021}
Y.~Navidtehrani, C.~Betegón, E.~Martínez-Pañeda, A simple and robust abaqus
  implementation of the phase field fracture method, Applications in
  Engineering Science 6 (2021) 100050.

\bibitem{TAFM2020}
P.~K. Kristensen, E.~Mart{\'{i}}nez-Pa{\~{n}}eda, {Phase field fracture
  modelling using quasi-Newton methods and a new adaptive step scheme},
  Theoretical and Applied Fracture Mechanics 107 (2020) 102446.

\bibitem{Jiang2012}
C.~P. Jiang, X.~F. Wu, J.~Li, F.~Song, Y.~F. Shao, X.~H. Xu, P.~Yan, {A study
  of the mechanism of formation and numerical simulations of crack patterns in
  ceramics subjected to thermal shock}, Acta Materialia 60~(11) (2012)
  4540--4550.

\bibitem{Sicsic2014}
P.~Sicsic, J.~J. Marigo, C.~Maurini, {Initiation of a periodic array of cracks
  in the thermal shock problem: A gradient damage modeling}, Journal of the
  Mechanics and Physics of Solids 63~(1) (2014) 256--284.

\bibitem{RUAN2023105169}
H.~Ruan, S.~Rezaei, Y.~Yang, D.~Gross, B.-X. Xu, A thermo-mechanical
  phase-field fracture model: Application to hot cracking simulations in
  additive manufacturing, Journal of the Mechanics and Physics of Solids 172
  (2023) 105169.

\bibitem{Tang2016}
S.~B. Tang, H.~Zhang, C.~A. Tang, H.~Y. Liu, {Numerical model for the cracking
  behavior of heterogeneous brittle solids subjected to thermal shock},
  International Journal of Solids and Structures 80 (2016) 520--531.

\bibitem{Chu2017}
D.~Chu, X.~Li, Z.~Liu, {Study the dynamic crack path in brittle material under
  thermal shock loading by phase field modeling}, International Journal of
  Fracture 208~(1-2) (2017) 115--130.

\bibitem{MANDAL2021113648}
T.~K. Mandal, V.~P. Nguyen, J.-Y. Wu, C.~Nguyen-Thanh, A.~{de Vaucorbeil},
  Fracture of thermo-elastic solids: Phase-field modeling and new results with
  an efficient monolithic solver, Computer Methods in Applied Mechanics and
  Engineering 376 (2021) 113648.

\bibitem{DelPiero1989}
G.~{Del Piero}, {Constitutive equation and compatibility of the external loads
  for linear elastic masonry-like materials}, Meccanica 24~(3) (1989) 150--162.

\bibitem{YOSHIOKA2016137}
K.~Yoshioka, B.~Bourdin, A variational hydraulic fracturing model coupled to a
  reservoir simulator, International Journal of Rock Mechanics and Mining
  Sciences 88 (2016) 137--150.

\bibitem{Zhou2018c}
S.~Zhou, X.~Zhuang, T.~Rabczuk, {A phase-field modeling approach of fracture
  propagation in poroelastic media}, Engineering Geology 240 (2018) 189--203.

\bibitem{IJP2021}
M.~Isfandbod, E.~Mart{\'{i}}nez-Pa{\~{n}}eda, {A mechanism-based multi-trap
  phase field model for hydrogen assisted fracture}, International Journal of
  Plasticity 144 (2021) 103044.

\bibitem{mandal2024computational}
T.~K. Mandal, J.~Parker, M.~Gagliano, E.~Mart{\'\i}nez-Pa{\~n}eda,
  Computational predictions of weld structural integrity in hydrogen transport
  pipelines, International Journal of Hydrogen Energy (2024).

\bibitem{Duddu2014}
R.~Duddu, {Numerical modeling of corrosion pit propagation using the combined
  extended finite element and level set method}, Computational Mechanics 54~(3)
  (2014) 613--627.

\bibitem{Gao2020}
H.~Gao, L.~Ju, R.~Duddu, H.~Li, {An efficient second-order linear scheme for
  the phase field model of corrosive dissolution}, Journal of Computational and
  Applied Mathematics 367 (2020) 112472.

\bibitem{Ernst2002}
P.~Ernst, R.~C. Newman, {Pit growth studies in stainless steel foils. I.
  Introduction and pit growth kinetics}, Corrosion Science 44~(5) (2002)
  927--941.

\end{thebibliography}


\end{document}